\documentclass{article}
\usepackage[english]{babel}
\usepackage[utf8]{inputenc}
\usepackage{color}

\usepackage{graphicx} %% figures
\usepackage{amsmath} %% maths
\usepackage{amssymb} %% maths
\usepackage{amsthm} %% maths
\usepackage{tabularx} %% tableaux largeur automatique
\usepackage{siunitx} %% unites
\usepackage{algorithm}
\usepackage{algorithmic} %% algos
\usepackage{caption} %% captions et subcaptions (a) (b) (c) etc.
\usepackage{subcaption} %% captions et subcaptions (a) (b) (c) etc.
\usepackage{filecontents} %% bibliographie dans le fichier tex
\usepackage{epstopdf}
\usepackage{authblk}

\everymath{\displaystyle}
\graphicspath{{img/}{Figures/}}

\providecommand{\keywords}[1]{\textbf{\textit{Keywords:}} #1}

\newtheorem{theorem}{Theorem}[section]

\newtheorem{definition}{Definition}[section]
\newtheorem{corollary}{Corollary}[definition]
\theoremstyle{remark}
\newtheorem*{notation}{Notation}

\title{A new level set-finite element formulation for anisotropic grain boundary migration}% Force line breaks with \\

\author[1]{J.~Fausty\thanks{corresponding author}}
\author[1]{B.~Murgas}
\author[1]{S.~Florez}
\author[1]{N.~Bozzolo}
\author[1]{M.~Bernacki}

\affil[1]{Mines-ParisTech, PSL-Research University, CEMEF – Centre de mise en forme des mat\'{e}riaux, CNRS UMR 7635, CS 10207 rue Claude Daunesse, 06904 Sophia Antipolis Cedex, France}%

\begin{document}
\maketitle

\begin{abstract}

  Grain growth in polycrystals is one of the principal mechanisms that take place during heat treatment of metallic components. This work treats an aspect of the anisotropic grain growth problem. By applying the first principles of thermodynamics and mechanics, an expression for the velocity field of a migrating grain boundary with an inclination dependent energy density is expressed. This result is used to generate the first, to the authors' knowledge, analytical solution (for both the form and kinetics) to an anisotropic boundary configuration. This new benchmark is simulated in order to explore the convergence properties of the proposed level-set finite element numerical model in an anisotropic setting. Convergence of the method being determined, another configuration, using a more general grain boundary energy density, is investigated in order to show the added value of the new formulation.

\end{abstract}


\section{Introduction}

During metal forming operations the microstructures of metallic components are modified by a host of phenomena ranging from solid state transformations to recrystallization \cite{humphreys2012recrystallization}. Perhaps the most critical effect of these mechanisms is the grain boundary motion they induce. Since the in-service properties of metal pieces depend on the material's microstructural characteristics (grain size, crystal orientation, composition, etc\ldots) \cite{suttonballuf2006}, it is important to study how these boundaries evolve under thermo-mechanical loads. Crystalline interfaces migrate differently during the different stages of annealing \cite{humphreys2012recrystallization}: deformation, recovery, recrystallization and grain growth can take place. These dynamics have been widely studied experimentally and numerically. Even so, the long characteristic times associated with grain growth allow investigators to decouple its effects from other processes. This is most likely the reason for which the theory of grain growth is the most established in monographs on the subject.

Grain boundary motion during grain growth is thought to be driven by the reduction of the interfacial free energy \cite{herring1999surface}. Classical models for grain growth in polycrystals use homogenized grain boundary properties to describe crystal interfaces (i.e. constant energy density, constant mobility, etc\ldots) \cite{anderson1984computer, gao1996real, lazar2011more, bernacki2011level, garcke1999multiphase}. However, at the mesoscopic scale, the grain boundary can be parameterized by five macroscopic crystalline parameters: a boundary plane unit normal vector and a misorientation element \cite{suttonballuf2006}. The main challenge in the current study of grain boundary motion is the dependence of intrinsic properties such as the grain boundary energy and mobility on these multiple structural parameters. Notwithstanding the difficulty in determining both the energy and mobility of a crystalline interface experimentally \cite{Holm01, Rohrer2010, Adams1997, Morawiec2000, Saylor2003} or numerically \cite{OlmstedI2009, OlmstedII2009}, the grain boundary configuration space is itself highly non-Euclidean. Indeed, current work towards elucidating the structure of this space has been oriented towards defining five dimensional identification spaces \cite{olmsted2009new} as well as defining higher dimensional embeddings into the unit octonions for example \cite{francis2019geodesic}.

In order to better predict microstructural evolution using numerical models, the intrinsic properties of the crystalline interface must be taken into account. However, many nuances of models taking into account different aspects of boundary variability can be developed. The authors have chosen to differentiate three classes of models and will refer to them using the terminology: isotropic, heterogeneous and anisotropic. In isotropic models, boundary properties are defined as constants for the entire system. While being able to reproduce mean value evolutions, such as mean grain size or even grain size distributions, rather efficiently, local heterogeneities in microstructures, such as the twin boundary, can not be modeled correctly \cite{anderson1984computer, gao1996real, lazar2011more, bernacki2011level, garcke1999multiphase}. Heterogeneous models may employ homogenized intrinsic properties along each grain boundary but differentiate different boundaries between each other \cite{rollett1989simulation, hwang1998simulation, upmanyu2002boundary, Fausty2018, zollner2019texture, kazaryan2002grain, miyoshi2016validation, chang2019effect, miyoshi2019accuracy, miessen2015advanced, Fausty2020}. As such, in a polycrystalline setting, the misorientation dependence of boundary properties can be modeled by these methods but not the inclination dependence. Fully anisotropic approaches attempt to take into account the five parameter dependence of grain boundary properties and as such constitute the most general of the three types.

In the journey towards these fully anisotropic models, able to account for general energy densities and mobilities, this work is constrained to treating the anisotropic grain boundary energy density in a one boundary setting. As such, the energy density function will be inclination dependent and the mobility will remain constant. This constraint on the energy density is rather easily scaled up to the polycrystalline case since a grain boundary, by definition, has a constant misorientation in fully recrystallized microstructures and thus any property variation can be fully attributed to changes in the boundary plane. As such, the developments here can be readily integrated into heterogeneous models, such as \cite{Fausty2018, chang2019effect}, in order to model anisotropic grain boundary energy densities. The homogeneity of the mobility is another matter however. Very few investigations deal with fully anisotropic mobilities mostly because the mobility of the boundary is really a derived notion which does not have a clear definition. Indeed, the grain boundary mobility, contrary to the thermodynamic definition of the energy density, is a kinetic parameter which is often fitted to produce observed migration rates. In this work a global grain boundary mobility definition will be proposed in relation to a normalized rate of energy dissipation. However, the question of using fully anisotropic and possibly tensorial values for the grain boundary mobility is still an open one.

As such, this manuscript, starting from a differential geometry description of the grain boundary and thermo-dynamical first principles, proposes an expression for the velocity of a migrating grain boundary in a general anisotropic energy density setting. This velocity is then applied to the transport equation of the level set method. In order to test this new formulation, an analytical benchmark using a collapsing ellipse is developed. Subsequently, a finite element level set numerical model is proposed to simulate this analytical test case. This new numerical model is then used to compute dynamics in a more general boundary energy density configuration.

\section{The interface model} \label{sec:BenchmarkDevelopment}

The model presented here is developed using elements from the field of differential geometry \cite{lee2003graduate, spivak2005comprehensive}. While more rudimentary mathematics can treat problems dealing with surfaces, the language of differential geometry seems like the correct one to treat the anisotropic problem. The necessary mathematical tools are briefly introduced before developing the formalism itself. While these tools might be well-known by experts, the readership specializing in physical metallurgy will, most likely, not be familiar with the terminology. For this reason, a consequential section of the text is devoted to definitions of known mathematical concepts.

\subsection{Tools and Definitions} \label{sec:TopoBasis}

\begin{definition}
  A smooth $n$-manifold $\mathcal{M} = (M, \mathcal{O}, \mathcal{A})$ is a triplet comprised of a set $M$, a given topology $\mathcal{O}$ on the set $M$ and a smooth atlas $\mathcal{A}$ made up of smooth charts.
\end{definition}

The $C^{n}$ manifold description of interfaces is chosen because it is technically the minimal structure with which one must endow a space in order to define derivatives. One could of course weaken the smooth condition to a $C^{2}$ or perhaps even $C^{1}$ constraint, however, for the sake of simplicity, smoothness ($C^{\infty}$) is considered here.

\begin{notation}
  Let $C^{\infty}(\mathcal{M})$ is the set of all smooth functions that can be defined on the smooth manifold $\mathcal{M}$.
\end{notation}

For the following definitions let $\mathcal{M}$ be a smooth manifold.

\begin{definition}
  The tangent space $T_{p}\mathcal{M}$ at the point $p \in \mathcal{M}$ is the vector space comprised of elements $X$ such that there exists $C$ a smooth curve of $\mathcal{M}$
\begin{gather*}
  \begin{array}{rl}
    C:& \mathbb{R} \rightarrow M\\
      & t \mapsto C(t)
  \end{array}
\end{gather*}
with $C(0) = p$ and
\begin{gather*}
  \begin{array}{rl}
    X:& C^{\infty}(\mathcal{M}) \rightarrow \mathbb{R}\\
      & f \mapsto Xf := \dfrac{d}{d t}(f \circ C)(0)
  \end{array}
\end{gather*}
\end{definition}

The elements of the tangent space, $X$, are also called \emph{tangent vectors}. Indeed, the elements of the tangent space to a point in the manifold and the classical notion of tangent vectors in Euclidean space are related. As an example using these definitions, if one chooses a chart $(U, x) \in \mathcal{A}$ such that $p \in U$ and a function $f \in C^{\infty}(\mathcal{M})$ then and element $X \in T_{p}\mathcal{M}$ acts on $f$ through its equivalent curve $C$

\begin{gather*}
  \begin{array}{rl}
    Xf &= \dfrac{d}{d t}(f \circ C)\\[0.3cm]
       &= \dfrac{d}{d t}(f \circ x^{-1} \circ x \circ C)
  \end{array}
\end{gather*}

which, using the multidimensional chain rule brings one too

\begin{gather*}
  Xf = \dfrac{d}{d t}(x^{i} \circ C) \partial_{i}(f \circ x^{-1})
\end{gather*}

where $x^{i}$ is the $i$th component function of the chart $x$, $\partial_{i}$ is the derivative operator of a multidimensional function with respect to its $i$th component and the \emph{Einstein summation convention} is in effect, which will be implied from here on unless stated otherwise. 

\begin{theorem}
One may construct an orthonormal basis for $T_{p}\mathcal{M}$ with the vectors $\left\{\frac{\partial}{\partial x^{i}}, i = 1,\ldots, n\right\}$ defined as

\begin{gather}
  \dfrac{\partial}{\partial x^{i}} f := \partial_{i}(f \circ x^{-1})
\end{gather}

As such, for any element of $X \in T_{p}\mathcal{M}$, one may define its components $\{X^{i} \in \mathbb{R}, i=1,\ldots,n\}$ in this basis, using

\begin{gather}
  X^{i} = \dfrac{d}{dt}(x^{i} \circ C)(0)
\end{gather}
where $C$ is the curve associated with $X$. As such, its decomposition is written

\begin{gather}
  X = X^{i}\dfrac{\partial }{\partial x^{i}}
\end{gather}

\end{theorem}

Seeing as $T_{p}\mathcal{M}$ is a vector space, it admits a dual space.

\begin{definition}
  The dual vector space $T^{*}_{p}\mathcal{M}$ or \emph{co}-tangent space to the tangent space $T_{p}\mathcal{M}$ is the space of \emph{linear} maps $\omega$ such that
  \begin{gather*}
    \begin{array}{rl}
      \omega: & T_{p}\mathcal{M} \rightarrow \mathbb{R}\\
              & X \mapsto \omega(X)
    \end{array}
  \end{gather*}
\end{definition}

More generally, the local tensor spaces are constructed from the tangent space and its dual.

\begin{definition}
  The space of $(q, s)$-tensors, $(q,s) \in \mathbb{N}^{2}$ , at $p\in \mathcal{M}$ is defined as
  \begin{gather*}
    T_{p,s}^{q}\mathcal{M} := \underset{s}{\underbrace{T^{*}_{p}\mathcal{M} \otimes \cdots \otimes T^{*}_{p}\mathcal{M}}}\otimes\underset{q}{\underbrace{T_{p}\mathcal{M} \otimes \cdots \otimes T_{p}\mathcal{M}}}
  \end{gather*}
  where $\otimes$ is the tensor product of spaces.
\end{definition}

From the tangent spaces at each point of $\mathcal{M}$ the tangent bundle can be constructed.

\begin{definition}
  Let $T\mathcal{M}$ be defined as
  \begin{gather*}
    T\mathcal{M} = \underset{p\in\mathcal{M}}{\bigcup}(p, T_{p}\mathcal{M})
  \end{gather*}
  such that the tangent bundle $(T\mathcal{M}, \mathcal{M}, \pi)$ is defined as
  \begin{gather*}
    T\mathcal{M} \overset{\pi}{\longrightarrow} \mathcal{M}
  \end{gather*}
  where $\pi$ is a continuous surjective map.
\end{definition}

Analogously, the $(q, s)$-tensor bundles $(T_{s}^{q}\mathcal{M}, \mathcal{M}, \pi_{s,q})$ are defined in the same manner.

\begin{definition}
  A section of a bundle $(E, B, \pi)$ is a continuous map $\sigma$ such that
  \begin{gather*}
    \begin{array}{rl}
      \sigma: & B \rightarrow E\\
              & \pi(\sigma(p)) = p
    \end{array}
  \end{gather*}
\end{definition}

Colloquially, the sections of the tangent bundle are called \emph{vector fields} and in the same manner sections of the tangent bundles are called \emph{tensor fields}.

\begin{notation}
  $\Gamma(T_{s}^{q}\mathcal{M})$ is the space of all smooth sections of the bundle $(T_{s}^{q}\mathcal{M}, \mathcal{M}, \pi_{s,q})$.
\end{notation}

\begin{definition}
  A Riemannian $n$-manifold $(\mathcal{M}, g)$ is a smooth $n$-manifold $\mathcal{M}$ equipped with a symmetric $(0,2)$-tensor field $g \in \Gamma(T_{2}^{0}\mathcal{M})$, called a metric, such that $\forall p \in M \; g(p)$ is a positive-definite tensor.
\end{definition}
  
The positive definiteness of $g$ means that for any $X \in T_{p}\mathcal{M}, X \ne 0$
\begin{gather*}
  g(p)(X, X) > 0
\end{gather*}
$\forall p \in M$.

Riemannian manifolds are of general interest since the metric structure defines inner products on the tangent spaces. As such, a Riemannian manifold is convenient for defining lengths of curves and more general measures of volume. Indeed, this metric structure is what allows one to define the Riemannian integral on the manifold.

\begin{definition}
  A differential $q$-form $\omega$ on a smooth manifold is a completely anti-symmetric $(0,q)$-tensor field.
\end{definition}

\begin{corollary}
\label{cor:Volumeform}
  The volume form $dM$ of an oriented Riemmannian $n$-manifold $(\mathcal{M}, g)$ is the differential $q$-form such that for a given chart $(U, x) \in \mathcal{A}$ the volume form may be expressed as
  \begin{gather*}
    dM = \sqrt{\det(g)}dx^{1}\wedge \cdots \wedge dx^{n}
  \end{gather*}
  where $\det(g)$ is the determinant of the matrix composed by the components of $g$ in the chart $(U, x)$, $\{dx^{i}, i=0,\ldots,n\}$ is the dual basis of the co-vector space and $\wedge$ is the exterior product of differential forms.
\end{corollary}

Using this machinery, any function can be integrated over the manifold. One is also able to define a relatively straightforward connection on the space called the \emph{Levi-Civita} connection.

\begin{definition}
  A connection $\nabla$ over a bundle $(E, B, \pi)$ is a set of linear maps
  \begin{align*}
    \nabla: & \Gamma(T_{s}^{q}B) \rightarrow \Gamma(T_{s}^{q}B \otimes T^{*}B)
  \end{align*}
  that respect the Leibniz rule, $f \in C^{\infty}(B), \sigma \in \Gamma(T_{s}^{q}B), \tau \in \Gamma(T_{i}^{j}B)$
  \begin{gather}
    \nabla(f\sigma) = \sigma \otimes df + f \nabla\sigma\\
    \nabla(\tau \otimes \sigma) = \nabla \tau \otimes \sigma + \tau \otimes \nabla\sigma
  \end{gather}
  where $df$ is the classic differential of a smooth function $df = \frac{\partial f}{\partial x^{i}}dx^{i}$.
\end{definition}

\begin{corollary}
  From a given connection $\nabla$, one may construct a covariant derivative 
  \begin{align*}
    \nabla_{\cdot}: &\Gamma(TB) \times \Gamma(T_{s}^{q}B) \rightarrow \Gamma(T_{q}^{q}B)\\
                    &\nabla_{\cdot}(X,\sigma) = \nabla_{X}\sigma = (\nabla \sigma)(X)
  \end{align*}
  where, when working in a chart, one may use
  \begin{align*}
    (\nabla_{X}\sigma)^{k\ldots}_{j\ldots} = (\nabla\sigma)^{k\ldots}_{j\ldots i} X^{i} = \nabla_{i}\sigma^{k\ldots}_{j\ldots} X^{i}
  \end{align*}
\end{corollary}

\begin{definition}
  The Levi-Civita connection $\nabla$ on a Riemannian manifold $(\mathcal{M}, g)$ is the unique connection on the tensor bundles which satisfies
  \begin{gather*}
    \nabla g = 0
  \end{gather*}
  and has no torsion.
\end{definition}

\subsection{An energetic embedded smooth manifold}

\begin{figure}
  \centering
  \includegraphics[scale=0.35]{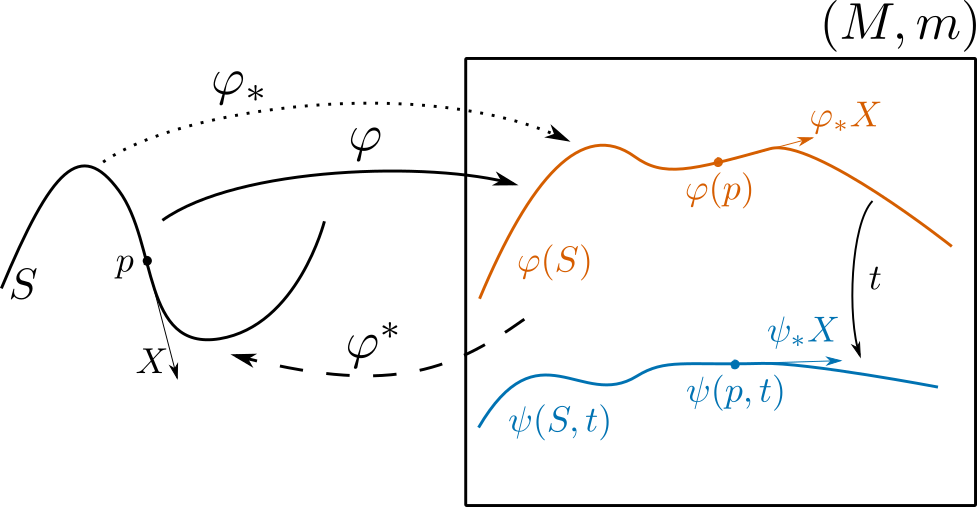}
  \caption{Diagram of the embedding $\varphi$.}\label{fig:embedding}
\end{figure}

Let $\mathcal{M} = (M, \mathcal{O}_{M}, \mathcal{A}_{M})$ be a Riemannian $n$-manifold with metric $m$ and $\mathcal{S} = (S, \mathcal{O}_{S}, \mathcal{A}_{S})$ is a smooth $s$-manifold with $n \ge s$. Let $\varphi$ be a smooth embedding from $\mathcal{S}$ to $\mathcal{M}$

 \begin{gather}
     \begin{array}{rl}
       \varphi:& S \rightarrow M\\
               & S \equiv_{homeo} \varphi(S)
     \end{array}
 \end{gather}

 where $\equiv_{homeo}$ describes a homeomorphism equivalence and Figure \ref{fig:embedding} provides an illustration. The embedding also provides a map from the tangent bundle of $\mathcal{S}$ to the tangent bundle of $\mathcal{M}$ called the \emph{push-forward}.

\begin{definition}
  The push-forward $\varphi_{*}$ of a map $\varphi$ from $\mathcal{S}$ to $\mathcal{M}$, two smooth manifolds, is the linear map such that
  \begin{gather*}
    \begin{array}{rl}
      \varphi_{*}: & T\mathcal{S} \rightarrow T\mathcal{M}\\
                   & (p, X) \mapsto (\varphi(p), \varphi_{*}X)\\
                   & (\varphi_{*}X) f := X(f \circ \varphi)
    \end{array}
  \end{gather*}
for $f \in C^{\infty}(\mathcal{M})$
\end{definition}

Much in the same manner, the embedding generates a map from the co-tangent bundles $T_{q}^{0}\mathcal{M}$ to $T_{q}^{0}\mathcal{S}$.

\begin{definition}
  The pull-back $\varphi^{*}$ of a map $\varphi$ from $\mathcal{S}$ to $\mathcal{M}$, two smooth manifolds, is the linear map such that
  \begin{align*}
      \varphi^{*}: & T_{q}^{0}\mathcal{M}|_{\varphi(S)} \rightarrow T_{q}^{0}\mathcal{S}\\
                   & (\varphi(p), \sigma) \mapsto (p, \varphi^{*}\sigma)\\
                   & (\varphi^{*}\sigma)(X^{(1)}, \ldots, X^{(q)}) = \sigma(\varphi_{*}X^{(1)}, \ldots, \varphi_{*}X^{(q)})
  \end{align*}
\end{definition}

Using the charts $(U, x) \in \mathcal{A}_{S}$ and $(V, Z) \in \mathcal{A}_{M}$ and using the convention by which objects in $\mathcal{M}$ are indexed by Greek letters and objects in $\mathcal{S}$ are indexed by Latin letters one can express the components of the pushforward of a vector $X \in T_{p}\mathcal{S}$ using its action on a function $f \in C^{\infty}(\mathcal{M})$

\begin{align*}
    (\varphi_{*}X) f &= (\varphi_{*}X)^{\alpha}\dfrac{\partial f}{\partial Z^{\alpha}}
\end{align*}

and

\begin{align*}
    (\varphi_{*}X) f &= X^{i}\dfrac{\partial (Z^{\alpha} \circ \varphi)}{\partial x^{i}} \dfrac{\partial f}{\partial Z^{\alpha}}
\end{align*}

 which, defining

\begin{align*}
  \varphi^{\alpha} : & S \rightarrow \mathbb{R}\\
                     & p \rightarrow Z^{\alpha}(\varphi(p))
\end{align*}

 leads to, through identification,

\begin{equation}
  (\varphi_{*}X)^{\alpha} = X^{i}\dfrac{\partial \varphi^{\alpha}}{\partial x^{i}}
\end{equation}

Using the pull-back one may define the metric $g$ on $\mathcal{S}$ and therefore turn $\mathcal{S}$ into a Riemannian manifold $(\mathcal{S}, g)$ with 

\begin{gather}
  g(p) = (\varphi^{*}m)(\varphi(p))
\end{gather}

which, using the same charts as above and two vectors $(X, Y) \in T_{p\in S}\mathcal{S}$

\begin{align*}
  g(X, Y) &= (\varphi^{*}m)(X, Y)\\
  g_{ij}X^{i}Y^{j} &= m(\varphi_{*}X, \varphi_{*}Y)\\
          &= m_{\alpha\beta}(\varphi_{*}X)^{\alpha}(\varphi_{*}Y)^{\beta}\\
          &= m_{\alpha\beta}\dfrac{\partial \varphi^{\alpha}}{\partial x^{i}}\dfrac{\partial \varphi^{\beta}}{\partial x^{j}}X^{i}Y^{j} .
\end{align*}

This results defines the components of the induced metric by identification

\begin{equation}
  g_{ij} = m_{\alpha\beta}\dfrac{\partial \varphi^{\alpha}}{\partial x^{i}}\dfrac{\partial \varphi^{\beta}}{\partial x^{j}}
\end{equation}

Now, let $\mathcal{B}$ be an internal property space. For example, $\mathcal{B}$, when applied to grain boundaries, would be the five-dimensional space created by the misorientation and inclination parameters $(M, n)$. Let

\begin{gather}
  S\mathcal{B} = \underset{p\in S}{\bigcup} (p, \mathcal{B}) = S \times \mathcal{B}
\end{gather}

and define the trivial property bundle $(S\mathcal{B}, \mathcal{S}, \pi_{B})$

\begin{gather}
  S\mathcal{B} \overset{\pi_{B}}{\longrightarrow} S
\end{gather}

such that a section $b \in \Gamma(S\mathcal{B})$ of the property bundle describes exactly the properties of the $s$-manifold at each point. If one was to define an energy density map

\begin{align*}
  \gamma: & \mathcal{B} \rightarrow \mathbb{R}^{+}
\end{align*}

then one could calculate the energy density at any point $p \in S$ through the property field as $\gamma(b(p))$. Given that $(\mathcal{S}, g)$ is now a Riemannian manifold, this energy density can be integrated in order to give the total interface energy $I$ of the embedding as

\begin{align*}
  I = \int_{S} (\gamma \circ b) dS
\end{align*}

The model developed here for the interface is thus a triple $(\mathcal{S}, \varphi, b)$ from which, with an energy density map $\gamma$, the total energy of the interface may be expressed. By design, this model puts no lower bound on $s$. Therefore, this structural model is readily generalized to objects that are not strictly interfaces but can be of lower dimension, such as lines if $n \ge 3$. This is an important aspect of this model, even if it might be out of the scope of this article, if ever one was to attempt to attribute properties and therefore energies to other defects in the polycrystal microstructure.

\subsection{Interface thermodynamics} \label{sec:LSImmersion}

Consider a closed thermodynamic system made up of a Riemannian $n$-manifold $(\mathcal{M}, m)$ of volume $V$, an embedded interface $(\mathcal{S}, \varphi, b)$ with a boundary energy density map $\gamma$, in a heat bath at absolute temperature $T$, a system entropy $\eta$ and a homogeneous pressure field $p$. One defines an idealized case with the following conditions: 

\begin{itemize}
	\item	isothermal heat treatment - $T$ and $p$ are constant
	\item	the time is parameterized so that $t \in [0; 1]$
	\item	and the system is closed
\end{itemize}

the change in the internal energy during the free evolution of the system is defined as

\begin{align}
  \dfrac{d U}{dt} = \dfrac{d I}{dt} + T\dfrac{d \eta}{dt} - p\dfrac{d V}{dt}
\end{align}

If one looks at the evolution of the Gibbs free energy:

\begin{align*}
  \dfrac{dG}{dt} &= \dfrac{d U}{dt} + \dfrac{d (pV)}{dt} - \dfrac{d (T\eta)}{dt}\\
                 &= \dfrac{d I}{dt} + V\dfrac{dp}{dt} - \eta \dfrac{dT}{dt},
\end{align*}

which considering the isobaric and isothermal conditions,

\begin{align}
  \dfrac{dG}{dt} = \dfrac{dI}{dt} \label{eq:freeenergy}
\end{align}

such that the change in free energy is exactly equal to the change in interface energy. Additionally, according to the second law of thermodynamics, the closed system must tend to minimize its free energy $G$ such that

\begin{gather}
  \dfrac{d I}{dt} \le 0.
\end{gather}

The principle of least action affirms that the energy dissipation must be maximal and thus $\frac{d I}{dt}$ must be minimal $\forall t \in [0; 1]$.

The flow of the interface, $\forall t \in [0; 1]$, is defined as 

\begin{align*}
  \psi: & S \times [0; 1] \rightarrow M\\
        & (p, t) \mapsto \psi(p, t)\\
        & \psi(p, 0) = \varphi(p)
\end{align*}

thus the embedding $\mathcal{S}$ is defined.

As the interface evolves only its geometry changes. This means that the misorientation remains constant. Following this statement, the property field $b$ should depend, in some manner, on the embedding. If we consider that the boundary is parameterized by the misorientation-inclination pair, $b = (M, n) \in \mathcal{B}$, the misorientation is invariant

\begin{gather}
  \dfrac{d M}{dt} = 0 .
\end{gather}

However, since the inclination $n$ of the boundary is a geometrical characteristic of the boundary it does change. The $n$ field depends exclusively on the embedding and they are related by means of the push-forward of the tangent vectors to the interface. For any $X \in T_{p}\mathcal{S}$ at any time $t$ the value of $n(\psi(p, t))$ is in completely determined by 

\begin{align*}
  \left\{
  \begin{array}{l}
    m(n, n) = 1\\
    m(n, \psi_{*}X) = 0
\end{array}
\right .
\end{align*}
or, in component form, 
\begin{align*}
  \left\{
    \begin{array}{l}
      m_{\alpha\beta}n^{\alpha}n^{\beta} = 1\\
      m_{\alpha\beta}n^{\alpha}\dfrac{\partial \psi^{\beta}}{\partial x^{i}}X^{i} = 0
    \end{array}
  \right .
\end{align*}

which leads to,

\begin{gather}
  m_{\alpha\beta}n^{\alpha}\dfrac{\partial \psi^{\beta}}{\partial x^{i}} = 0, \quad \forall i = 1,\ldots, s
\end{gather}

In the context of the one boundary, and thus one misorientation, the following simplification holds

\begin{gather}
  \gamma\left(M, n\left(\ldots, \dfrac{\partial \psi^{\alpha}}{\partial x^{i}}, \ldots\right)\right) = \gamma\left(\ldots, \dfrac{\partial \psi^{\alpha}}{\partial x^{i}}, \ldots\right) .
\end{gather}

For $q = \psi(p, t)$ and with $f \in C^{\infty}(\mathcal{M})$ the velocity field is defined as

\begin{align*}
  (v f)(q) &= \dfrac{d}{dt}(f \circ \psi(p, \cdot))(t)\\
           &= \dfrac{d \psi^{\alpha}}{dt}(p, t)\dfrac{\partial f}{\partial Z^{\alpha}}(q)\\
           &= v^{\alpha}\dfrac{\partial f}{\partial Z^{\alpha}}
\end{align*}

such that, by identification

\begin{gather}
  v^{\alpha}(\psi(p, t)) = \dfrac{d \psi^{\alpha}}{dt}(p, t).
\label{eqn:VelComp}
\end{gather}

Using the statement in corollary \ref{cor:Volumeform}, equation \eqref{eqn:VelComp} and considering the Levi-Civita connection $\nabla$ of $(\mathcal{S}, g)$, knowing that $\frac{\partial \psi^{\alpha}}{\partial x^{i}} = \nabla_{i}\psi^{\alpha}$, the energy dissipation may be defined as

\begin{align*}
  \dfrac{d I}{dt} &= \int_{S}\dfrac{d}{dt}\left(\gamma dS\right)\\
                  &= \int_{S} \dfrac{1}{\sqrt{\det(g)}}\dfrac{\partial (\gamma \sqrt{\det(g)})}{\partial \nabla_{i}\psi^{\alpha}}\nabla_{i}v^{\alpha}dS.
\end{align*}

Expressing 

\begin{align*}
  \dfrac{\partial (\gamma \sqrt{\det(g)})}{\partial \nabla_{i}\psi^{\alpha}} 
  &= \sqrt{\det(g)}\dfrac{\partial \gamma}{\partial \nabla_{i}\psi^{\alpha}} + \gamma \dfrac{\partial \sqrt{\det(g)}}{\partial \nabla_{i}\psi^{\alpha}},
\end{align*}

using Jacobi's formula to define the derivative of the determinant of matrix and defining the components of an inverse metric tensor as $(g^{-1})^{ij} = g^{ij}$, the energy dissipation may be rewritten as 

\begin{gather}
  \dfrac{dI}{dt} = \int_{S} \left(\dfrac{\partial \gamma}{\partial \nabla_{i}\psi^{\alpha}} + \gamma g^{iq}m_{\sigma\alpha}\nabla_{q}\psi^{\sigma}\right)\nabla_{i}v^{\alpha} dS
\end{gather}

One can define the boundary of $\mathcal{S}$ as $\partial \mathcal{S}$ and apply Stokes' theorem such that

\begin{align*}
  \dfrac{d I}{dt} = &\int_{\partial S}g_{ik}\tau^{k}\left(\dfrac{\partial \gamma}{\partial \nabla_{i}\psi^{\alpha}} + \gamma g^{iq}m_{\sigma\alpha}\nabla_{q}\psi^{\sigma}\right)v^{\alpha} d\partial S\\
                    & - \int_{S}\nabla_{i}\left(\dfrac{\partial \gamma}{\partial \nabla_{i}\psi^{\alpha}} + \gamma g^{iq}m_{\sigma\alpha}\nabla_{q}\psi^{\sigma}\right) v^{\alpha} dS
\end{align*}

where $\tau$ is the outside pointing unitary normal field to $\partial \mathcal{S}$.

In order to encapsulate the quantities of interest, we define the following restricted vector fields, $B \in \Gamma(T\mathcal{M}|_{\varphi(\partial S)})$ with components

\begin{align}
  B^{\alpha} &= m^{\alpha\beta}g_{ik}\tau^{k}\left(\dfrac{\partial \gamma}{\partial \nabla_{i}\psi^{\beta}} + \gamma g^{iq}m_{\sigma\beta}\nabla_{q}\psi^{\sigma}\right)
\end{align}

and $A \in \Gamma(T\mathcal{M}|_{\varphi(S)})$ with

\begin{align}
  A^{\alpha} = m^{\alpha\beta}\nabla_{i}\left(\dfrac{\partial \gamma}{\partial \nabla_{i}\psi^{\beta}} + \gamma g^{iq}m_{\sigma\beta}\nabla_{q}\psi^{\sigma}\right) \label{eq:anisovector}
\end{align}

such that

\begin{equation}
  \dfrac{d I}{dt}(t) = \int_{\partial S} m(B, v))|_{\psi(p, t)} d\partial S - \int_{S} m(A, v)|_{\psi(p, t)} dS
\end{equation}

Given the one interface restriction used in this work, the boundary of the interface $\partial \mathcal{S}$ can only be either empty $\partial \mathcal{S} = \emptyset$ or part of the boundary of the base manifold $\psi(\partial \mathcal{S}) \in \partial \mathcal{M}$. Using the $\partial \mathcal{S} = \emptyset$ case, the boundary term disappears and

\begin{gather}
  \dfrac{d I}{dt} = -\int_{S}m(A, v) dS,
\end{gather}

As such, the velocity of the boundary is that which minimizes the previous expression. The bilinear form

\begin{align}
  \langle \cdot, \cdot \rangle: & \Gamma(T\mathcal{M}|_{\psi(S, t)})^{2} \longrightarrow \mathbb{R}\\
                                & (X, Y) \mapsto \int_{S} m(X, Y) dS
\end{align}

defines an inner product on the $\Gamma(T\mathcal{M}|_{\psi(S,t)})$ space, turning it into a Hilbert space. As such, using the Cauchy-Schwarz inequality, one may show that the velocity field that minimizes the energy dissipation has the form

\begin{gather}
  v = \mu A \label{eq:velocity}, \quad \dfrac{d I}{dt} = - \mu \langle A, A \rangle
\end{gather}

with $\mu \in \mathbb{R}$ being classically the mobility of the boundary.

By replacing the interfacial energy dissipation term in equation \eqref{eq:freeenergy} with its equivalent expression $-\mu\langle A, A \rangle$ one determines a definition of the mobility parameter as

\begin{gather}
  \mu = -\dfrac{1}{\langle A, A \rangle}\dfrac{d G}{dt} .
  \label{eqn:MobilityDefinition}
\end{gather}

The mobility is thus proportional to a normalized value of the energy dissipation of the system. Thus, the mobility of the boundary appears as a kinetic parameter related to the capacity of the system to dissipate energy in the form of heat or work (by a contraction due to the excess volume of the boundaries for example). As such, the mobility of the grain boundaries may have more to do with the boundary conditions imposed on the system then previously imagined.

\subsection{From embeddings to level-set fields}

\begin{definition}
  A level-set map or function $\phi$ is a smooth scalar field over the smooth manifold $\mathcal{M}$ such that, given an embedding $\varphi: S \rightarrow M$
  \begin{gather}
    \phi(\varphi(p)) = 0
  \end{gather}
  $\forall p \in S$.
\end{definition}

Most often, one defines the level-set function as a signed distance function to the interface such that with

\begin{align*}
  d: & M \times M \rightarrow \mathbb{R}^{+}\\
     & (p, q) \mapsto \underset{C(p,q)}{\min}\int_{C(p, q)}dC
\end{align*}

where $C(p,q)$ is any curve from $p$ to $q$, one may then fix

\begin{align*}
  \phi(q) = \pm d(q, \varphi(S)) := \pm \underset{p\in S}{\min} \, d(q, \varphi(p))
\end{align*}

where one makes a choice of sign over the domains that the interface separates. The evolution of the interface is simulated by solving the transport equation 

\begin{equation}
  \dfrac{\partial \phi}{\partial t} + v \cdot \tilde{\nabla} \phi = 0
\end{equation}

everywhere in $\mathcal{M}$, where $\tilde{\nabla}$ is the Levi-Civita connection on the Riemannian manifold $(\mathcal{M}, m)$. One may replace the velocity field with the expression developed in the previous paragraphs after a suitable extension of the fields defined on $\mathcal{M}|_{\psi(S, t)}$ to the entire manifold. As such, 

\begin{gather}
  \dfrac{\partial \phi}{\partial t} + \mu A^{\alpha} \tilde{\nabla}_{\alpha}\phi = 0, \label{eq:bettertransport}
\end{gather}

Using the following identity

\begin{equation}
  \nabla_{j}\nabla_{i}\varphi^{\alpha}\tilde{\nabla}_{\alpha}\phi = - \nabla_{i}\varphi^{\alpha}\nabla_{j}\varphi^{\beta}\tilde{\nabla}_{\beta}\tilde{\nabla}_{\alpha}\phi, \label{eq:2ndderivs}
\end{equation}

derived from the orthogonality condition of tangent vectors and the gradient of the level-set, one may express the transport equation in a fully level-set form

\begin{align}
  \dfrac{\partial \phi}{\partial t} -\mu\left(\gamma m^{\alpha\beta} + \dfrac{\partial^{2} \gamma}{\partial \tilde{\nabla}_{\beta}\phi \partial \tilde{\nabla}_{\alpha} \phi}\right)\tilde{\nabla}_{\alpha}\tilde{\nabla}_{\beta}\phi = 0\label{eq:fulltransport}
\end{align}

where the full derivation is reported in \ref{app:Mfds2LS}. Of course, some direct analogies can be made with the derivations proposed in \cite{herring1999surface}.

\subsection{Constraints on the anisotropic grain boundary energy density function} \label{sec:GammaConditions}

Let $D \in \Gamma(T^{2}_{0}\mathcal{M})$ be the symmetrized tensor such that

\begin{align}
  D^{\alpha\beta} := \gamma m^{\alpha\beta} + \dfrac{1}{2}\left(\dfrac{\partial^{2} \gamma}{\partial \tilde{\nabla}_{\beta}\phi \partial \tilde{\nabla}_{\alpha} \phi} + \dfrac{\partial^{2} \gamma}{\partial \tilde{\nabla}_{\alpha}\phi \partial \tilde{\nabla}_{\beta} \phi}\right) \label{eq:Ddefine}
\end{align}

where the tensor can be symmetrized because $\tilde{\nabla}_{\alpha}\tilde{\nabla}_{\beta}\phi$ is already symmetric. Now, equation \eqref{eq:fulltransport} is a purely diffusive equation with $\mu D$ as a diffusive coefficient tensor. As such, the well-posedness of the problem depends largely on the positive definiteness of $D$. For solutions to be unique, one must have $\forall \omega \in \Gamma(T^{*}\mathcal{M}), \omega \ne 0$

\begin{align}
  D(\omega, \omega) > 0 \label{eq:pdcondition}
\end{align}

and, therefore,

\begin{align}
  D^{\alpha\beta}\omega_{\alpha}\omega_{\beta} > 0. \label{eq:pdcomponents}
\end{align}

Given the arbitrariness of $\omega$, applying \eqref{eq:pdcomponents} to the basis vectors of the dual tangent spaces at each point, one quickly obtains (not using the summation convention)

\begin{align*}
  D_{\alpha\alpha}\omega_{\alpha}^{2} > 0\\
  D_{\alpha\alpha} > 0.
\end{align*}

More complex conditions exist for the off-diagonal components. For example, for $n = 2$ and $s = 1$, one can show that

\begin{align*}
  |D^{12}| &< \underset{(\omega_{1}, \omega_{2})}{\min} \dfrac{1}{2}\left|\dfrac{D^{11}\omega_{1}\omega_{1} + D^{22}\omega_{2}\omega_{2}}{\omega_{1}\omega_{2}}\right|
\end{align*}

which admits a unique minimum

\begin{align}
  |D^{12}| &< \sqrt{D^{11}D^{22}} .\label{eq:dxy}
\end{align}

As such, given that the $D$ tensor depends entirely on the grain boundary energy function $\gamma$, these constraints are actually directly transferable to the $\gamma$ function. Thus, in order to preserve uniqueness of the grain boundary flow, the anisotropy of the $\gamma$ function is restricted to maps which satisfy these conditions. While determining the space of functions that satisfy these relations would be a valuable discovery for the community, this kind of development is out of the scope of this article.

\section{An elliptical benchmark}

To the authors' knowledge, no analytical test case exists for the anisotropic one boundary setting of grain growth. Indeed, while the shrinking sphere is a viable benchmark for the isotropic case and the ``Grim Reaper'' \cite{Garcke1999} is very useful for testing heterogeneous models, no equivalent configuration has been developed for more general grain boundary energy densities. Theoretical studies have proven that minimal energy surfaces can be constructed for virtually any inclination dependent energy density function \cite{HoffmanI1972} using Wulff shapes, the kinetics with which these shapes should evolve, in a isolated grain undergoing coarsening for example, are completely unknown. Thus, these semi-analytical benchmarks remain incomplete cases for numerical testing. This section is devoted to generating such a completely analytical solution to the problem with constrained kinetics as well as definite morphology.

\subsection{The setting}

\begin{figure}
  \centering
  \includegraphics[scale=0.4]{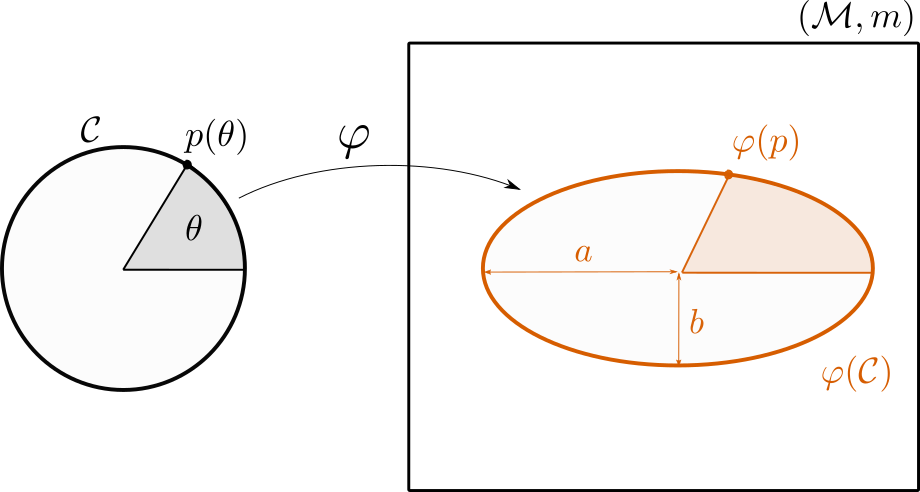}
  \caption{Ellipse embedding of the circle $\mathcal{C}$ into $\mathcal{M}$}\label{fig:ellipse}
\end{figure}

Consider a circle $\mathcal{C} = ([0; 2\pi], \mathcal{O}_{C}, \mathcal{A}_{C})$ as a smooth manifold with the circle topology and smooth structure and the Riemannian manifold $\mathcal{M} = (\mathbb{R}^{2}, \mathcal{O}_{std}, \mathcal{A}_{std}, m)$ equipped with the standard topology and differentiable structures and the flat metric $m$. Using the chart $([0; 2\pi], \theta) \in \mathcal{A}_{C}$ and the Cartesian chart $(\mathbb{R}^{2}, (x, y))\in \mathcal{A}_{std}$ one may construct the following elliptical embedding

\begin{align*}
  \varphi :& [0; 2\pi] \longrightarrow \mathbb{R}^{2}\\
           & \theta \mapsto (a\cos\theta, b\sin\theta)
\end{align*}

where $(a,b) \in \mathbb{R}^{2}$ and Figure \ref{fig:ellipse} illustrates this embedding.

All of the relevant geometrical information may thus be extracted from the embedding. The pushforward of the tangent space

\begin{align}
  \dfrac{\partial \varphi^{x}}{\partial \theta} &= -a\sin\theta\\
  \dfrac{\partial \varphi^{y}}{\partial \theta} &= b\cos\theta
\end{align}

and the induced metric tensor

\begin{align*}
  g_{ij} &= m_{\alpha\beta}\nabla_{i}\varphi^{\alpha} \nabla_{j}\varphi^{\beta}\\
  g_{\theta\theta} &= (\nabla_{\theta}\varphi^{x})^{2} + (\nabla_{\theta}\varphi^{y})^{2}
\end{align*}

such that

\begin{align}
  g_{\theta\theta} = a^{2}\sin^{2}\theta + b^{2}\cos^{2}\theta\label{eq:metric}
\end{align}

The Levi-Civita connection $\nabla$ is thus defined by

\begin{align*}
  \nabla_{\theta}g_{\theta\theta} &= 0\\
  \dfrac{\partial g_{\theta\theta}}{\partial \theta} - 2\Gamma^{\theta}_{\theta\theta}g_{\theta\theta} &= 0\\
  \Gamma_{\theta\theta}^{\theta} &= \dfrac{1}{2g_{\theta\theta}}\dfrac{\partial g_{\theta\theta}}{\partial \theta}
\end{align*}

where $\Gamma_{ij}^{k}$ are Christoffel symbols. Therefore,

\begin{align}
  \Gamma_{\theta\theta}^{\theta} = \dfrac{(a^{2} - b^{2})\cos\theta\sin\theta}{a^{2}\sin^{2}\theta + b^{2}\cos^{2}\theta}
\end{align}

\subsection{A solution}

Now consider the boundary energy

\begin{align}
  \gamma(\theta) = G^{\theta\theta}g_{\theta\theta}\label{eq:ellipsegamma}
\end{align}

where $G$ is a $(2, 0)$-tensor field of $\mathcal{C}$ whose only component $G^{\theta\theta} \in \mathbb{R}$ is actually a constant in this chart. As such, using equations \eqref{eq:velocity} and \eqref{eq:anisovector} the velocity field of the minimizing energy flow is

\begin{align*}
  v^{\alpha} = \mu m^{\alpha\beta}\nabla_{i}\left(\dfrac{\partial \gamma}{\partial \nabla_{i}\psi^{\beta}} + \gamma g^{iq}m_{\sigma\beta}\nabla_{q}\varphi^{\sigma}\right)
\end{align*}

where, replacing with the expression for $\gamma$ in equation \eqref{eq:ellipsegamma}, one has

\begin{align*}
  v^{\alpha} &= \mu m^{\alpha\beta}\nabla_{i}\left(\dfrac{\partial G^{sk}g_{ks}}{\partial \nabla_{i}\psi^{\beta}} +  G^{sk}g_{ks}g^{iq}m_{\sigma\beta}\nabla_{q}\varphi^{\sigma}\right)\\
             &= \mu m^{\alpha\beta}\nabla_{i}\left(2G^{si}m_{\beta\zeta}\nabla_{s}\varphi^{\zeta} +  G^{sk}g_{ks}g^{iq}m_{\sigma\beta}\nabla_{q}\varphi^{\sigma}\right)\\
             &= \mu m^{\alpha\beta}\nabla_{\theta}\left(2G^{\theta\theta}m_{\beta\zeta}\nabla_{\theta}\varphi^{\zeta} +  G^{\theta\theta}g_{\theta\theta}g^{\theta\theta}m_{\sigma\beta}\nabla_{\theta}\varphi^{\sigma}\right)\\
             &= 3\mu G^{\theta\theta}\nabla_{\theta}\nabla_{\theta}\varphi^{\alpha}
\end{align*}

using 

\begin{align*}
  \nabla_{\theta}\nabla_{\theta}\varphi^{\alpha} = \dfrac{\partial^{2}\varphi^{\alpha}}{\partial\theta^{2}} - \Gamma_{\theta\theta}^{\theta}\dfrac{\partial \varphi^{\alpha}}{\partial \theta}
\end{align*}

one arrives at

\begin{align*}
  \left( \begin{array}{c}
      v^{x}\\
      v^{y}
  \end{array}\right) = 
  -3\mu G^{\theta\theta}\left( \begin{array}{c}
      a\cos\theta\\
      b\sin\theta
  \end{array}\right) - \dfrac{(a^{2} - b^{2})\cos\theta\sin\theta}{a^{2}\sin^{2}\theta + b^{2}\cos^{2}\theta}\left( \begin{array}{c}
      -a\sin\theta\\
      b\cos\theta
  \end{array}\right) 
\end{align*}

However, any tangential terms in the velocity, such as the second term in the above equation, have no influence on the flow of the interface such that the flow generated by the velocity field above is equivalent to the flow generated by

\begin{align}
  \left( \begin{array}{c}
      v^{x}\\
      v^{y}
  \end{array}\right) = 
  -3\mu G^{\theta\theta}\left( \begin{array}{c}
      a\cos\theta\\
      b\sin\theta
  \end{array}\right) 
\end{align}

Thus, turning $\varphi$ into a flow $\varphi:[0; 2\pi] \times [0; 1] \rightarrow \mathbb{R}^{2}$, one has

\begin{align*}
  \dfrac{d\varphi^{\alpha}}{dt}(\theta, t) = -3\mu G^{\theta\theta}\varphi^{\alpha}(\theta, t)
\end{align*}

for which there is only one solution

\begin{align*}
  \varphi^{\alpha}(\theta, t) = \varphi^{\alpha}(\theta, 0)e^{-3\mu G^{\theta\theta}t}
\end{align*}

leading to

\begin{align*}
  \left( \begin{array}{c}
      \varphi^{x}(\theta, t)\\
      \varphi^{y}(\theta, t)
  \end{array}\right) = 
  e^{-3\mu G^{\theta\theta}t}\left( \begin{array}{c}
      a\cos\theta\\
      b\sin\theta
  \end{array}\right)
\end{align*}

Now given that the minimizing energy flow of the embedding is just the original embedding multiplied by a time dependent function, the flow is actually simply shrinking the ellipse in a homothetic manner to the center $(0, 0)$ point of $\mathcal{M}$. Thus, assuming $a > b$, the eccentricity $e$ is a constant of the flow

\begin{align}
  e = \sqrt{1 - \left(\dfrac{\varphi^{y}(\frac{\pi}{2}, t)}{\varphi^{x}(0, t)}\right)^{2}} = \sqrt{1 - \left(\dfrac{e^{-3\mu G^{\theta\theta}t}b}{e^{-3\mu G^{\theta\theta}t}a}\right)^{2}} = \sqrt{1 - \left(\dfrac{b}{a}\right)^{2}}
\end{align}

and the scalar velocity of any point of the ellipse is

\begin{align}
  v(\theta, t) = \sqrt{(v^{x})^{2} + (v^{y})^{2}} = 3\mu G^{\theta\theta}e^{-3\mu G^{\theta\theta}t}\sqrt{a^{2}\cos^{2}\theta + b^{2}\sin^{2}\theta}
\end{align}

with, in particular,

\begin{align}
  v(0, t) = 3\mu G^{\theta\theta}e^{-3\mu G^{\theta\theta}t}a\\
  v(\frac{\pi}{2}, t) = 3\mu G^{\theta\theta}e^{-3\mu G^{\theta\theta}t}b
\end{align}

\section{The numerical model and test case applications}

In order to numerically simulate grain boundary configurations, a numerical model capable of representing boundary dynamics must be developed. As such, the following paragraphs are devoted to first reporting on the level set \cite{Merriman1994, Osher1988} finite element method applied to this kind of boundary transport. Subsequently, the analytical shrinking ellipse test case is simulated and convergence of the numerical model is studied. Finally, a more general grain boundary energy is applied to a circular case in order to compare the classical isotropic formulation for the velocity field with the expression proposed in this work.

\subsection{The finite element model}

In order to solve the minimizing energy flow for the level set function using the Finite Element (FE) method, the problem must first expressed in a weak form, then it can be discretized in both time and space.

Consider the transport equation \eqref{eq:fulltransport} and the definition in equation \eqref{eq:Ddefine} where the relevant fields have already been extended from the smooth manifold $\mathcal{S}$ to the enclosing manifold $\mathcal{M}$ and $\mu$ is known. With any test function $\omega \in H^{1}(\mathcal{M})$ a weak form of the equation can be derived as

\begin{align*}
  \int_{\mathcal{M}}\dfrac{\partial \phi}{\partial t}\omega dM - \int_{\mathcal{M}}\mu D^{\alpha\beta}\tilde{\nabla}_{\alpha}\tilde{\nabla}_{\beta}\phi\omega dM &= 0\\
  \int_{\mathcal{M}}\dfrac{\partial \phi}{\partial t}\omega dM + \int_{\mathcal{M}}\tilde{\nabla}_{\alpha}(\mu D^{\alpha\beta}\omega)\tilde{\nabla}_{\beta}\phi dM - \int_{\partial \mathcal{M}}\tilde{\nabla}_{\alpha}(\mu D^{\alpha\beta}\omega\tilde{\nabla}_{\beta}\phi)d\partial M &= 0
\end{align*}

such that
\begin{align}
  \int_{\mathcal{M}}\dfrac{\partial \phi}{\partial t}\omega dM + \int_{\mathcal{M}}\mu D^{\alpha\beta}\tilde{\nabla}_{\alpha}\omega\tilde{\nabla}_{\beta}\phi dM + \int_{\mathcal{M}}\mu \tilde{\nabla}_{\beta}D^{\beta\alpha}\omega\tilde{\nabla}_{\alpha}\phi dM &= 0\label{eq:weakform}
\end{align}

With respectively three distinct terms: the time derivative, a diffusive term and a convective contribution. 

In this numerical framework, the Riemannian manifold $\mathcal{M}$ is meshed using an unstructured simplicial grid generated using Gmsh \cite{Geuzaine2009}. Thus, the smooth Riemannian manifold $\mathcal{M}$ is approximated by a $C^{1}$ by parts manifold $\bar{\mathcal{M}}$ and any initially smooth field is approximated by a field whose component functions are in $H^{1}$ (i.e. a P1 field). As such, the level set field is approximated by a linear by parts (inside each cell) field $\bar{\phi}$. The details of the algorithm used to compute the distance function can be found in \cite{Shakoor2015amm}.

\begin{figure}
  \centering
  \begin{subfigure}{\textwidth}
    \centering
    \includegraphics[scale=0.2]{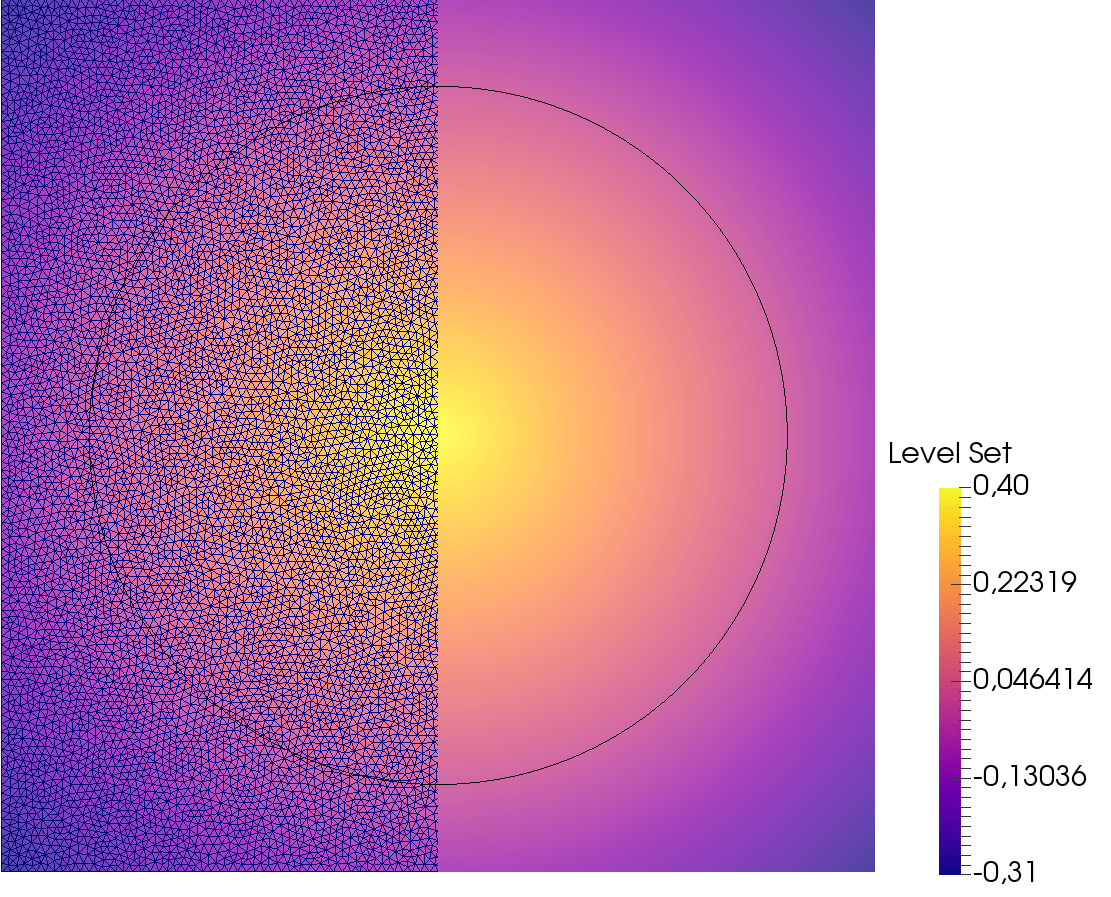}
    \caption{The level set field $\phi$}
  \end{subfigure}
  \begin{subfigure}{\textwidth}
    \centering
    \includegraphics[scale=0.2]{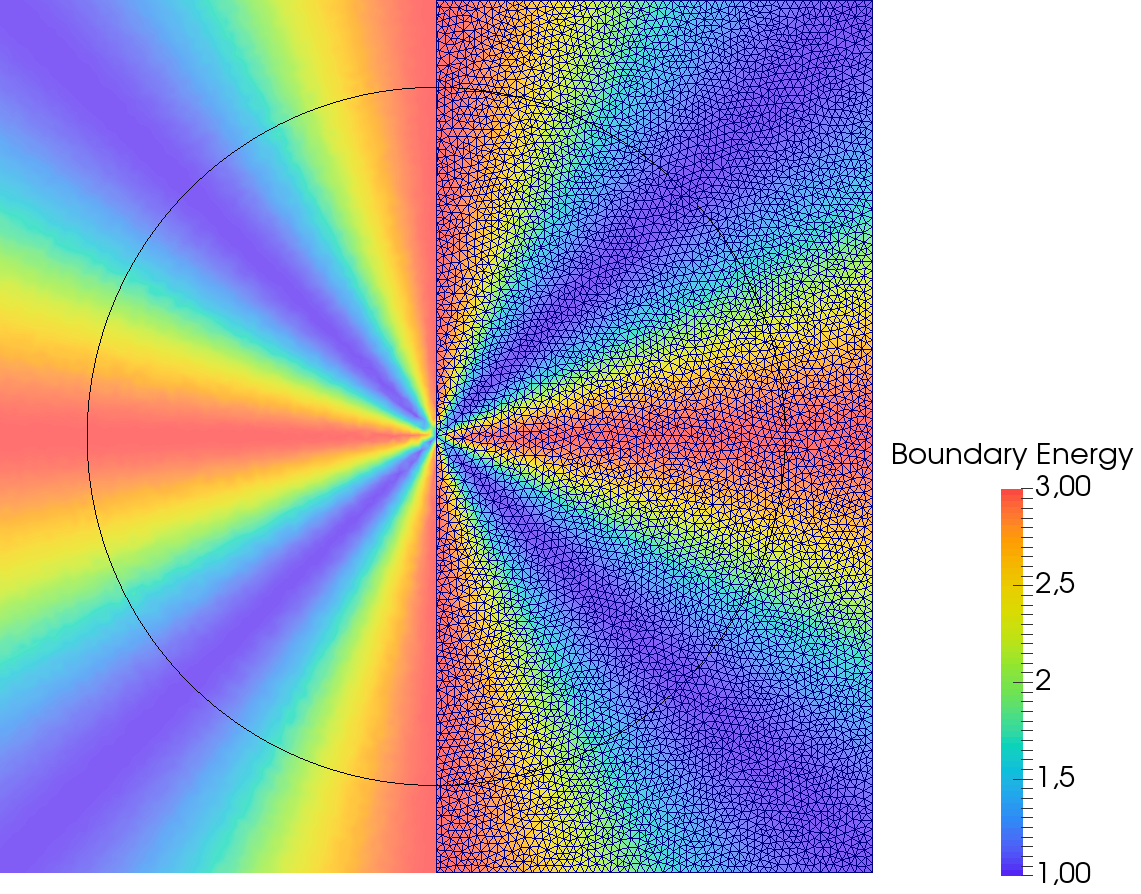}
    \caption{The boundary energy field $\gamma$}
  \end{subfigure}
\caption{Image of the $\phi$ and $\gamma$ fields defined on an unstructured mesh. The iso-zero value of the level set is represented in black and $\gamma = 2 + \cos(4 \arccos(X \cdot \nabla \phi))$ where $X$ is the unit vector field in the direction of the x axis.}\label{fig:LSGamma}
\end{figure}

Thus, given a boundary energy map $\gamma: \mathcal{B} \rightarrow \mathbb{R}^{+}$, with $\mathcal{B}$ the boundary property space, the $C^{1}$ geometry dependence of $\gamma$ can easily be evaluated at each node of the mesh $\bar{\mathcal{M}}$. Considering that $\mathcal{B}$ is only parameterized by the normal to the boundary $n$ for a given boundary, both values for $\gamma$ and $\frac{\partial^{2}\gamma}{\partial \nabla \phi^{2}}$ can be evaluated everywhere on the mesh. As such, the level set field induces a natural discretized extension of both $\gamma$ and $\frac{\partial^{2}\gamma}{\partial \nabla\phi^{2}}$ from $\varphi(S)$ to the entire discretized space $\bar{\mathcal{M}}$. Outside the interface the $\gamma$ field has no physical meaning. However, this extension is necessary for solving the problem in a FE setting. The interpolated values of the fields at the interface $\varphi(S)$ are also guaranteed to be the correct values given the linear by parts interpolation of $\bar{\phi}$. Figure \ref{fig:LSGamma} illustrates the construction for a circle and a particular choice of $\gamma(n)$.

The $\tilde{\nabla} \cdot D$ is computed numerically on the mesh using a Superconvergent Patch Recovery method inspired from \cite{Belhamadia2004} to obtain P1 fields. As such, both the diffusive tensor $D$ and the convective velocity are introduced explicitly into the formulation so as to create linearised approximations of the equation \eqref{eq:weakform}. Thus, solving the problem is completely linear without need for non-linear solvers or algorithms. 

In this work a Generalized Minimal Residual (GMRES) type solver along with an Incomplete LU (ILU) type preconditionner, both linked from the PetsC open source libraries, are used unless specified otherwise. The system is assembled using typical P1 FE elements with a Streamline Upwind Petrov-Galerkin (SUPG) stabilization for the convective term \cite{Brooks1982}. The boundary conditions used are classical von Neumann conditions which guarantees the orthogonality of the level sets to the boundary of the domain. The discretization of time is obtained using a fully implicit backward Euler method with time step $\Delta t$.

Because the resolution of the transport equation does not conserve the distance property of the level set field, the solution is reinitialized using the algorithm developed in \cite{Shakoor2015}. Also, since the geometry of the interface evolves after each time increment, all the other fields must also be recomputed from the reinitialized level set at each step of the simulation. The complete procedure for the minimizing interface energy flow simulation is reported in Algorithm \ref{alg:minE}. 
%Where $\mathbb{M}$ and $\mathbb{B}$ could be deduced from equation~\eqref{eq:weakform} and are defined by

%\begin{align*}
  %\mathbb{M}_{ik} &= \sum_{c \in \mathcal{E}}\int_{c}(\omega_{i}\tilde{\omega}_{k} + \Delta t \mu (D^{\alpha\beta}\tilde{\nabla}_{\alpha}\omega_{k}\tilde{\nabla}_{\beta}\omega_{i} + (P^{\alpha\beta}\tilde{\nabla}_{\beta}\gamma + \tilde{\nabla}_{\beta}D^{\beta\alpha})\tilde{\omega}_{k}\tilde{\nabla}_{\alpha}\omega_{i})) dM\\
  %\mathbb{B}_{k} &= \phi^{i}_{n}\sum_{c\in\mathcal{E}}\int_{c}\omega_{i}\omega_{k} dM
%\end{align*}

%the resolution of the discretized problem becomes a linear algebra system of the form $\mathbb{M}u = \mathbb{B}$ with $u = [\phi^{i}_{n+1}]$ a vector where the $i$th component is the value of the level set function.

\begin{algorithm}
  \caption{Minimizing Interface Energy Flow}\label{alg:minE}
  \begin{algorithmic}[1]
    \STATE \textbf{Data:} Initial Embedding, $\bar{\mathcal{M}}$, $\Delta t$, $t_{end}$
    \STATE Compute the initial Level Set and unit normal fields
    \STATE Calculate $\gamma$ and $D$ fields and their derivatives
    \STATE $t = 0$
    \WHILE {$t < t_{end}$}
    	\STATE Assemble the FE system
    	\STATE Solve the FE system
    	\STATE $t = t + \Delta t$
    	\STATE Reinitialize the Level Set and unit normal fields
    	\STATE Update the $\gamma$ and $D$ fields and their derivatives
    \ENDWHILE
  \end{algorithmic}
\end{algorithm}

\subsection{The shrinking ellipse}

One now has an embedding and a way to represent it as a level set field $\phi$ on an unstructured mesh. One also has the FE formulation needed to simulate the dynamics of the minimizing energy flow of the interface. However, the boundary energy $\gamma = G^{\theta\theta}g_{\theta\theta}$ is not readily computable on the finite element mesh since it does not explicitly depend on the normal to the interface. Using equation \eqref{eq:metric} such that

\begin{align*}
  g_{\theta\theta} &= (b^{2}\dfrac{a^{2}}{b^{2}}\sin^{2}\theta + b^{2}\cos^{2}\theta)\\
                   &= b^{2}(\dfrac{a^{2}}{b^{2}}\sin^{2}\theta + \cos^{2}\theta)
\end{align*}

which, if one considers $r(t) = \dfrac{a(t)}{b(t)}$, which should remain constant throughout the simulation if looking at the large and small axes of the ellipse $a(t), b(t)$ at each instant, then $\gamma$ can easily be extended throughout the mesh using

\begin{align*}
  \dfrac{n^{y}}{n^{x}} &= r\tan\theta\\
  \theta &= \arctan\left(\dfrac{1}{r}\dfrac{n^{y}}{n^{x}}\right)
\end{align*}

with

\begin{align}
  \gamma(\theta, t) = b(t)^{2}(r(t)^{2}\sin^{2}\theta + \cos^{2}\theta)
\end{align}

Given the definition of the level set field, $\phi$ takes maximal values at the points within the ellipse furthest away from the interface, i.e. the center of the ellipse. Seeing as $b(t)$ is the smallest of both ellipse axes and the level set is minimal distance valued, the value of the level set at the center of the ellipsis should be the value of the small axis. Therefore

\begin{align}
  b(t) = \max_{q \in M} \phi(q, t) \label{eq:bmeasure}
\end{align}

Also, implicit in the calculations in the previous section is the fact that

\begin{align}
  \dfrac{\partial^{2} \gamma}{\partial \tilde{\nabla}_{\alpha} \phi \partial \tilde{\nabla}_{\beta}\phi} = 2\gamma m^{\alpha\beta}\label{eq:frustratingCondition}
\end{align}

so that knowing the extension of the boundary energy $\gamma$ is sufficient for calculating $D^{\alpha\beta} = 3\gamma m^{\alpha\beta}$.

Thus the boundary energy field $\gamma$ can be computed at each iteration of the simulation. Using $\mu G^{\theta\theta} = 1$, the simulation can be run on any arbitrary mesh with arbitrary mesh size $h$ using any time step $\Delta t$.

\begin{figure}
  \centering
  \begin{subfigure}{0.48\textwidth}
    \centering
    \includegraphics[scale=0.15]{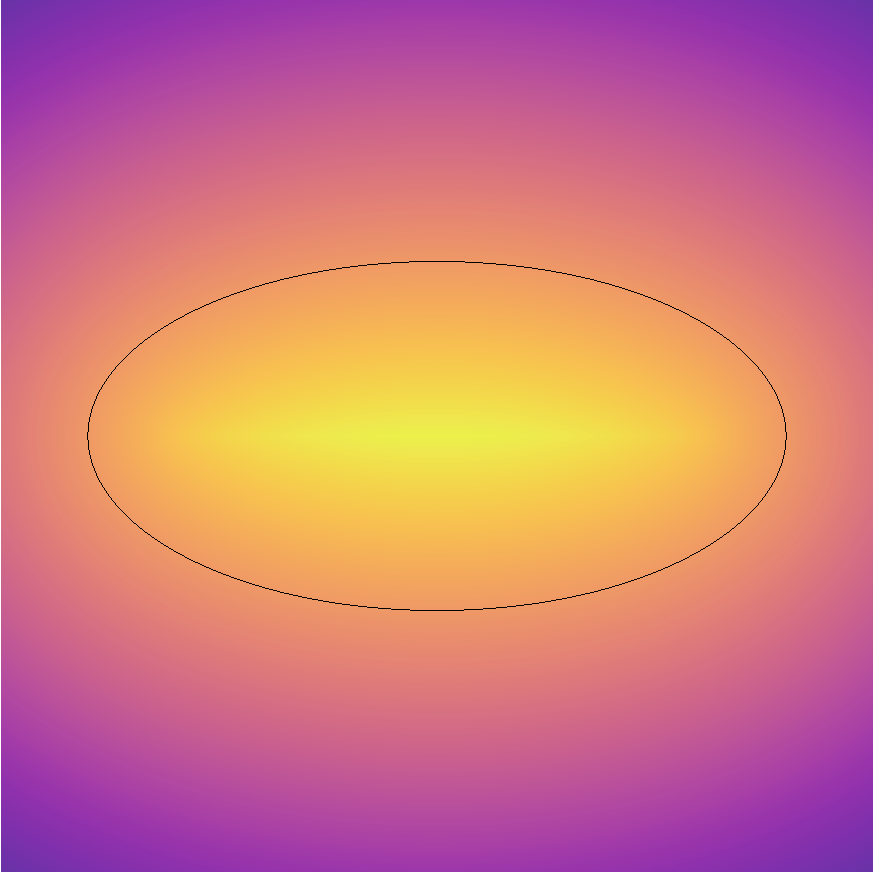}
    \caption{$t = 0$}
  \end{subfigure}
  \begin{subfigure}{0.48\textwidth}
    \centering
    \includegraphics[scale=0.15]{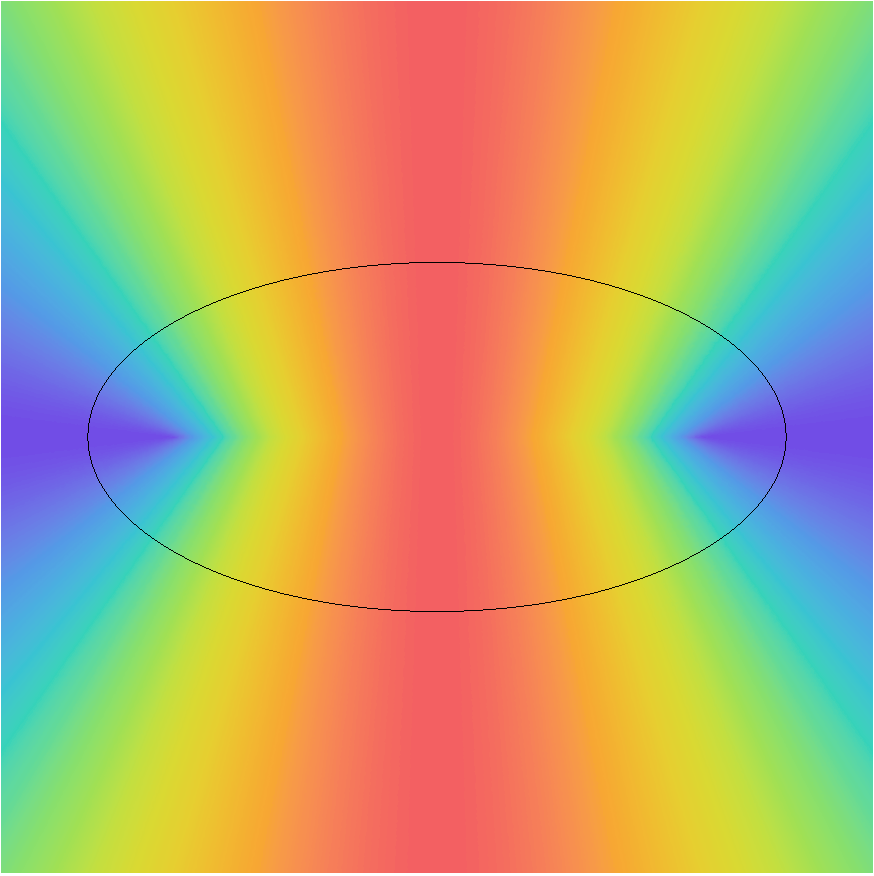}
    \caption{$t = 0$}
  \end{subfigure}
  \begin{subfigure}{0.48\textwidth}
    \centering
    \includegraphics[scale=0.15]{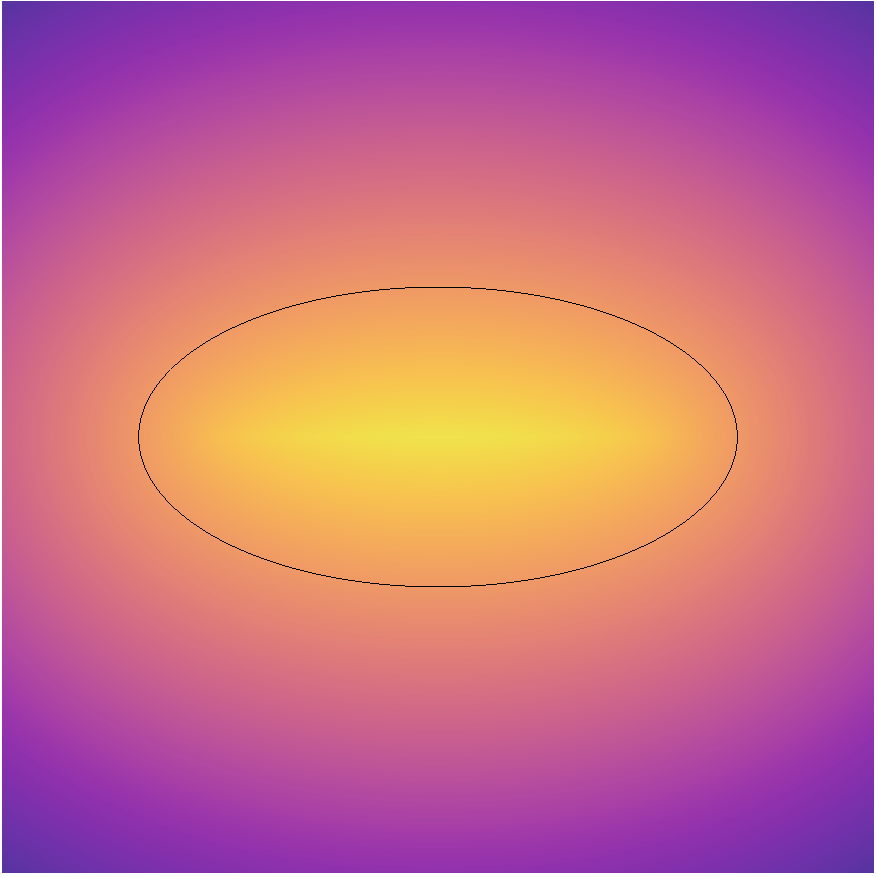}
    \caption{$t = 5e-3$}
  \end{subfigure}
  \begin{subfigure}{0.48\textwidth}
    \centering
    \includegraphics[scale=0.15]{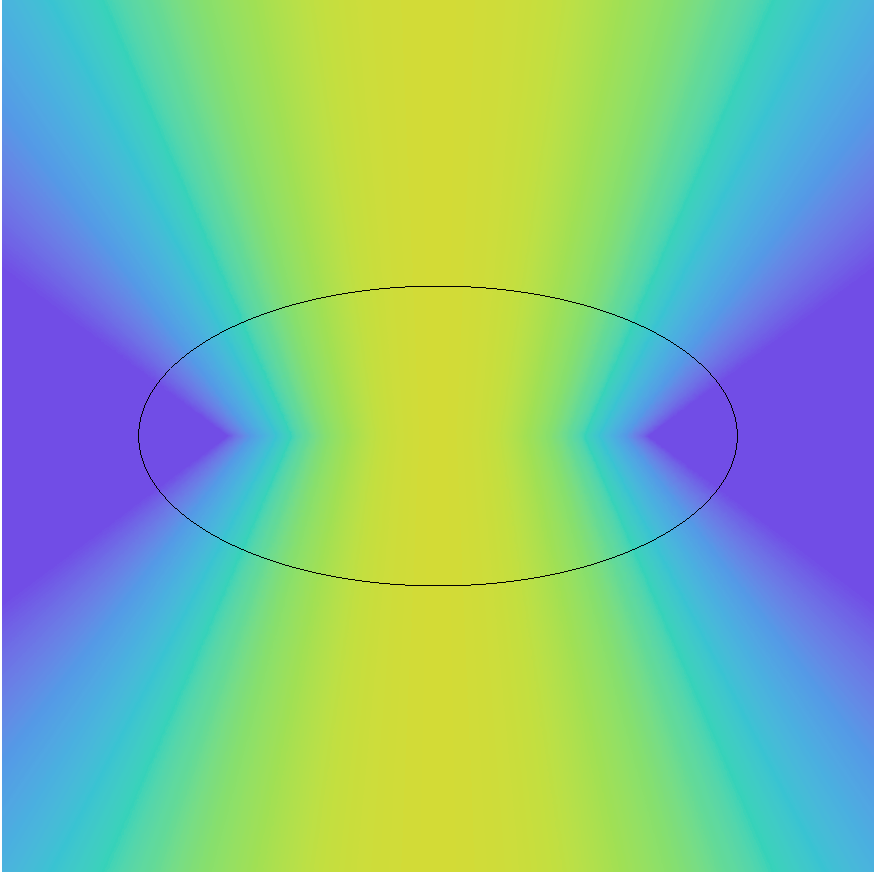}
    \caption{$t = 5e-3$}
  \end{subfigure}
  \begin{subfigure}{0.48\textwidth}
    \centering
    \includegraphics[scale=0.15]{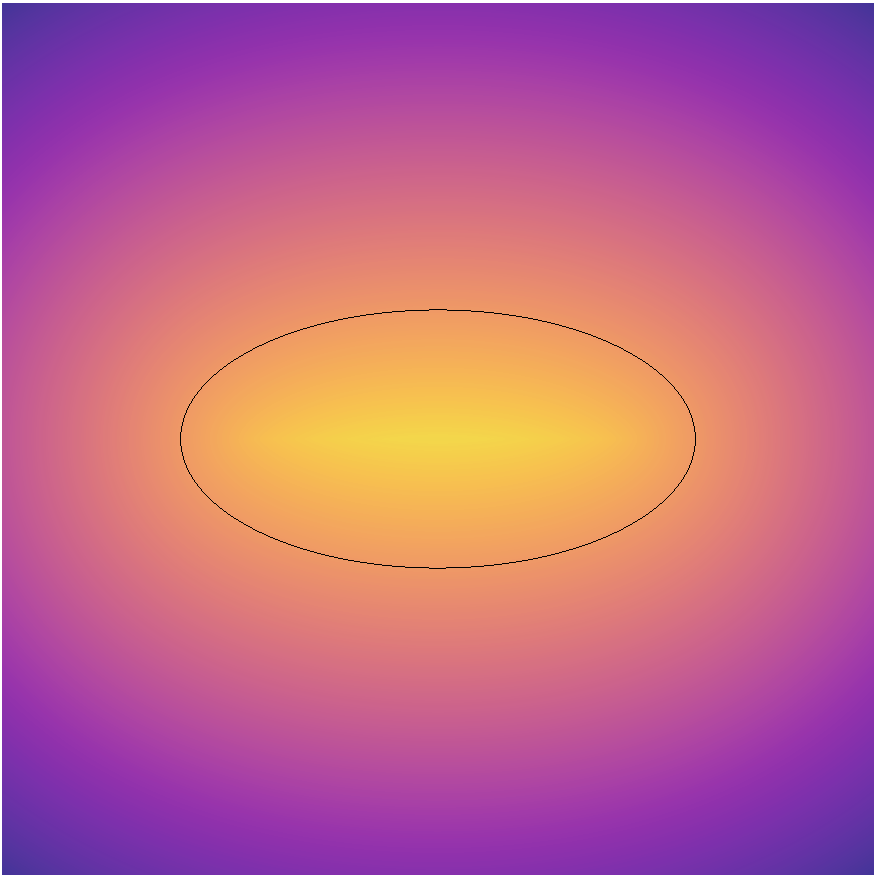}
    \caption{$t = 1e-3$}
  \end{subfigure}
  \begin{subfigure}{0.48\textwidth}
    \centering
    \includegraphics[scale=0.15]{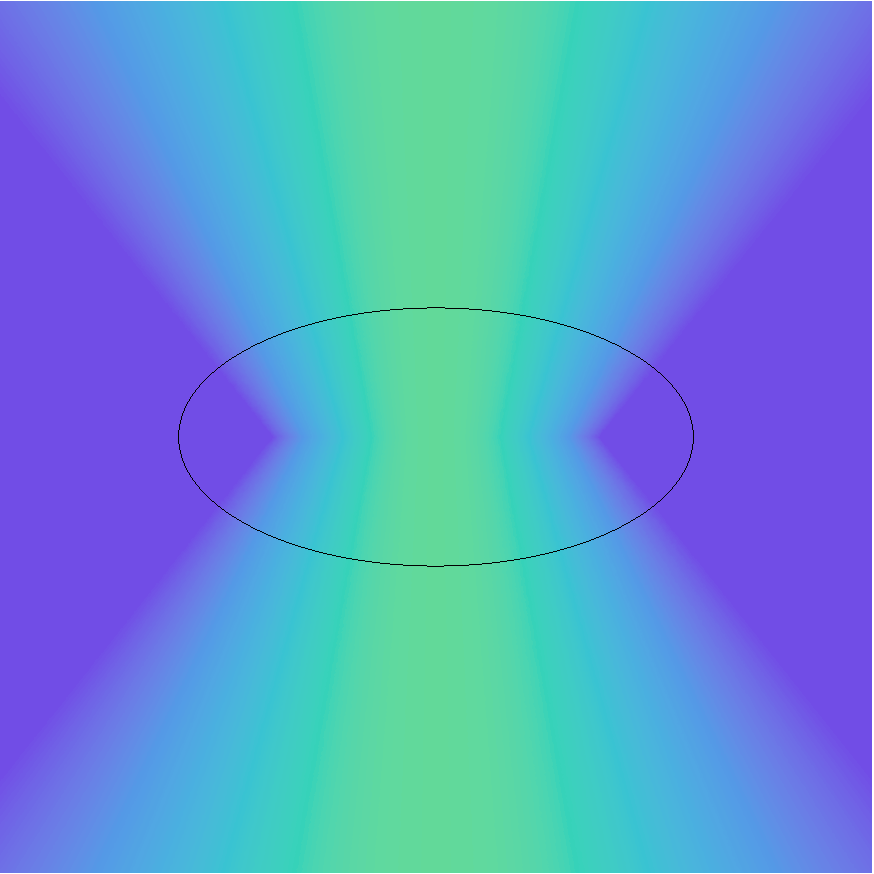}
    \caption{$t = 1e-3$}
  \end{subfigure}
  \begin{subfigure}{0.48\textwidth}
    \centering
    \includegraphics[scale=0.2]{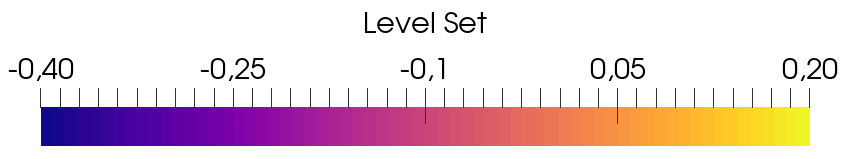}
  \end{subfigure}
  \begin{subfigure}{0.48\textwidth}
    \centering
    \includegraphics[scale=0.2]{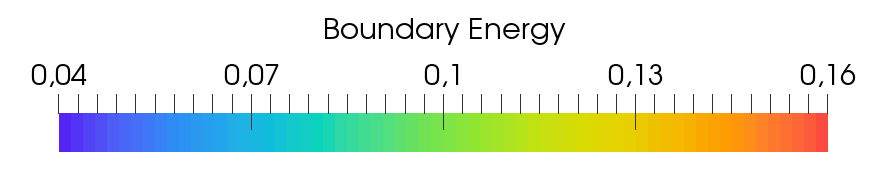}
  \end{subfigure}
  \caption{Time evolution of the level set $\phi$ and boundary energy $\gamma$ fields for the ellipse shrinkage test case. The iso-zero value of the level-set field is in black. The mesh size is $h = 3e-3$ and the time step is $\Delta t = 5e-4$ and the ellipse axes ratio is $r = 2$.}\label{fig:mosaicLS}
\end{figure}

\begin{figure}
  \centering
  \begin{subfigure}{\textwidth}
    \centering
    \includegraphics[scale=0.3]{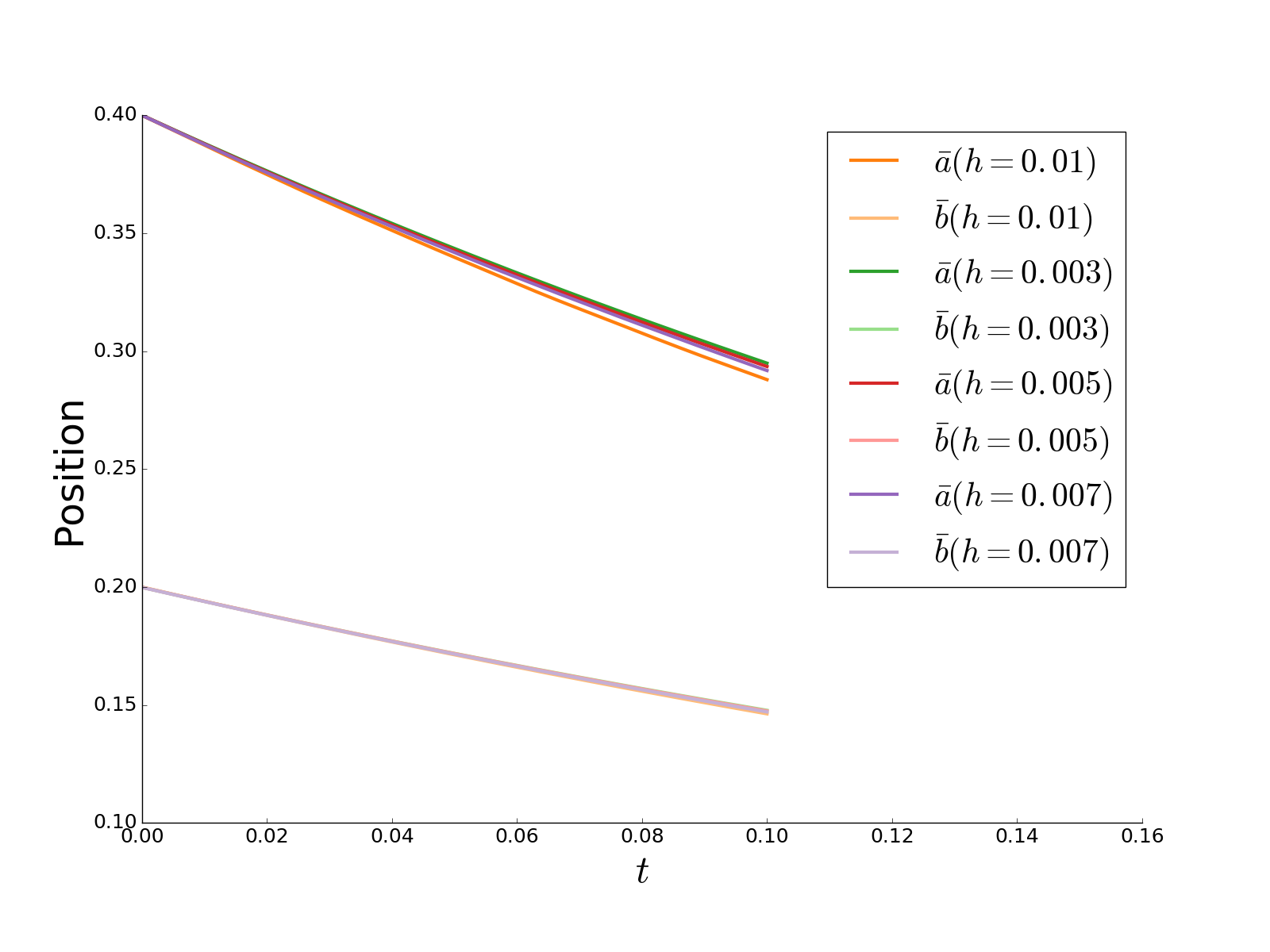}
    \caption{Both $a$ and $b$ as a function of simulated time $t$}
  \end{subfigure}
  \begin{subfigure}{\textwidth}
    \centering
    \includegraphics[scale=0.3]{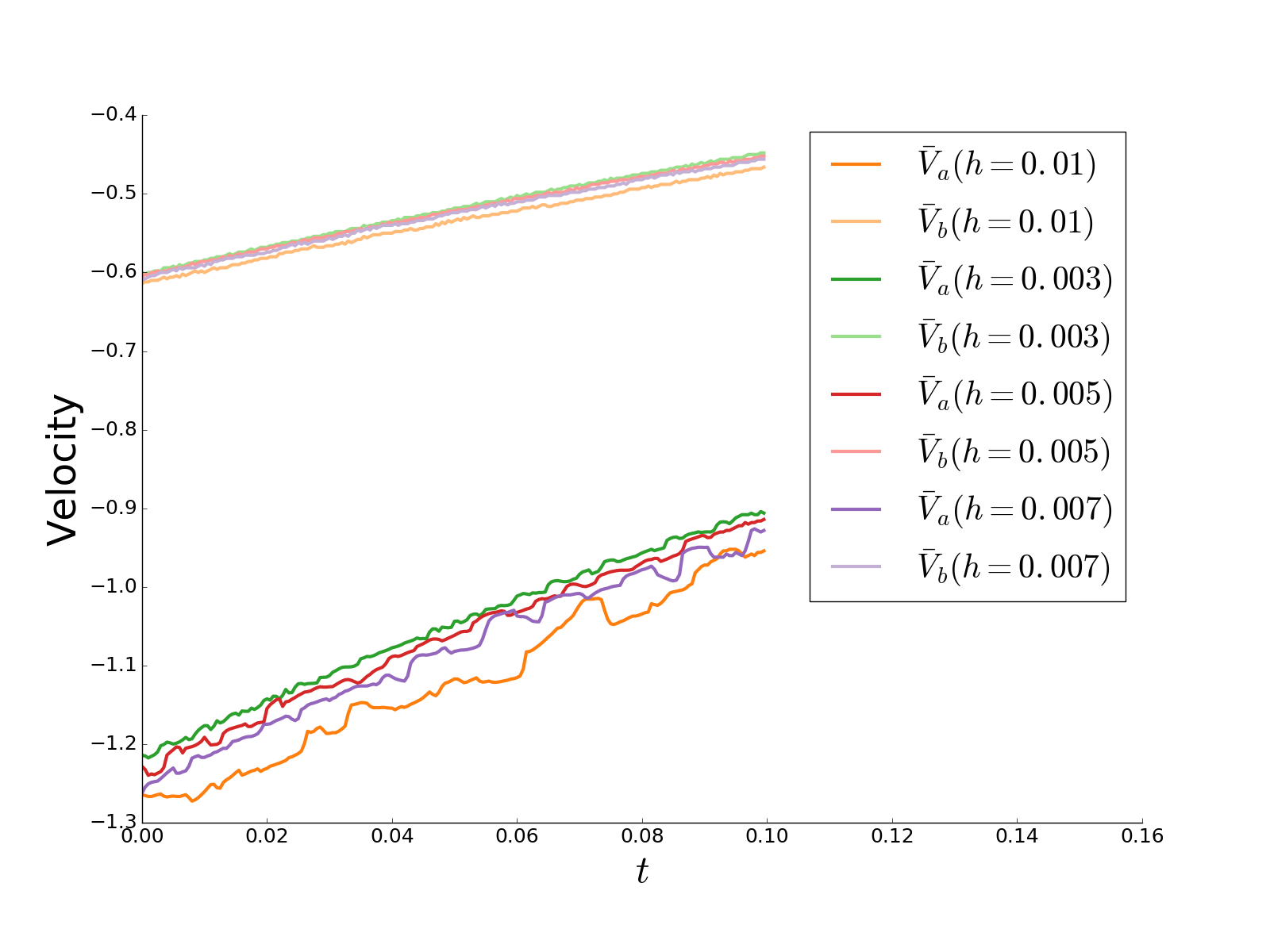}
    \caption{Both $V_{a}$ and $V_{b}$ as a function of simulated time $t$}
  \end{subfigure}
  \caption{Sensitivity of the trajectory and velocity to the mesh size $h$ parameter study with $\Delta t = 5e-4$, $r=2$ and $a(t = 0) = 0.4$ on a $1 \times 1$ size mesh.}\label{fig:ellipsemeshstudytv}
\end{figure}

\begin{figure}
  \centering
  \begin{subfigure}{\textwidth}
    \centering
    \includegraphics[scale=0.3]{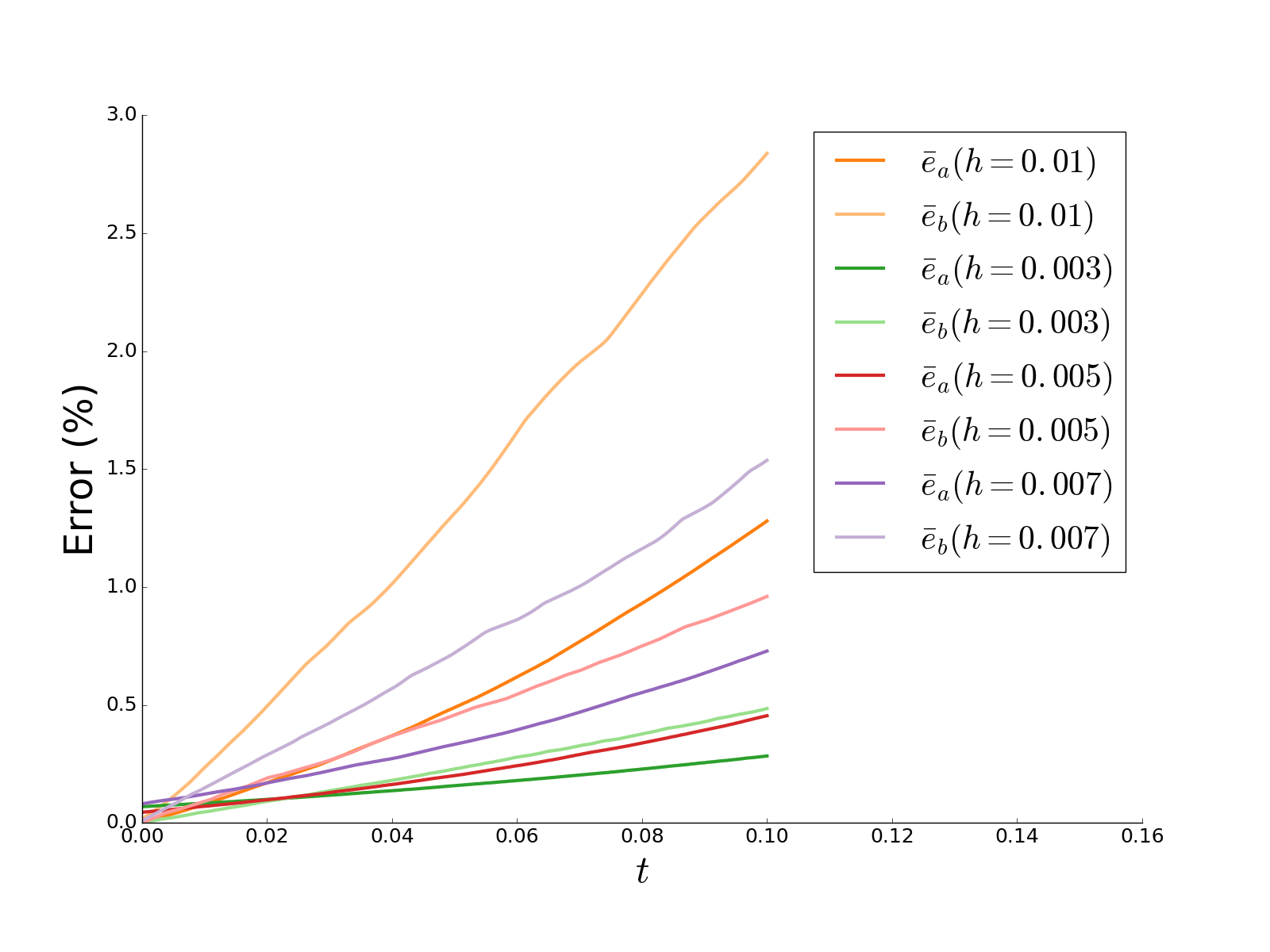}
    \caption{Both $e_{a}$ and $e_{b}$ errors committed on the positions as a function of simulated time $t$}
  \end{subfigure}
  \begin{subfigure}{\textwidth}
    \centering
    \includegraphics[scale=0.3]{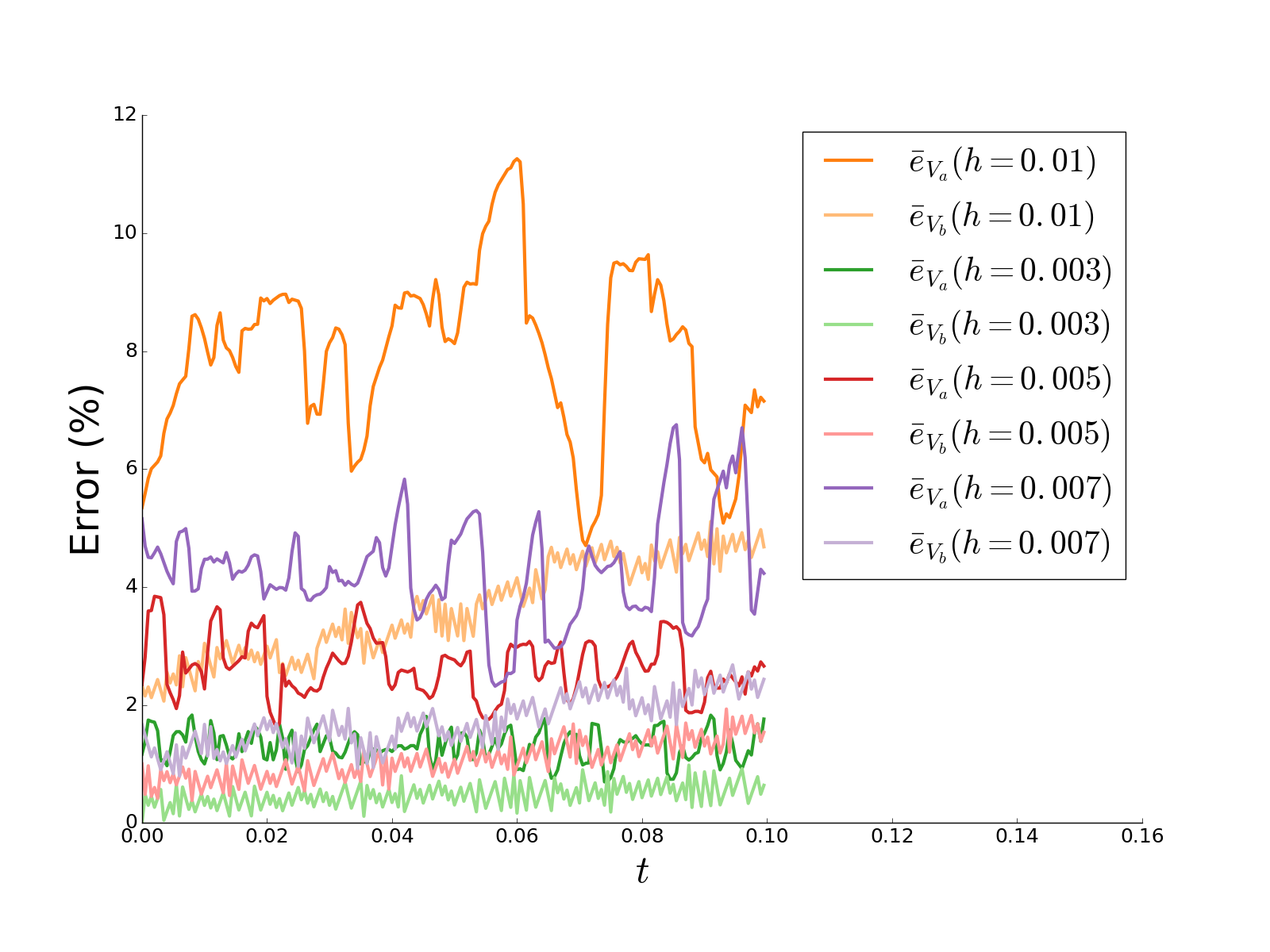}
    \caption{Both $e_{V_{a}}$ and $e_{V_{b}}$ errors committed on the velocities as a function of simulated time $t$}
  \end{subfigure}
  \caption{Sensitivity of the errors to the mesh size $h$ parameter study with $\Delta t = 5e-4$, $r=2$ and $a(t = 0) = 0.4$ on a $1 \times 1$ size mesh.}\label{fig:ellipsemeshstudyerrors}
\end{figure}

\begin{figure}
  \centering
  \begin{subfigure}{\textwidth}
    \centering
    \includegraphics[scale=0.3]{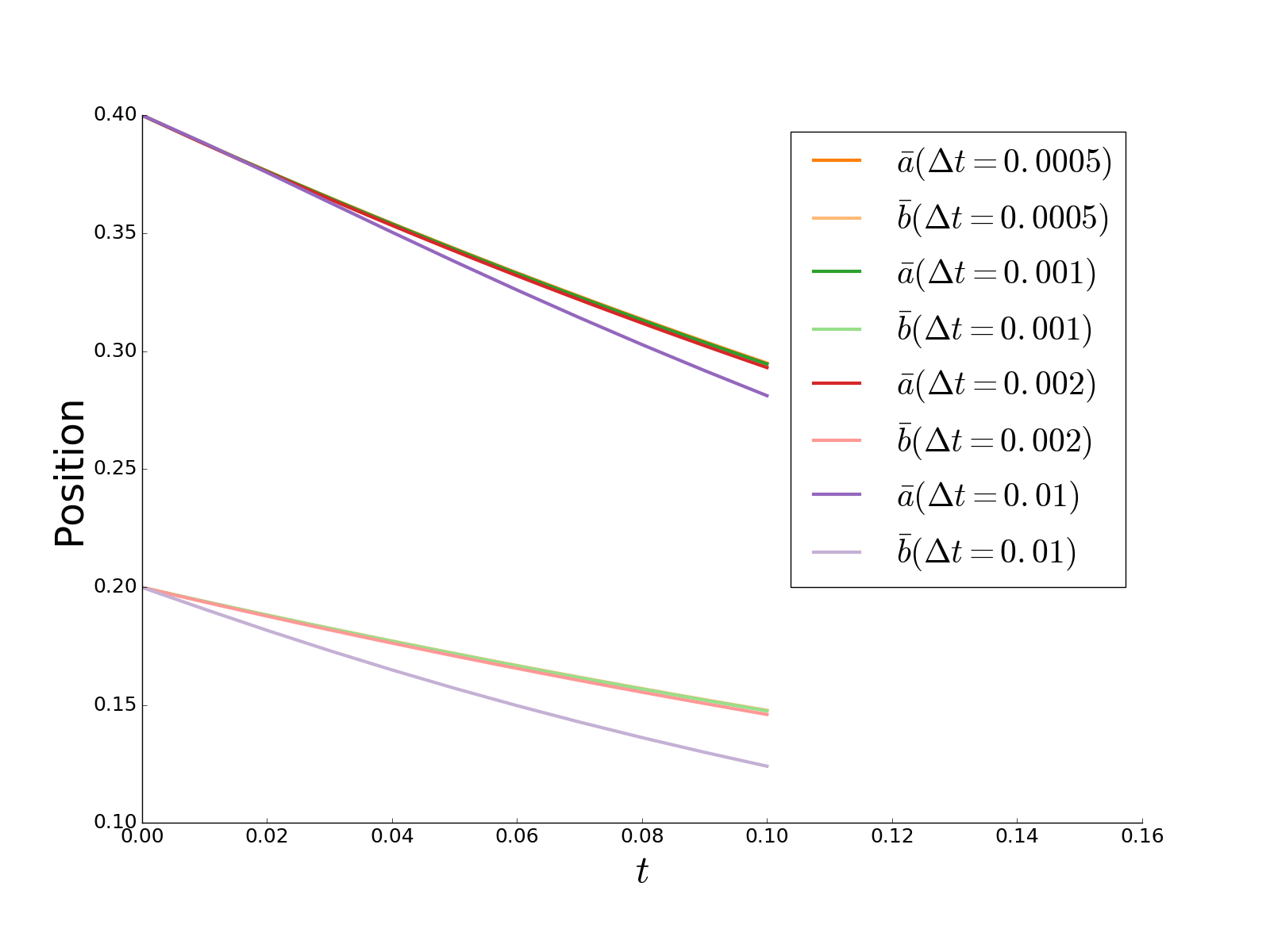}
    \caption{Both $a$ and $b$ as a function of simulated time $t$}
  \end{subfigure}
  \begin{subfigure}{\textwidth}
    \centering
    \includegraphics[scale=0.3]{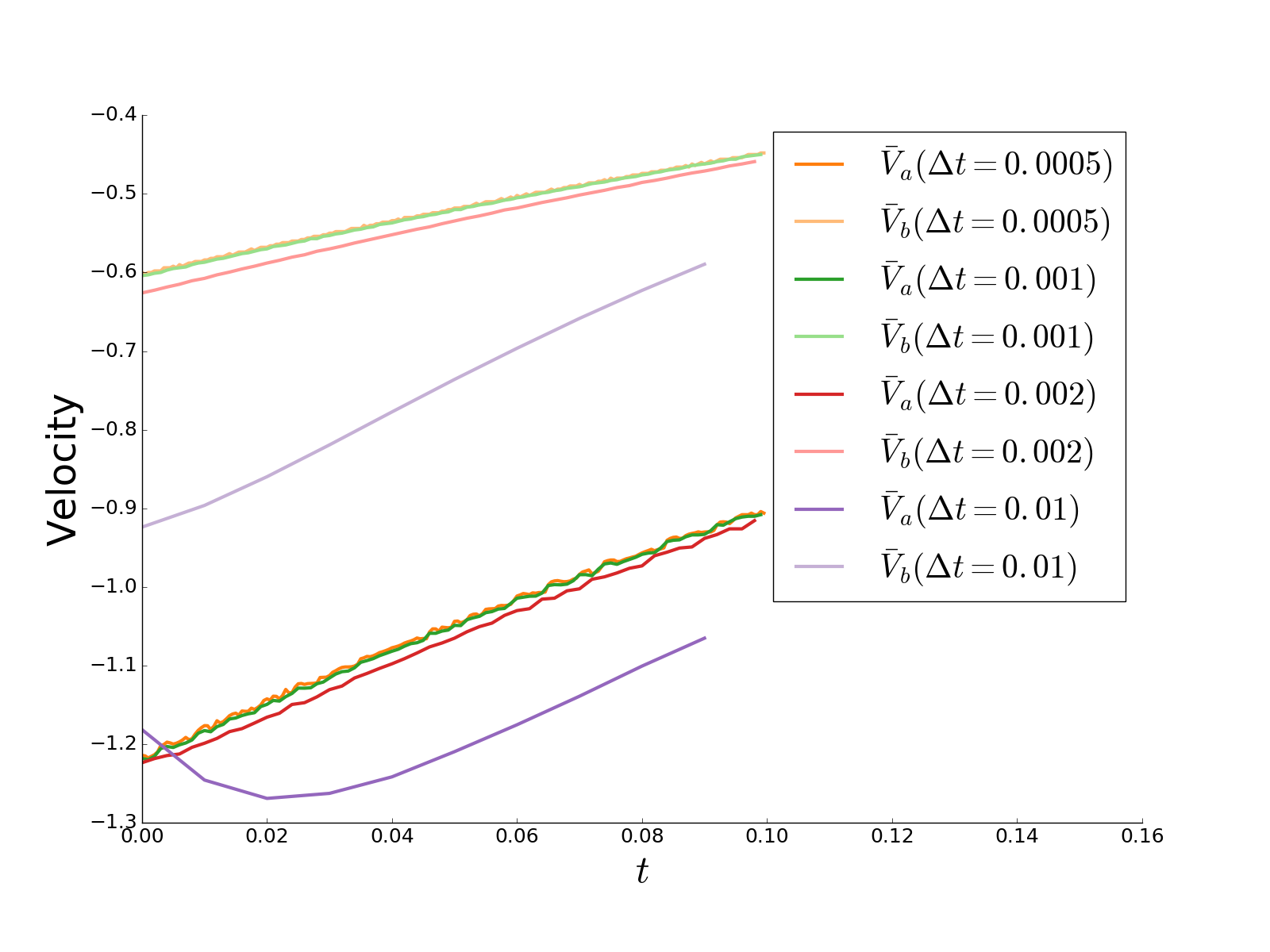}
    \caption{Both $V_{a}$ and $V_{b}$ as a function of simulated time $t$}
  \end{subfigure}
  \caption{Sensitivity of the trajectory and velocity to the time step $\Delta t$ parameter study with $h = 3e-3$, $r=2$ and $a(t = 0) = 0.4$ on a $1 \times 1$ size mesh.}\label{fig:ellipsetimestudytv}
\end{figure}

\begin{figure}
  \centering
  \begin{subfigure}{\textwidth}
    \centering
    \includegraphics[scale=0.3]{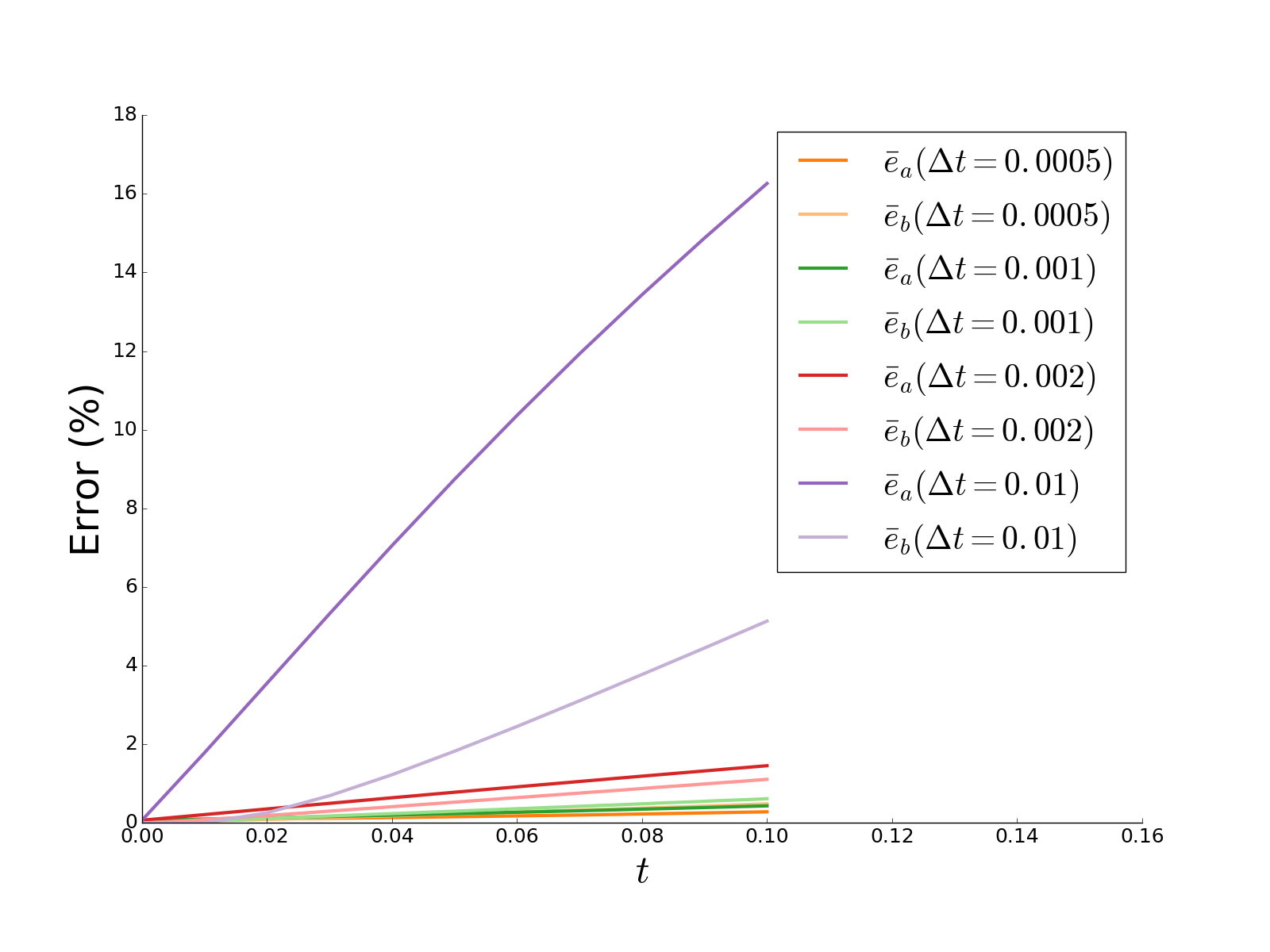}
    \caption{Both $e_{a}$ and $e_{b}$ errors committed on the positions as a function of simulated time $t$}
  \end{subfigure}
  \begin{subfigure}{\textwidth}
    \centering
    \includegraphics[scale=0.3]{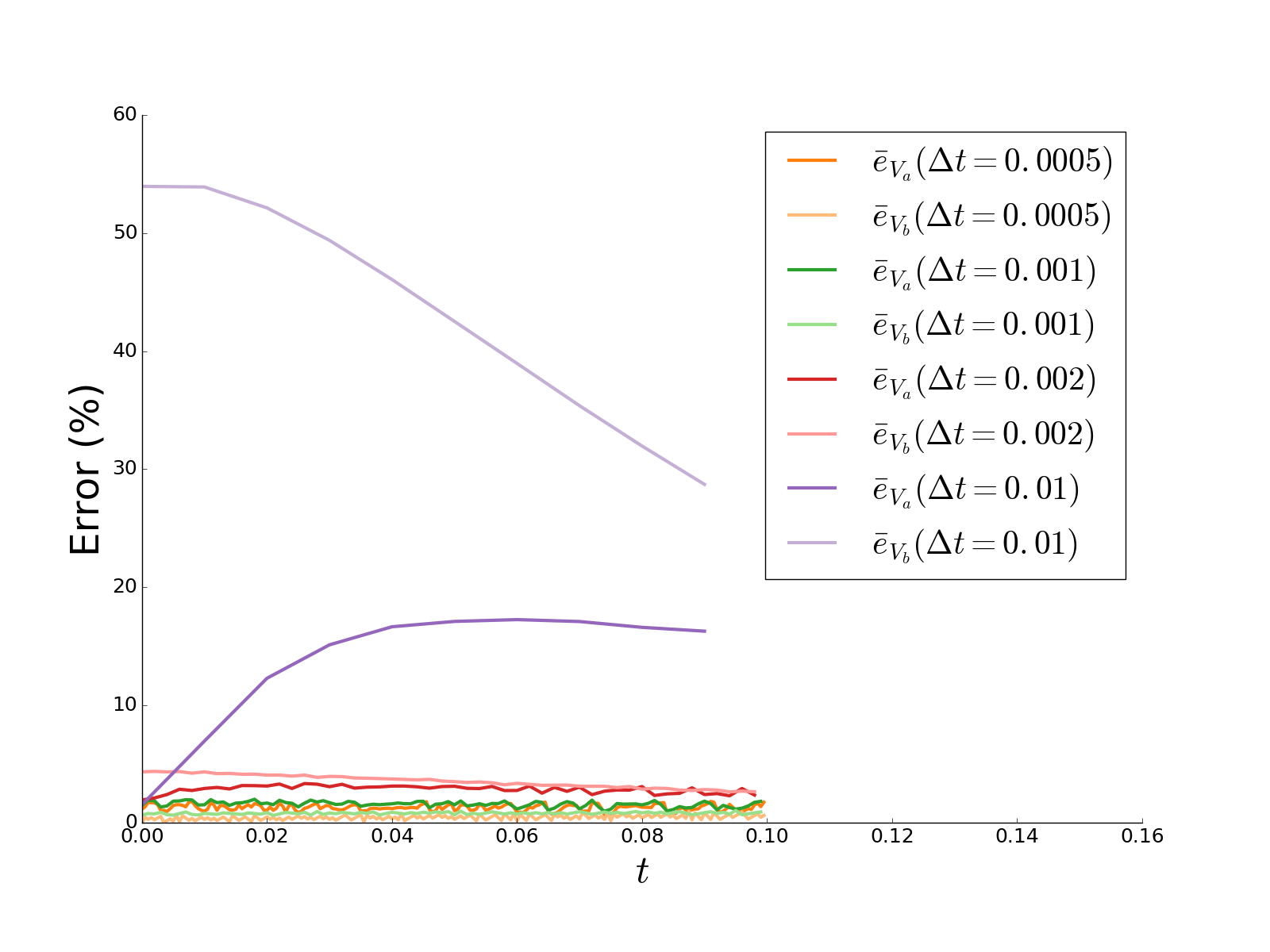}
    \caption{Both $e_{V_{a}}$ and $e_{V_{b}}$ errors committed on the velocities as a function of simulated time $t$}
  \end{subfigure}
  \caption{Sensitivity of the errors to the time step $\Delta t$ parameter study with $h = 3e-3$, $r=2$ and $a(t = 0) = 0.4$ on a $1 \times 1$ size mesh.}\label{fig:ellipsetimestudyerrors}
\end{figure}

Figure \ref{fig:mosaicLS} illustrates the time evolution of the level set field for an isotropic unstructured $1 \times 1$ mesh with $h = 3e-3$, $\Delta t = 5e-4$, $a(t=0) = 0.2$ and $r = 2$. A sensitivity analysis has been conducted with respect to the isotropic mesh size $h$ and time step $\Delta t$ whose results are reported in Figures \ref{fig:ellipsemeshstudytv}, \ref{fig:ellipsemeshstudyerrors}, \ref{fig:ellipsetimestudytv} and \ref{fig:ellipsetimestudyerrors}. The data is evaluated by looking at the time evolution of both $b$ and $a$ as well as their time derivatives $V_{b} = \frac{d b}{dt}$ and $V_{a} = \frac{d a}{dt}$. The $b$ value is evaluated using equation \eqref{eq:bmeasure} while the $a(t)$ parameter is evaluated at each time step by

\begin{align}
  a = a(t = 0) + \phi(x = a(t = 0), y = 0)\label{eq:ameasure}
\end{align}

The values are compared with their analytical analogs in order to compute errors. The convention in the legend is that bared quantities are measured while non-bared quantities are the analytical counterparts.



\begin{figure}
  \centering
  \begin{subfigure}{\textwidth}
    \centering
    \includegraphics[scale=0.3]{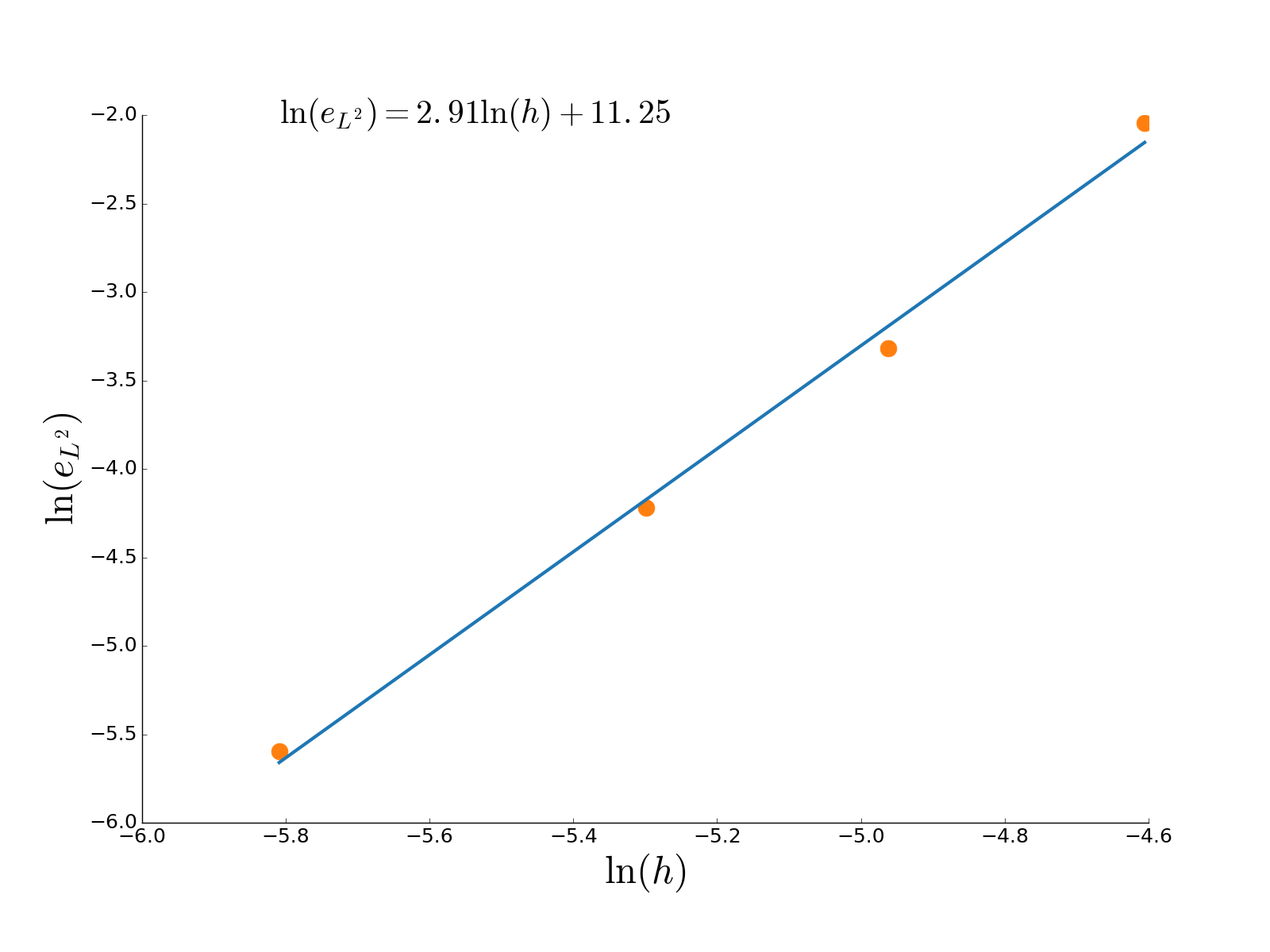}
    \caption{$\ln(e_{L^{2}}) = f(\ln(h))$}
  \end{subfigure}
  \begin{subfigure}{\textwidth}
    \centering
    \includegraphics[scale=0.3]{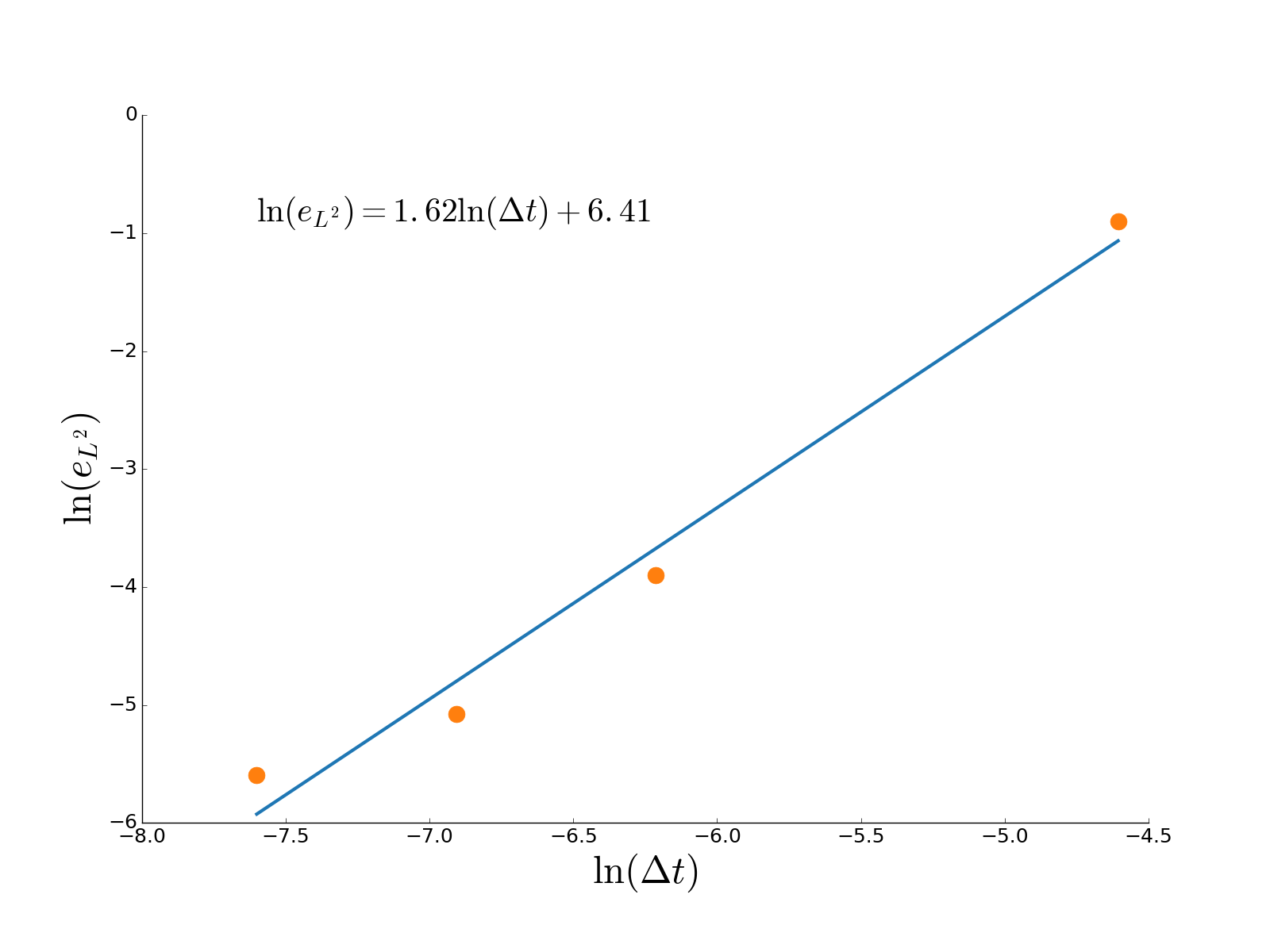}
    \caption{$\ln(e_{L^{2}}) = f(\ln(\Delta t))$}
  \end{subfigure}
  \caption{Evolution of the $\ln(e_{L^{2}})$ as a function of $h$ (for $\Delta t = 5e-4$) and $\Delta t$ (for $h = 3e-3$) with $r = 2$, $a(t=0) = 0.4$ in a $1\times1$ domain.}\label{fig:ellipsel2errors}
\end{figure}

Each simulation can be given a scalar error value by computing the $L^{2}$ error between the analytical evolution of $b(t)$ and the measured values

\begin{align}
  e_{L^{2}} = \int_{0}^{t_{end}}(b(t) - \bar{b}(t))^{2}dt
\end{align}

which can be approximated using a trapezoidal rule. Figure \ref{fig:ellipsel2errors} depicts the evolution of the logarithm of this $L^{2}$ error with respect to both $h$ and $\Delta t$.

Figures \ref{fig:ellipsemeshstudytv}, \ref{fig:ellipsemeshstudyerrors}, \ref{fig:ellipsetimestudytv}, \ref{fig:ellipsetimestudyerrors} and \ref{fig:ellipsel2errors} clearly establish convergence of the method towards the analytical solution as both the time step $\Delta t$ and mesh size $h$ become smaller. While it may seem that the simulation is actually less accurate in predicting the larger axis $a$, this can actually be attributed to the method of calculating $\bar{a}$ described in equation \eqref{eq:ameasure} which is much less precise than the measure of $b$. 

\begin{figure}
  \centering
  \begin{tabular}{ >{\centering\arraybackslash}m{1cm} m{4cm} m{4cm} m{4cm} m {0.5cm}}
    $r=\frac{8}{3}$ & \includegraphics[scale=0.12]{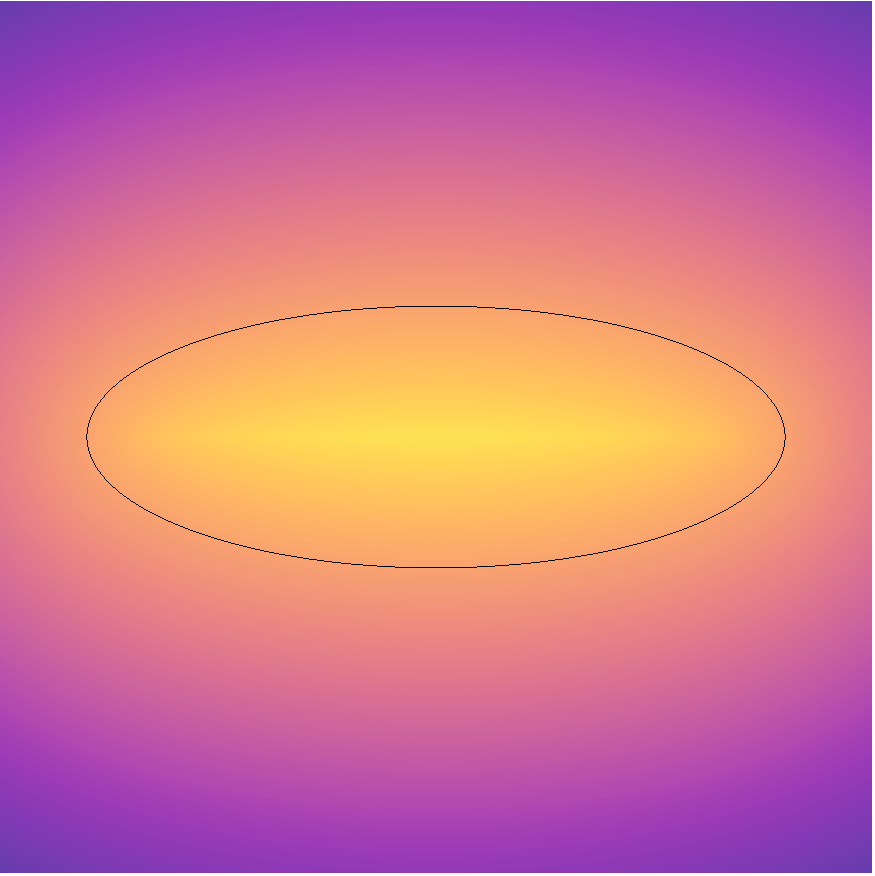} & \includegraphics[scale=0.12]{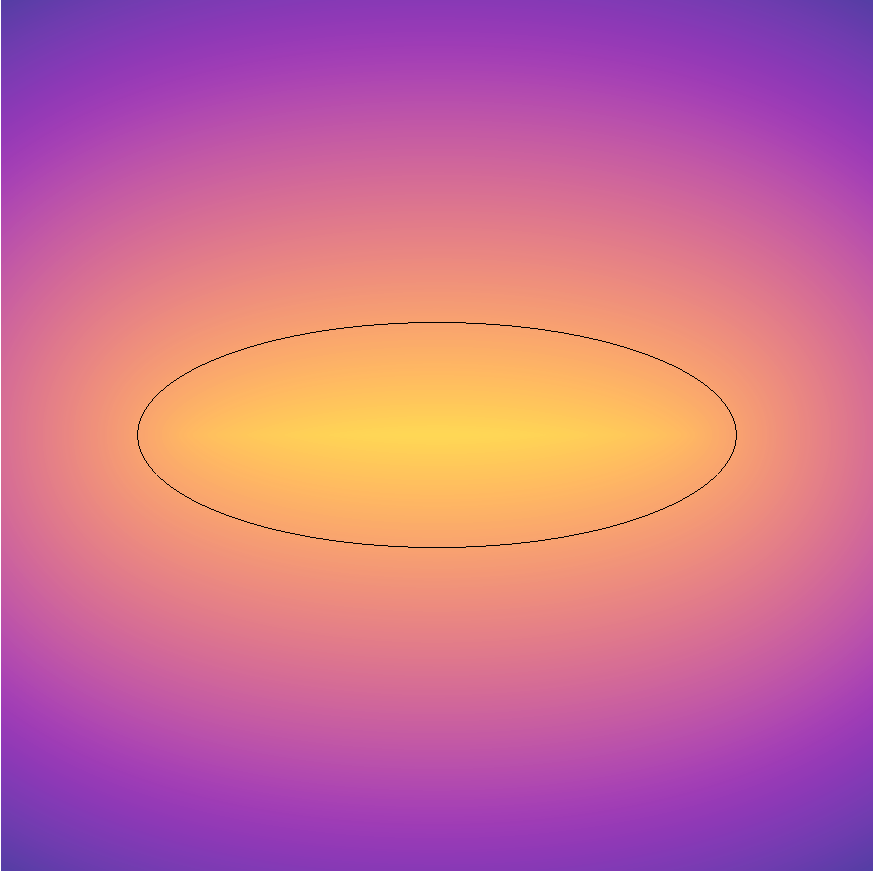} & \includegraphics[scale=0.12]{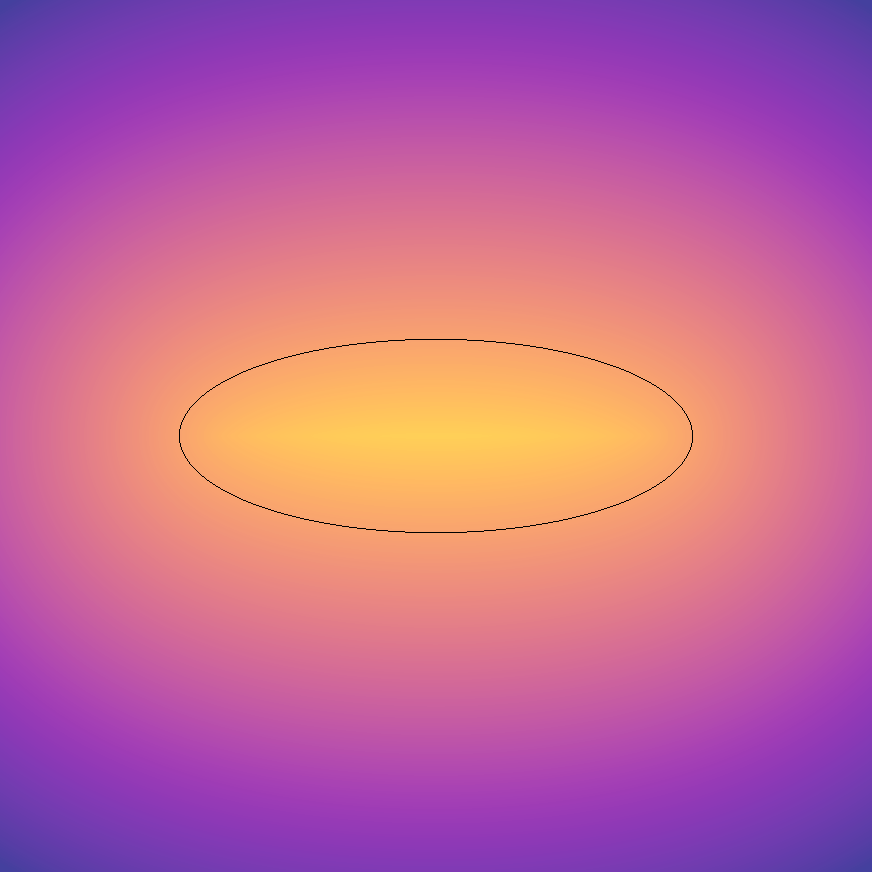} &\\
    $r=4$ & \includegraphics[scale=0.12]{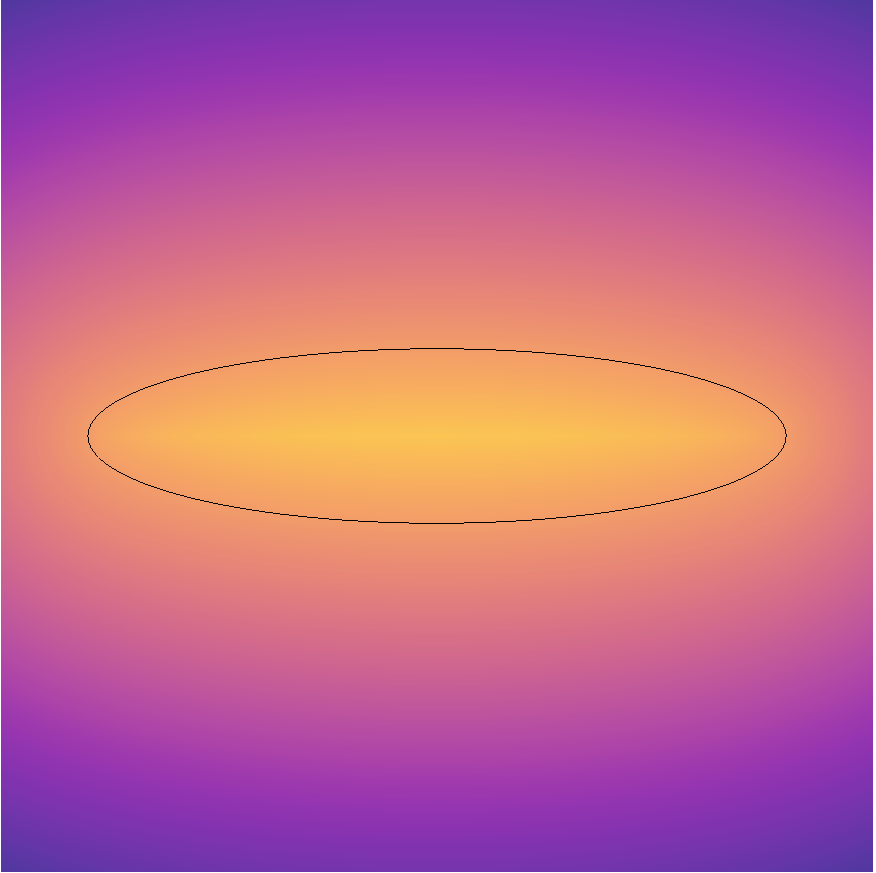} & \includegraphics[scale=0.12]{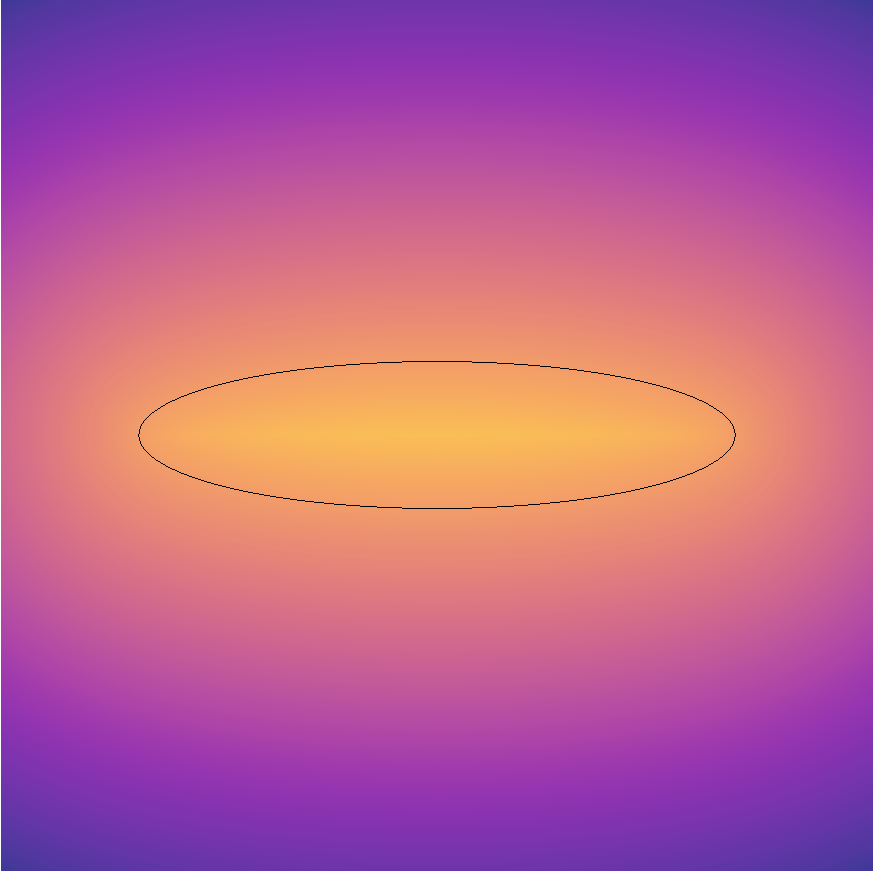} & \includegraphics[scale=0.12]{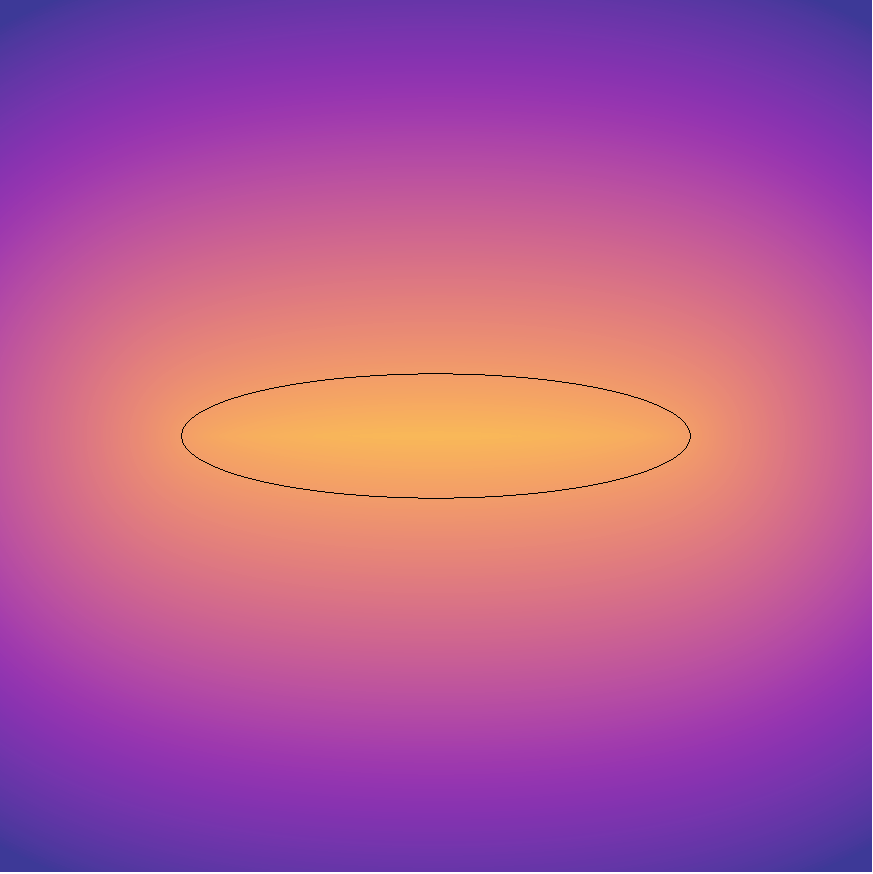} &\\
    $r=5$ & \includegraphics[scale=0.12]{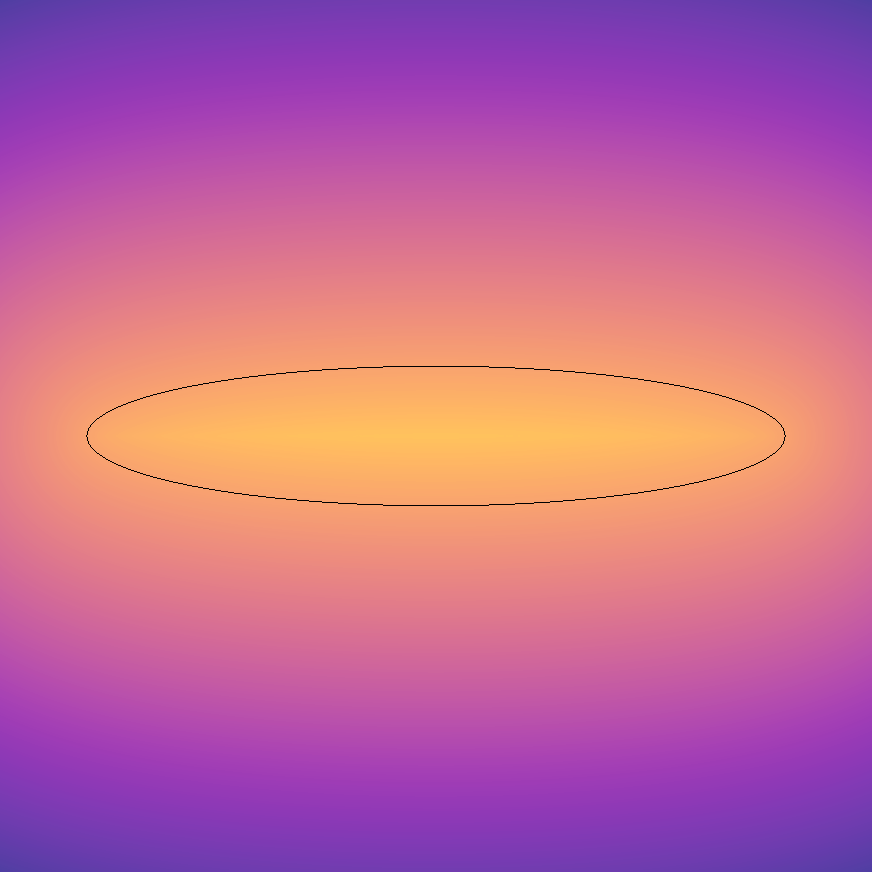} & \includegraphics[scale=0.12]{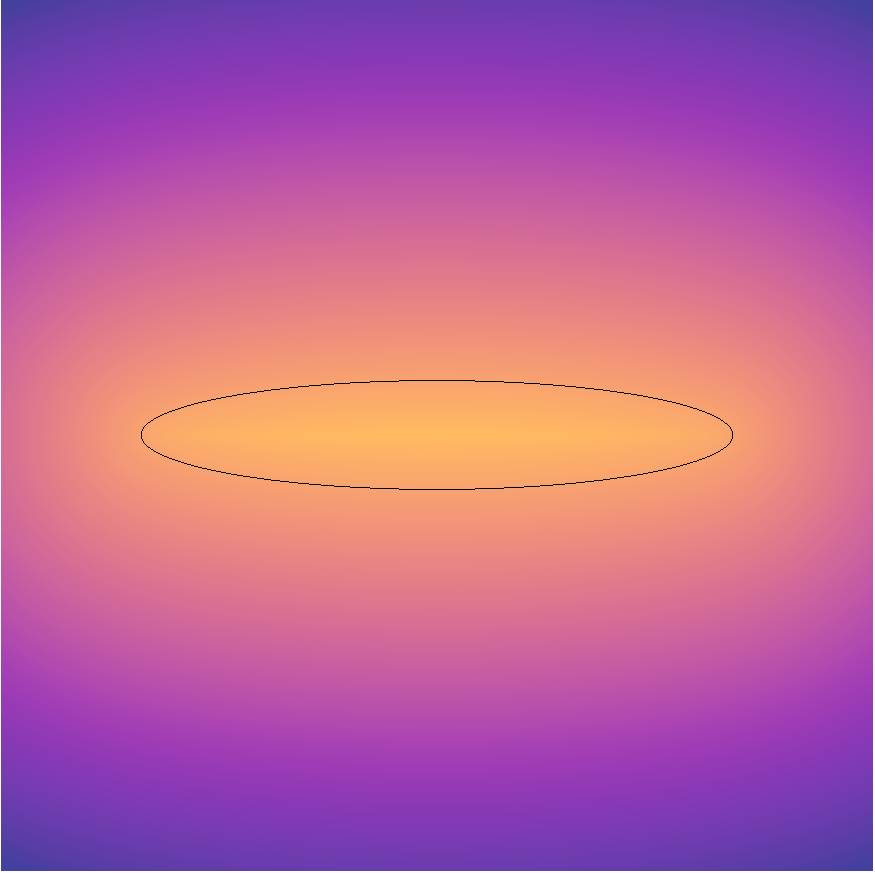} & \includegraphics[scale=0.12]{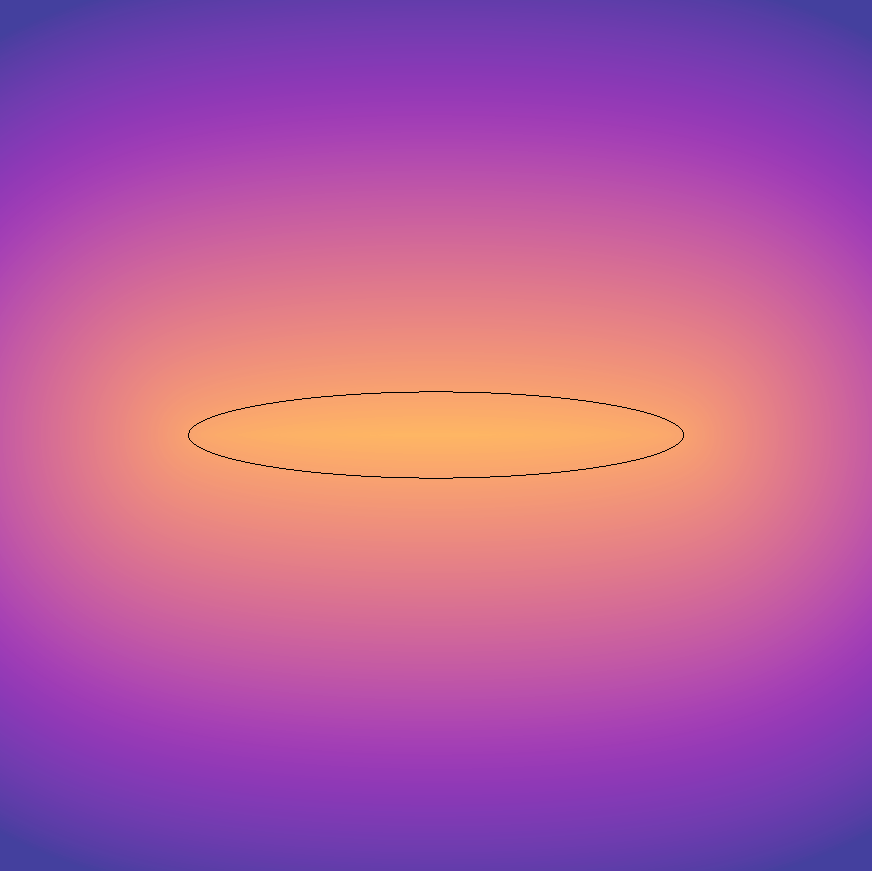} &\\
    $r=8$ & \includegraphics[scale=0.12]{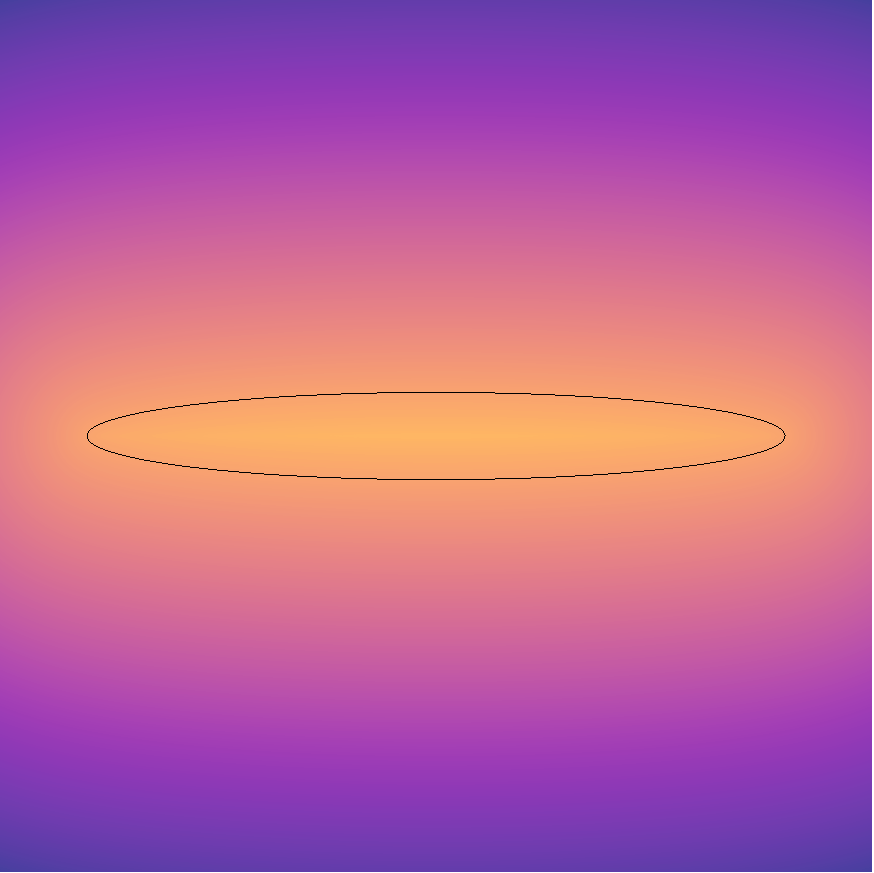} & \includegraphics[scale=0.12]{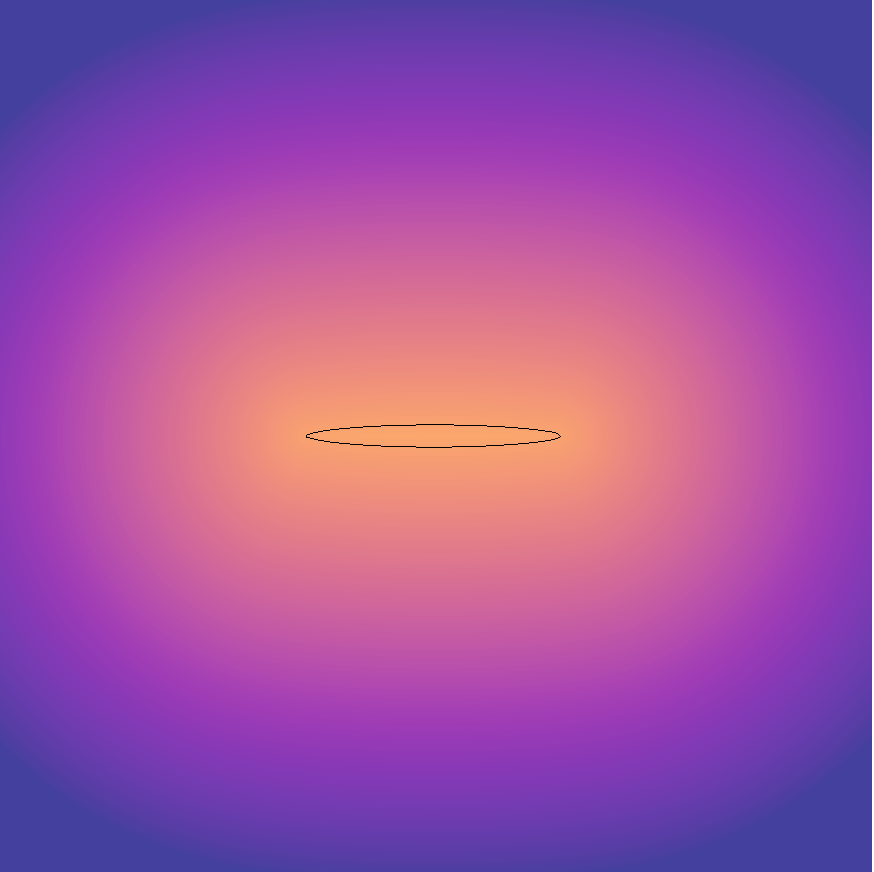} & \includegraphics[scale=0.12]{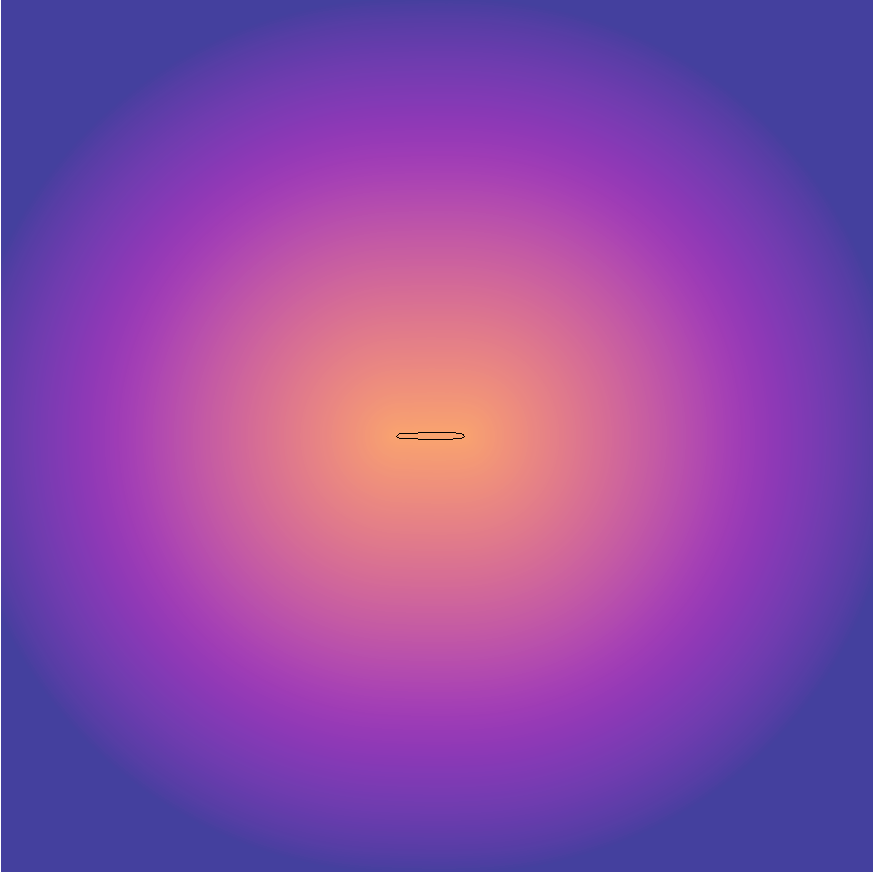} & \includegraphics[scale=0.2]{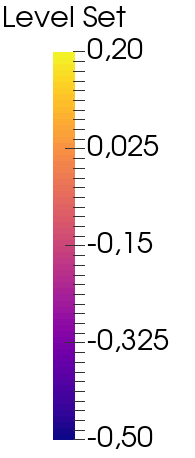}\\
          & \hspace{1.5cm}$t = 0$ & \hspace{1cm}$t = 5e-3$ & \hspace{1cm}$t = 1e-2$
  \end{tabular}
  \caption{Time evolution of the level set $\phi$ for the ellipse shrinkage test case for different ellipse ratios. The iso-zero value of the level-set field is in black. The mesh size is $h = 3e-3$ and the time step is $\Delta t = 5e-4$.}\label{fig:ratioMosaic}
\end{figure}

\begin{figure}
  \centering
  \begin{subfigure}{\textwidth}
    \centering
    \includegraphics[scale=0.3]{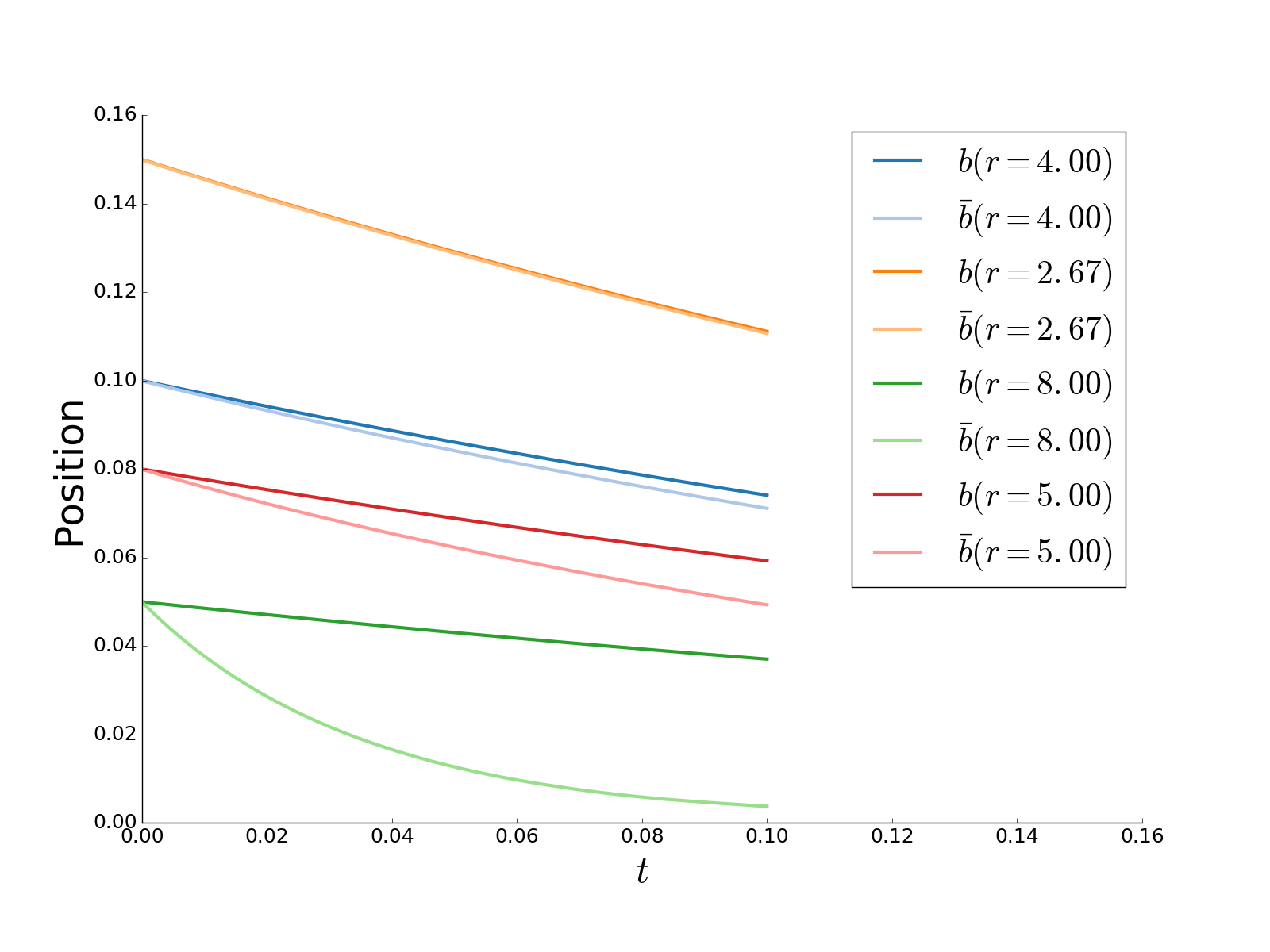}
    \caption{$b$ as a function of simulated time $t$}
  \end{subfigure}
  \begin{subfigure}{\textwidth}
    \centering
    \includegraphics[scale=0.3]{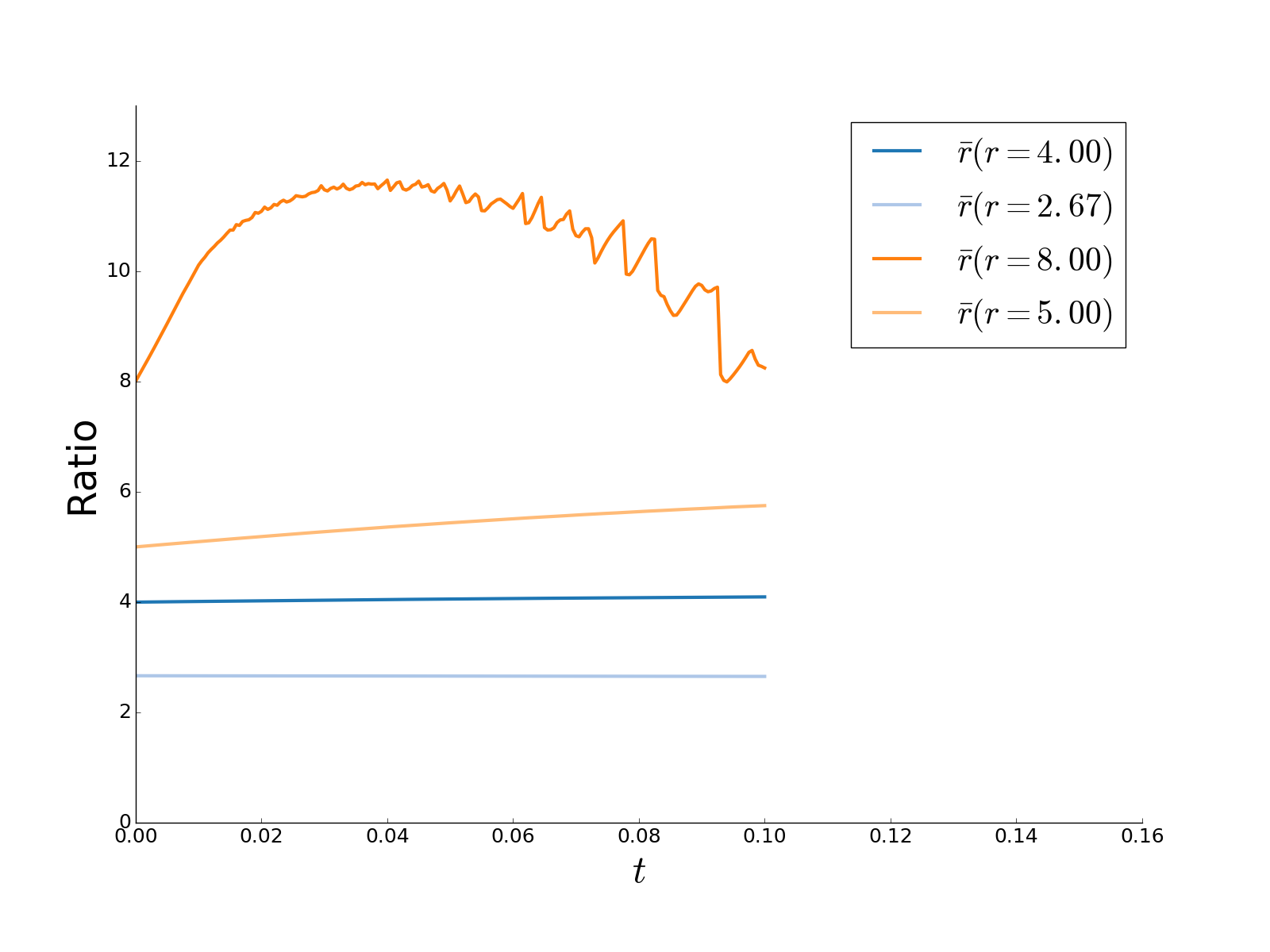}
    \caption{$r$ as a function of simulated time $t$}
  \end{subfigure}
  \caption{Sensitivity of the trajectory and measured ratio $\bar{r}$ to the initial ratio $r$ parameter study with $h = 3e-3$, $\Delta t = 5e-4$, and $a(t = 0) = 0.4$ on a $1 \times 1$ size mesh.}\label{fig:ellipseratiostudypos}
\end{figure}

\begin{figure}
  \centering
  \begin{subfigure}{\textwidth}
    \centering
    \includegraphics[scale=0.3]{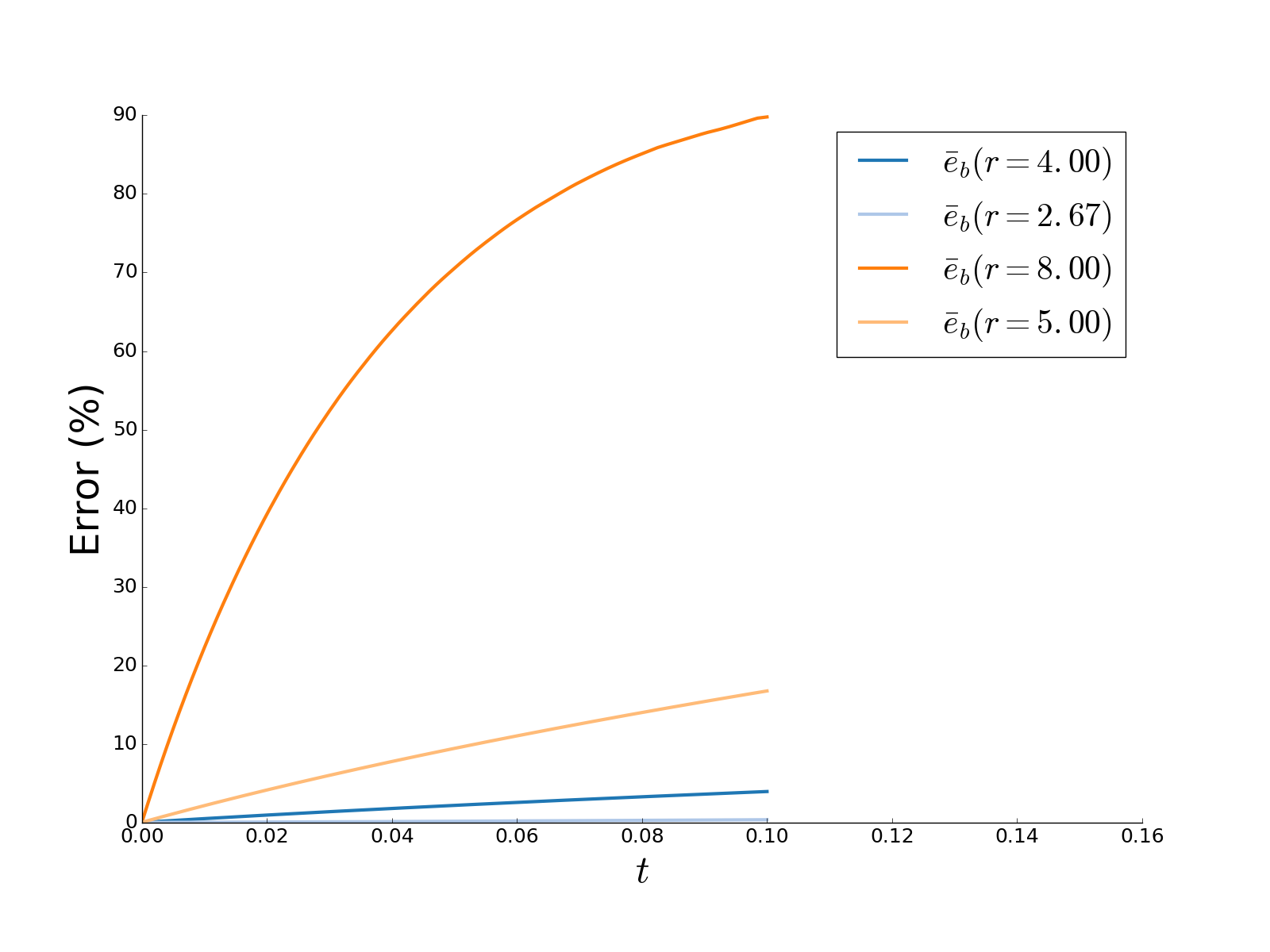}
    \caption{$e_{b}$ errors committed on the positions as a function of simulated time $t$}
  \end{subfigure}
  \begin{subfigure}{\textwidth}
    \centering
    \includegraphics[scale=0.3]{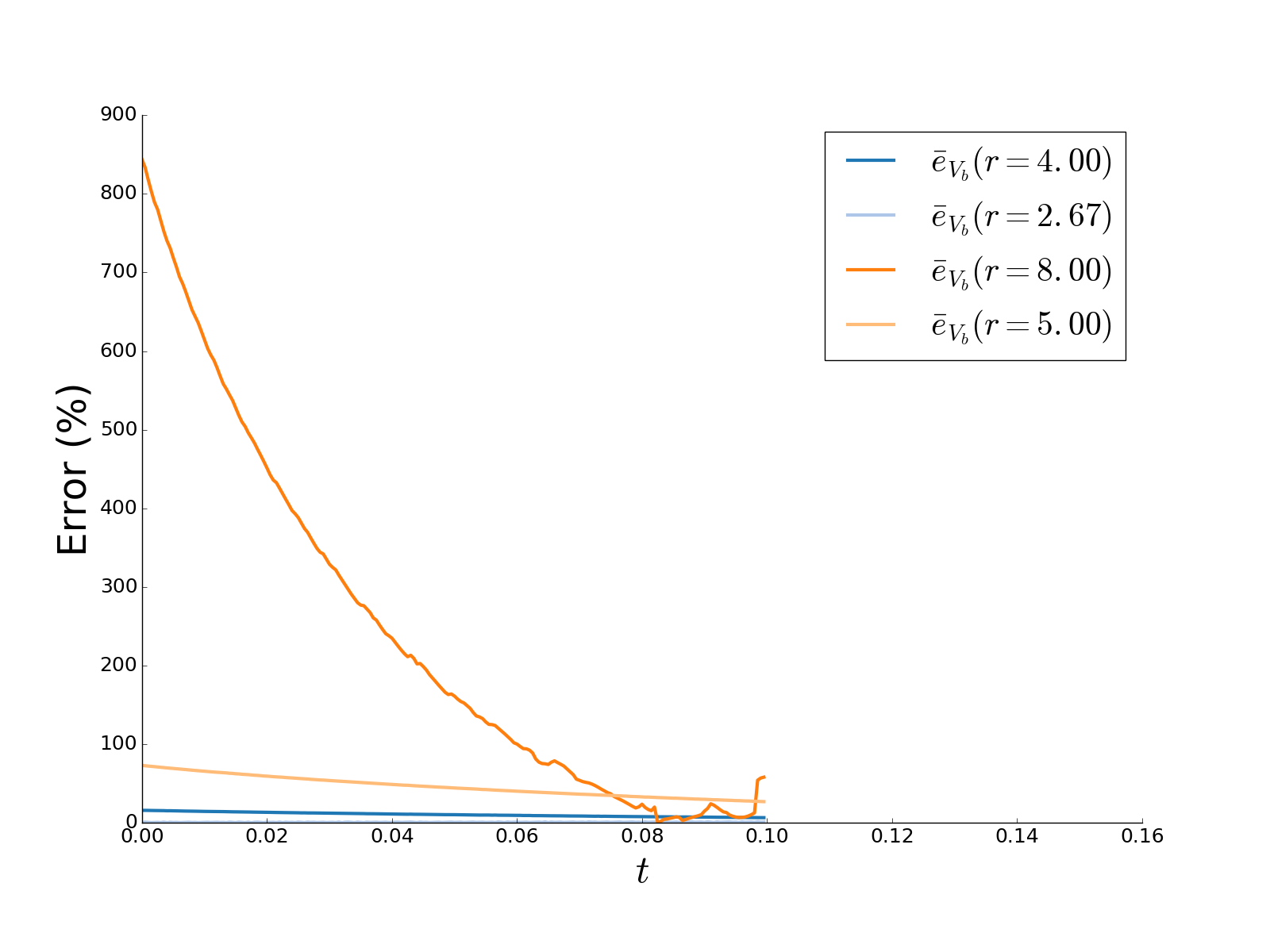}
    \caption{$e_{V_{b}}$ the errors committed on the velocities as a function of simulated time $t$}
  \end{subfigure}
  \caption{Sensitivity of the errors to the ellipse ratio $r$ parameter study with $h = 3e-3$, $\Delta t = 5e-4$ and $a(t = 0) = 0.4$ on a $1 \times 1$ size mesh.}\label{fig:ellipseratiostudyerrors}
\end{figure}

For ellipses with ratio $r=2$ one may expect the numerical formulation to give adequate approximations of the minimizing energy flow with a convergence rate of approximately $3$ in space and $1.5$ in time. However, one may remain dubious in terms of ellipses with even stronger axis ratios $r > 2$. Figures \ref{fig:ratioMosaic}, \ref{fig:ellipseratiostudypos} and \ref{fig:ellipseratiostudyerrors} report some results that have been obtained for $r = \{\frac{8}{3}, 4, 5, 8\}$ using $h = 3e-3$, $\Delta t = 5e-4$, $a(t = 0) = 0.4$ and a $1 \times 1$ domain.

While, in a qualitative sense, in Figure \ref{fig:ratioMosaic} the simulations give sensible results. For the ratios tested here, the level set fields remain elliptical while shrinking. However, in a quantitative sense, in Figures \ref{fig:ellipseratiostudypos} and \ref{fig:ellipseratiostudyerrors} one may observe that the errors committed during the simulation increase with increasing ellipse ratio $r$. Indeed, the mesh size used for these simulation is not sufficient to accurately describe the curvatures of the ellipses in the highest ratio cases. These simulations prove that in order to describe strong geometrical features and their evolution accurately, the mesh size must be sufficiently refined. The results could be greatly improved by using adaptive remeshing algorithms throughout the simulations to capture the strongest features of the geometry. In any case, the numerical parameters ($h, \Delta t$) must be adapted to the geometry of the problem in order to obtain sensible results.

Overall, the numerical formulation is adept at simulating the shrinking ellipse test case and converging towards the analytical solution when refining the discretization.

\subsection{A more general anisotropic case}

While no doubt relevant to the evaluation of the numerical formulation for the minimizing energy flow, the ellipse shrinkage case cannot truly distinguish between a velocity that does not include the anisotropic terms

\begin{align*}
  D_{iso}^{\alpha\beta} = \gamma m^{\alpha\beta}
\end{align*}

and one that does

\begin{align*}
  D_{aniso}^{\alpha\beta} = \gamma m^{\alpha\beta} + \dfrac{\partial^{2} \gamma}{\partial \tilde{\nabla}_{\alpha} \phi \partial \tilde{\nabla}_{\beta}\phi}
\end{align*}

even if one does compute a boundary energy density field that depends on the geometry $\gamma(\tilde{\nabla} \phi)$. This is because of equation \eqref{eq:frustratingCondition} where, for the boundary energy used in the ellipse shrinkage, case $D_{aniso} = 3D_{iso}$ which can be rectified in practice by a scaling of the mobility or of the time parameter. So, while the ellipse shrinkage case would differ by a factor of $3$ in comparing the cases, the geometry of the interface flow would be the same.

As such, in order to observe the added benefits of including the anisotropic term to the formulation, one may study a test case where the analytical solution is unknown but the anisotropic term modifies the velocity differently then the isotropic term. One may then compare simulations where the $D_{iso}$ is used to results where the $D_{aniso}$ is employed for the same boundary energy density functions $\gamma$ and the same initial geometries. 

\begin{figure}
  \centering
  \includegraphics[scale=0.3]{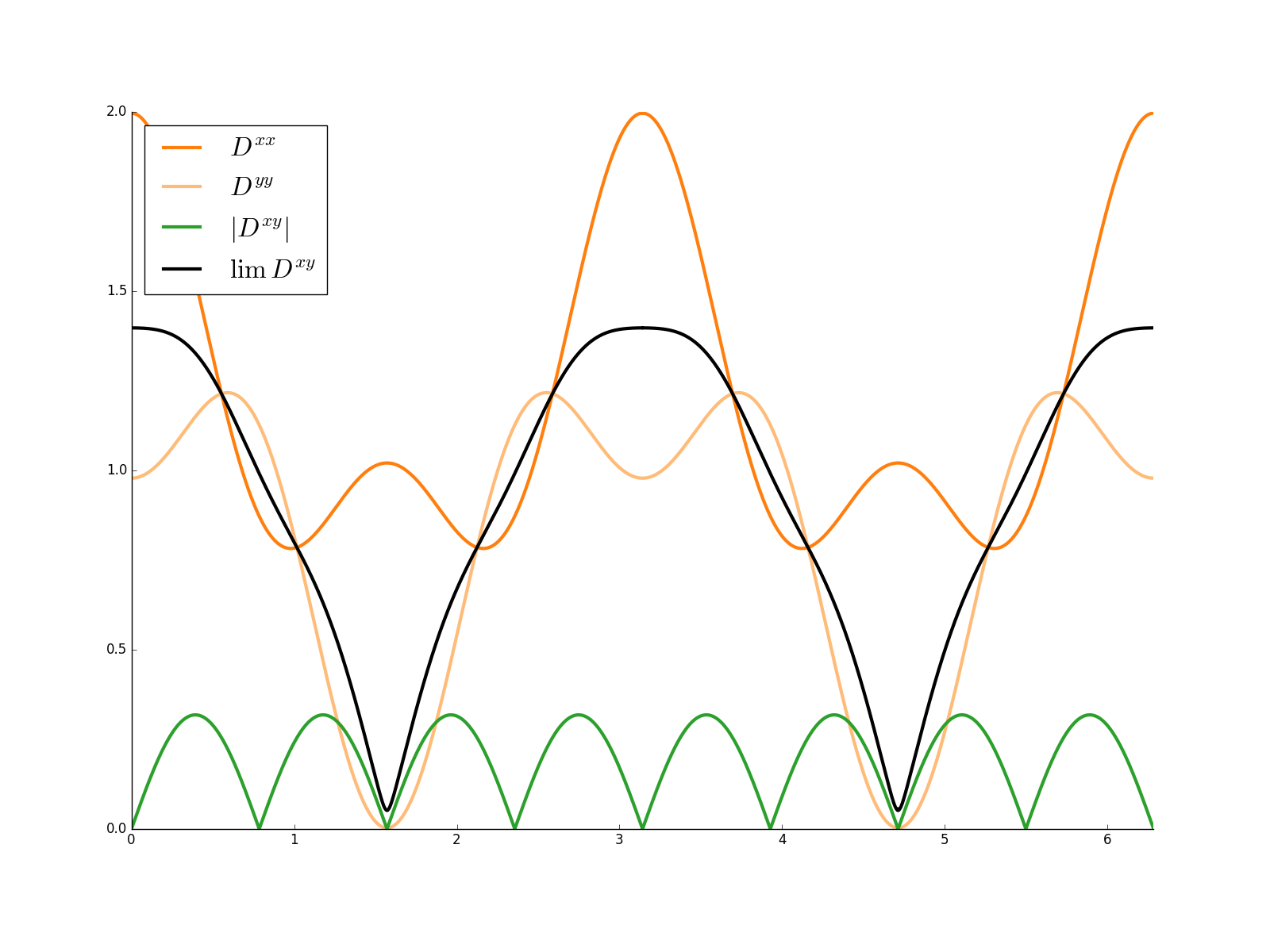}
  \caption{Components $D_{aniso}^{xx}$, $D_{aniso}^{yy}$ and $D_{aniso}^{xy}$ as a function of $\lambda \in [0, 2\pi]$. The limit expressed in the inequality \eqref{eq:dxy} is also shown for comparison as $\lim D^{xy}$.}\label{fig:Daniso}
\end{figure}

Considering, with $\cos(\lambda) = n^{x}$, 

\begin{align}
  \gamma(\lambda) = 1 + \dfrac{1}{377}(\cos(6\lambda)- 9\cos(2\lambda))
\end{align}

 one may show that the positive definiteness of the $D$ tensor is assured for any boundary. Figure \ref{fig:Daniso} illustrates the components of the $D_{aniso}$ tensor as a function of $\lambda$. Graphically, $D^{xy}_{aniso}$ is strictly inferior to the required limit.

\begin{figure}
  \centering
  \begin{subfigure}{0.48\textwidth}
    \centering
    \includegraphics[scale=0.13]{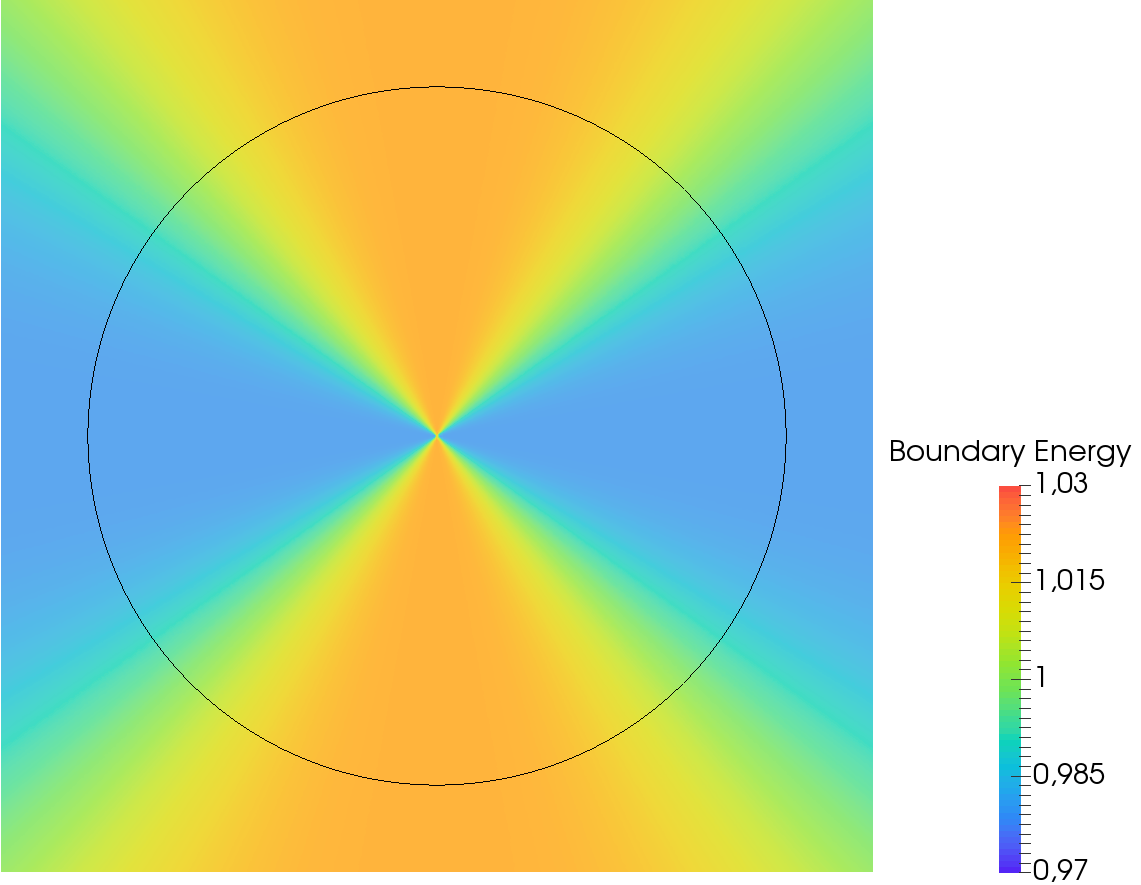}
    \caption{$\gamma$}
  \end{subfigure}
  \begin{subfigure}{0.48\textwidth}
    \centering
    \includegraphics[scale=0.13]{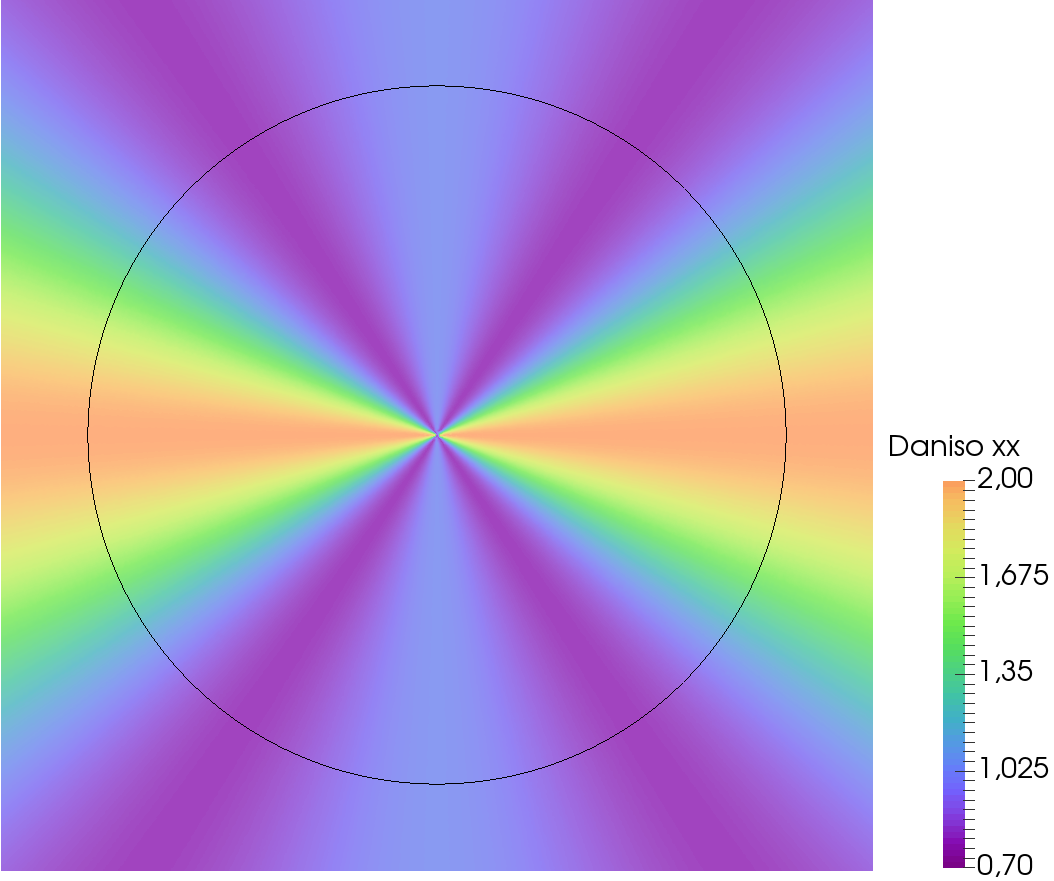}
    \caption{$D^{xx}_{aniso}$}
  \end{subfigure}
  \begin{subfigure}{0.48\textwidth}
    \centering
    \includegraphics[scale=0.13]{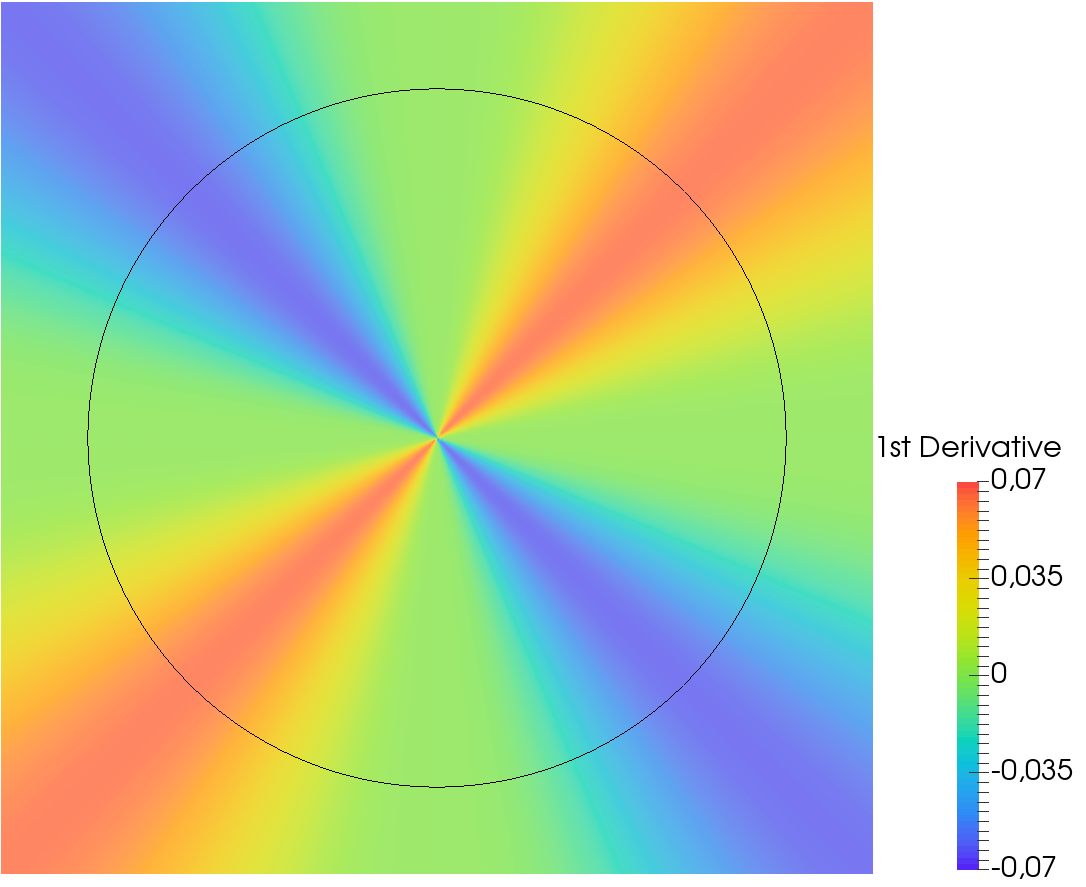}
    \caption{$\frac{\partial \gamma}{\partial \lambda}$}
  \end{subfigure}
  \begin{subfigure}{0.48\textwidth}
    \centering
    \includegraphics[scale=0.13]{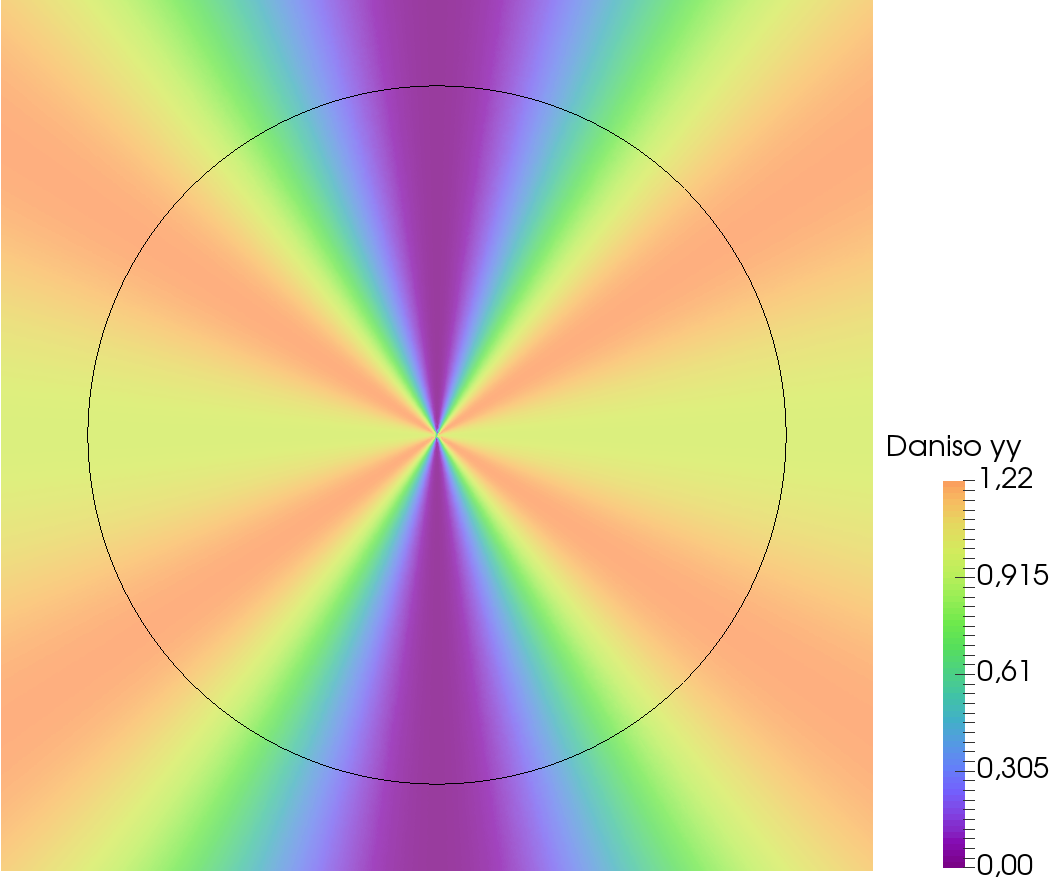}
    \caption{$D^{yy}_{aniso}$}
  \end{subfigure}
  \begin{subfigure}{0.48\textwidth}
    \centering
    \includegraphics[scale=0.13]{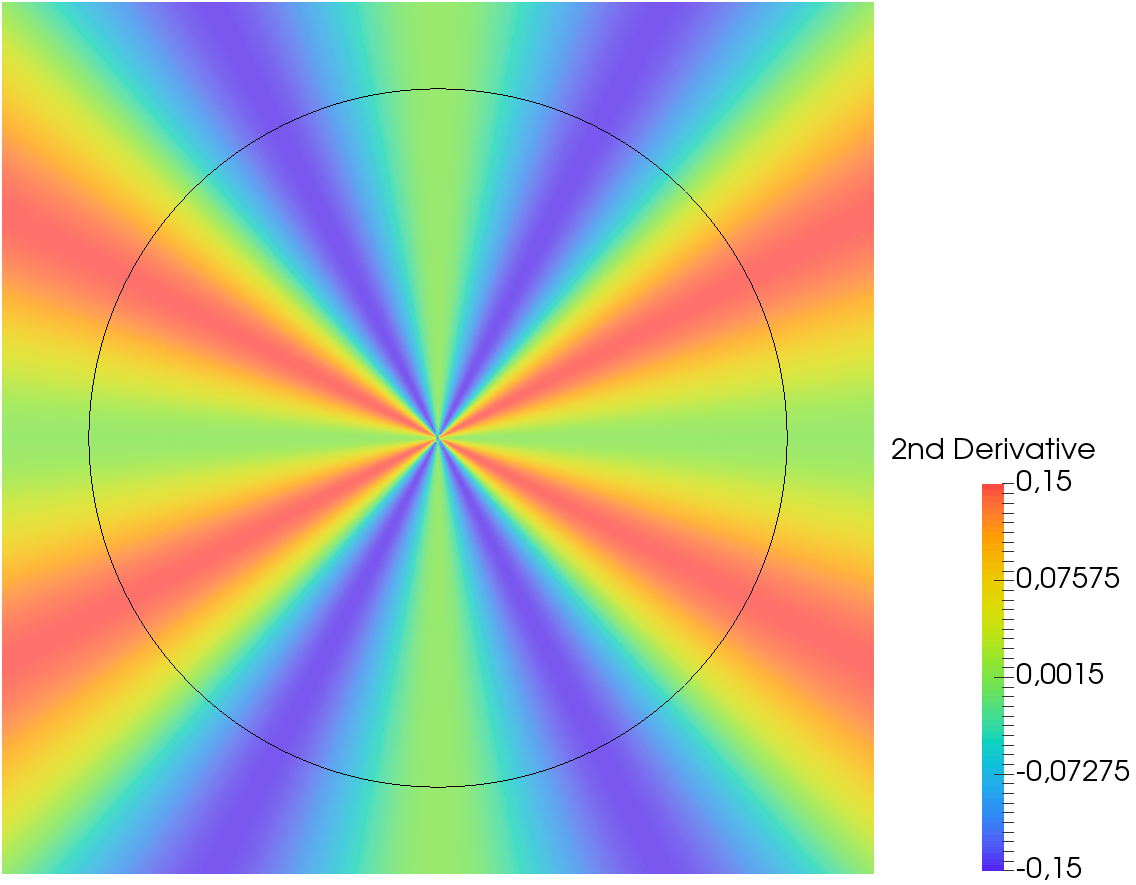}
    \caption{$\frac{\partial^{2} \gamma}{\partial \lambda^{2}}$}
  \end{subfigure}
  \begin{subfigure}{0.48\textwidth}
    \centering
    \includegraphics[scale=0.13]{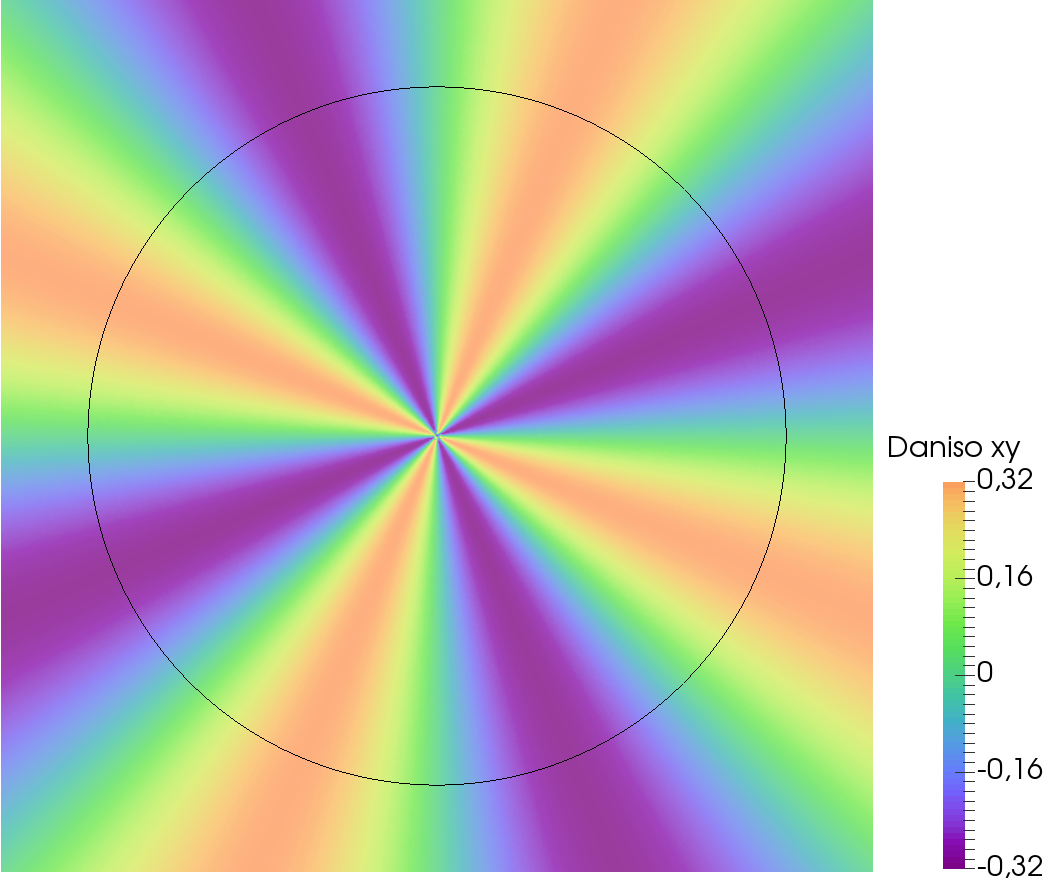}
    \caption{$D^{xy}_{aniso}$}
  \end{subfigure}
  \begin{subfigure}{0.48\textwidth}
    \centering
    \includegraphics[scale=0.13]{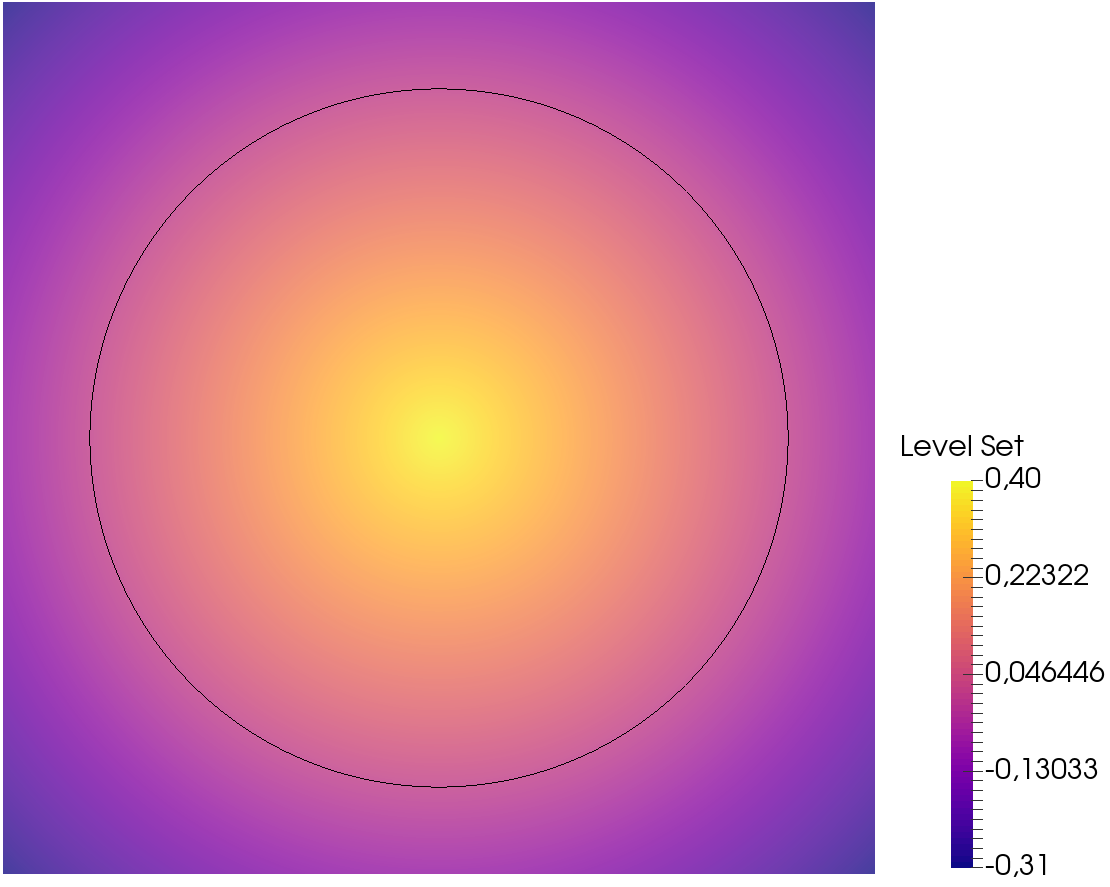}
    \caption{$\phi$}
  \end{subfigure}
  \caption{Initial values of the level set field $\phi$, boundary energy field $\gamma$, its derivatives and the components of $D_{aniso}$. The iso-zero value of the level set is in black.}\label{fig:anisoCircleInitial}
\end{figure}

Having the grain boundary energy function $\gamma$ and thus being able to calculate $D_{aniso}$, one may consider once again the circle $\mathcal{C} = ([0; 2\pi], \mathcal{O}_{C}, \mathcal{A}_{C})$ and the Riemannian manifold $\mathcal{M} = (\mathbb{R}^{2}, \mathcal{O}_{std}, \mathcal{A}_{std}, m)$. However, the initial embedding $\varphi$ is more direct

\begin{align*}
  \varphi :& [0; 2\pi] \longrightarrow \mathbb{R}^{2}\\
           & \theta \mapsto (R\cos\theta, R\sin\theta)
\end{align*}

where $R \in \mathbb{R}^{+}/\{0\}$ is the radius of the embedded circle. The initial conditions for both the level set field and the grain boundary energy field as well as its derivatives are represented in Figure \ref{fig:anisoCircleInitial} for $R = 0.4$.

\begin{figure}
  \centering
  \begin{tabular}{ >{\centering\arraybackslash}m{2cm} m{4cm}}
    $t=0$ & \includegraphics[scale=0.13]{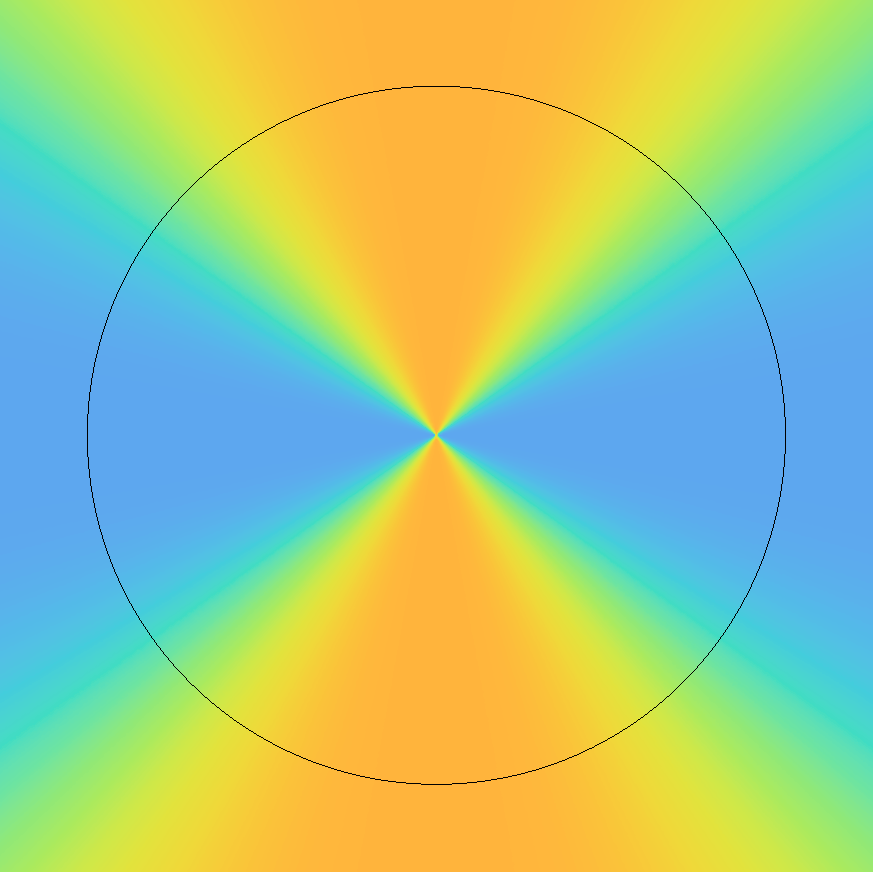}
  \end{tabular}
  \begin{tabular}{ >{\centering\arraybackslash}m{2cm} m{4cm} m{4cm}}
    $t = 2.5e-3$ & \includegraphics[scale=0.13]{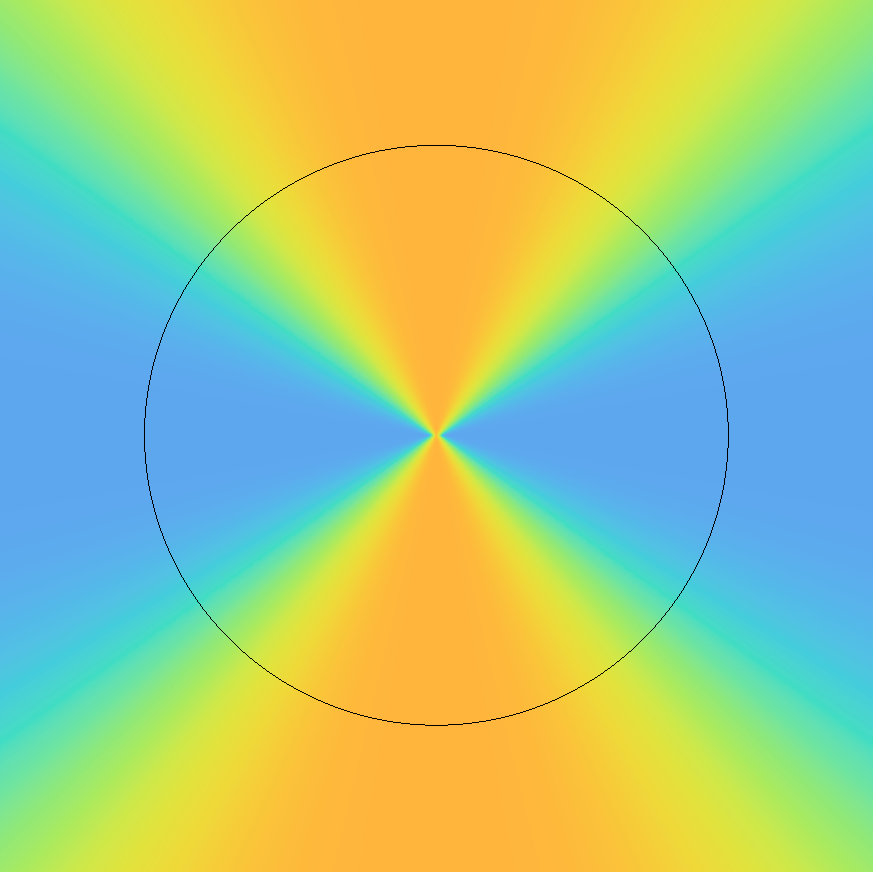} & \includegraphics[scale=0.13]{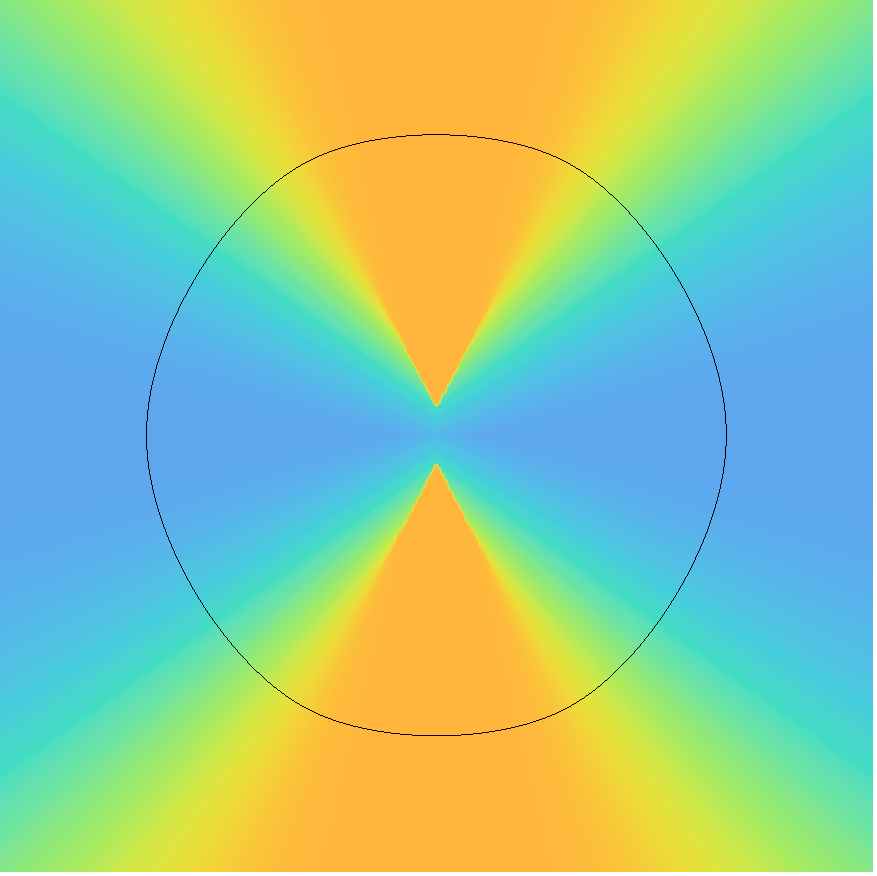} \\
    $t = 5e-3$ & \includegraphics[scale=0.13]{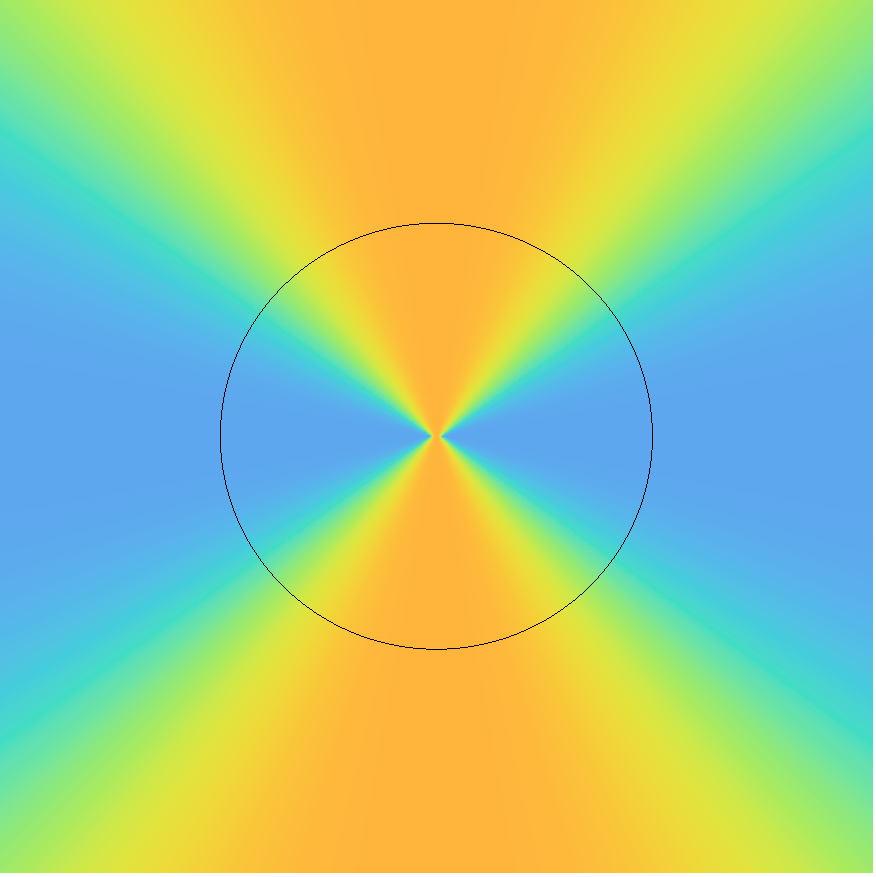} & \includegraphics[scale=0.13]{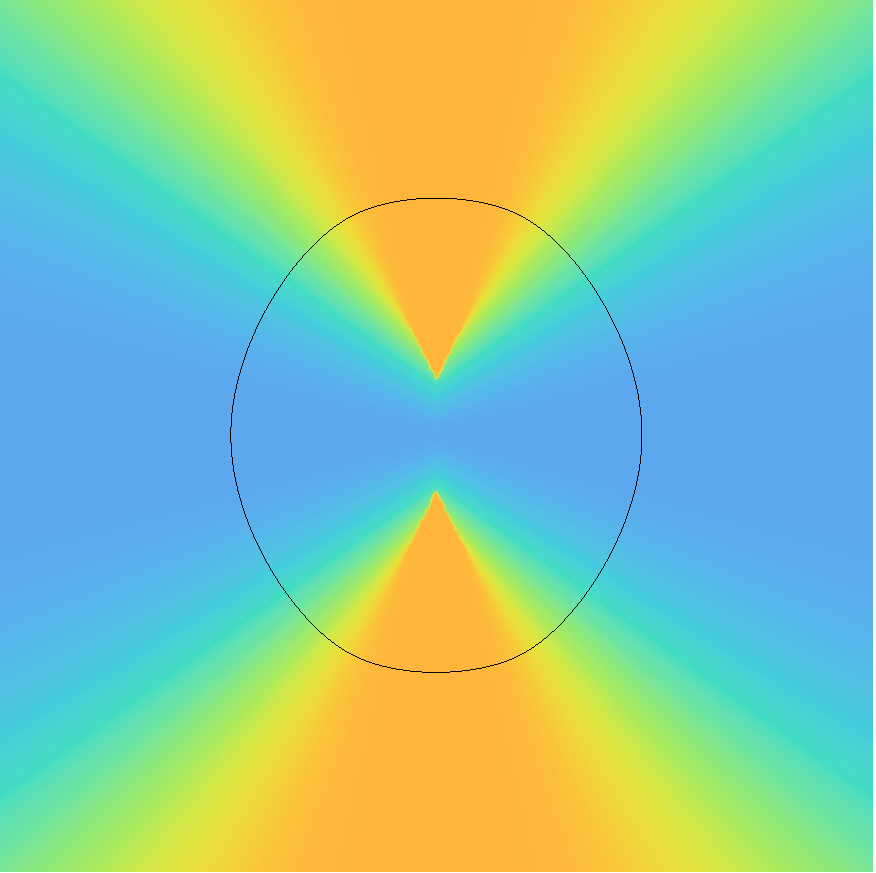} \\
    $t = 7.5e-3$ & \includegraphics[scale=0.13]{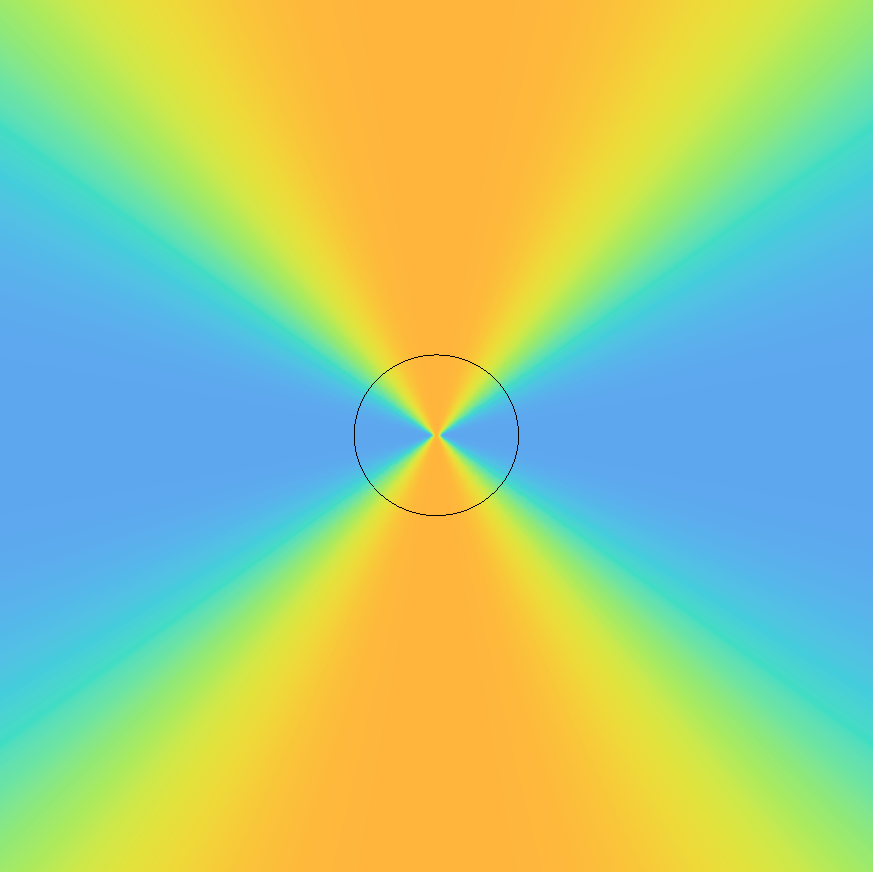} & \includegraphics[scale=0.13]{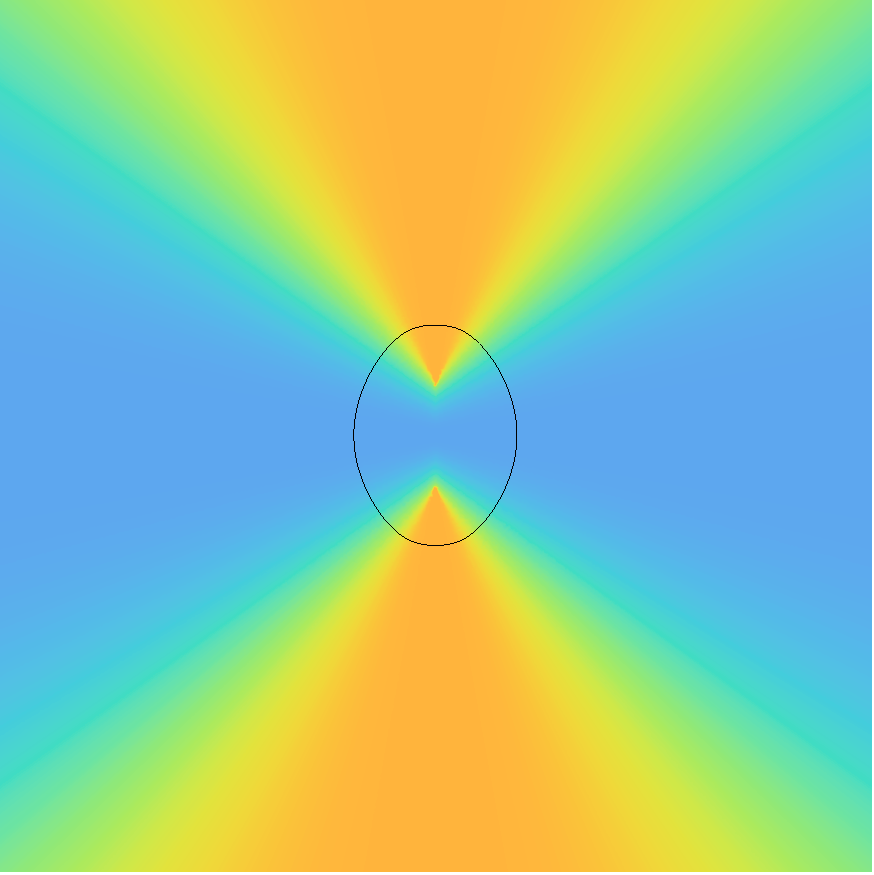}\\ 
                 &\hspace{1.5cm} $D_{iso}$ & \hspace{1.5cm}$D_{aniso}$
  \end{tabular}
  \begin{tabular}{ >{\centering\arraybackslash}m{2cm} m{9.5cm}}
    & \includegraphics[scale=0.3]{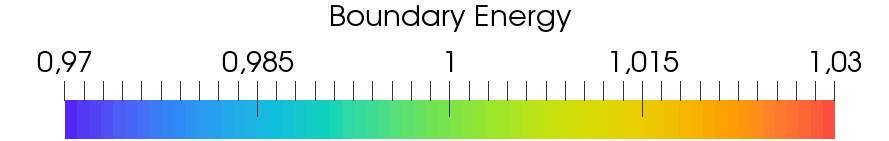}
  \end{tabular}
  \caption{Time evolution of the grain boundary energy field $\gamma$ and the iso-zero value of the level set for the circle shrinkage test case run with $D_{iso}$ and $D_{aniso}$. The iso-zero value of the level-set field is in black. The mesh size is $h = 3e-3$ and the time step is $\Delta t = 5e-4$.}\label{fig:DMosaic}
\end{figure}

The test case was run for both $D_{iso}$ and $D_{aniso}$ on a $1 \times 1$ size isotropic mesh with $h = 3e-3$ and $\Delta t = 5e-4$. The results of the form evolution of the circle as well as the evolution of the grain boundary energy field are presented in Figure \ref{fig:DMosaic}. The $D_{iso}$ and $D_{aniso}$ tensors generate very different boundary flows. While the $D_{iso}$ case tends to remain circular until disappearing, the $D_{aniso}$ case takes on a very distinctive form. The persistence of circularity of the $D_{iso}$ case is most likely due to the very small variations in the boundary energy of the order of only $3\%$.

However, the most efficient of the two simulations in terms of energy dissipation is thus the closer to reality since the principle of minimal action is in effect. Thus, the parameter of most relevance to comparing the two simulations is the energy efficiency of the geometry obtained in each step of the simulation, defined here as

\begin{align}
  \Lambda = \left(\dfrac{\int_{C}\gamma dC}{\int_{C}dC}\right)^{-1}\label{eq:efficiencyDef}.
\end{align}

with respect to the smooth manifold $\mathcal{C}$.

\begin{figure}
  \centering
  \includegraphics[scale=0.4]{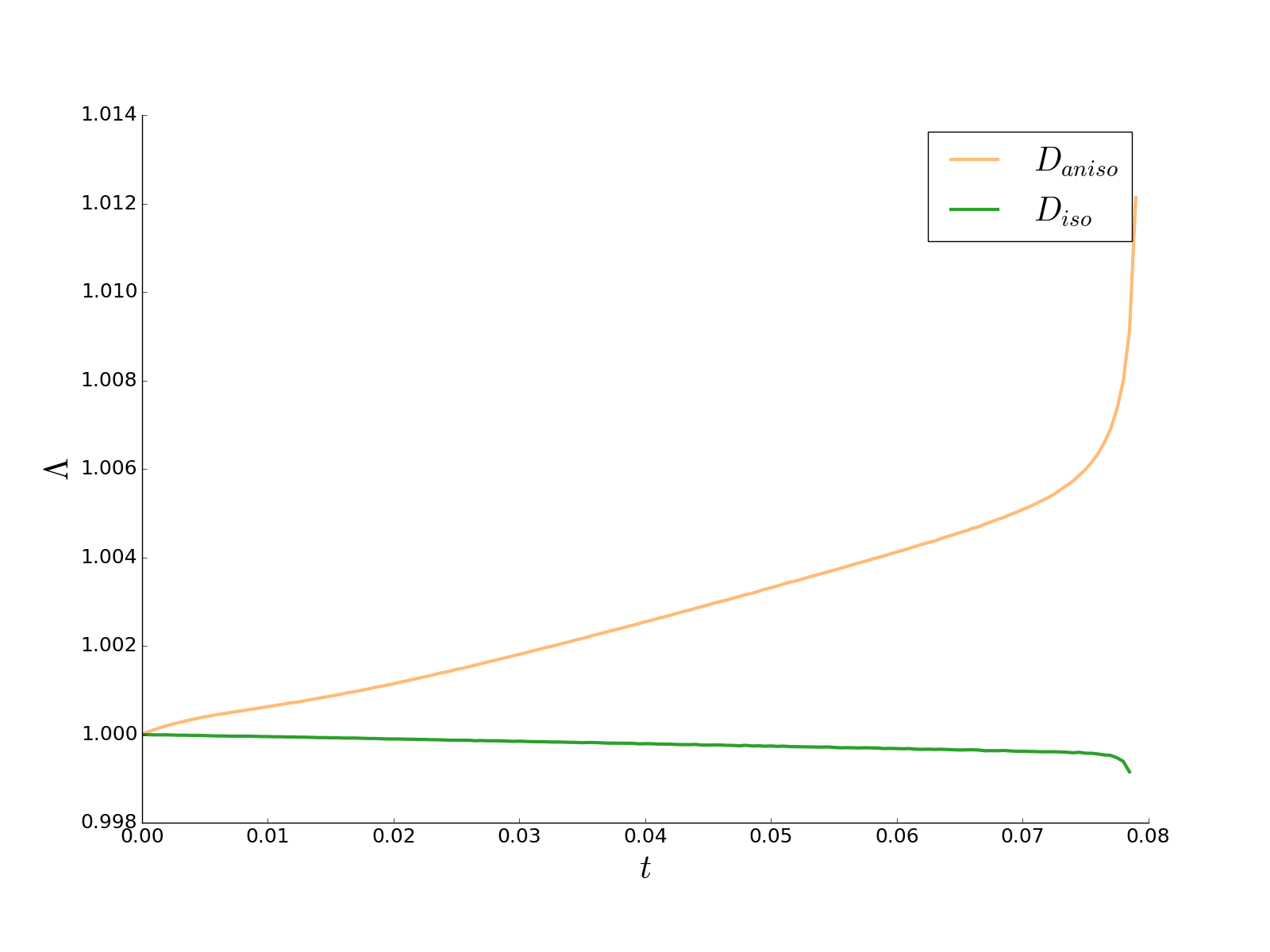}
  \caption{Computed energy efficiency $\Lambda$ as a function of time $t$ for circle shrinkage test cases run with $D_{iso}$ and $D_{aniso}$.}\label{fig:AnisoEefficiency}
\end{figure}

Figure \ref{fig:AnisoEefficiency} shows the evolution of the computed energy efficiency $\Lambda$ for both simulations. Clearly, the energy efficiency of the form developed by the $D_{aniso}$ flow is better than that of the $D_{iso}$ flow from the start of the simulation to the disappearance of the boundary. While not being a direct proof of the validity of the $D_{aniso}$ formulation, these test cases show that the full $D_{aniso}$ formulation is definitely more adept then the $D_{iso}$ formulation for the minimizing energy flow problem.

\section{Conclusions}

This work has contributed to developing a framework for simulating anisotropic grain growth. By studying the anisotropic energy density one boundary problem the authors have managed to give an expression for the velocity field of a migrating interface. An anisotropic analytical benchmark case based on a shrinking ellipse has been proposed. Using a more general energy density in a circle configuration the authors have shown the improvement that the current formalism delivers as compared to the classical formulations. To the authors' knowledge, no other investigations have treated this issue in such an applied setting.

Future studies will be dedicated to generalizing this formalism to the polycrystal case where multiple junctions may be found in great number. Also, given the dimensionless nature of the mathematical framework, one may attempt to generate a 3D analogue to the ellipse case and using the same numerical model. This new model may also serve as a foundation for including an anisotropic tensorial mobility value into the calculations by way of the base manifold's metric tensor $m$. Moving away from the level-set method, the formalism may also be applied to front tracking approaches \cite{Florez2020}. Additionally, the conditions expressed in the inequality \eqref{eq:pdcondition} seem to be more general and could possibly be used to discriminate between anisotropic grain boundary energy densities that can be found in the literature. Finally, the methodology used here was applied to the dynamics of grain boundaries but can be generalized to studies with arbitrarily energetic interfaces, such as in fluid dynamics, where the energy density of interfaces may depend on temperature or other parameters exhibiting spatial gradients giving rise to Marangoni effects.

\section*{Acknowledgements}
The authors thank the SAFRAN company and the ANR for their financial support through the OPALE ANR industrial Chair (ANR-14-CHIN-0002). The authors would also like to thank the ArcelorMittal, ASCOMETAL, AUBERT \& DUVAL, CEA, SAFRAN, FRAMATOME, TIMET, Constellium and TRANSVALOR companies for their financial support through the DIGIMU consortium and ANR industrial Chair (ANR-16-CHIN-0001).

\section*{Data availability}
The raw data required to reproduce these findings cannot be shared at this time as the data also forms part of an ongoing study. The processed data required to reproduce these findings cannot be shared at this time as the data also forms part of an ongoing study.

\appendix
\section{Level-set setting for interface dynamics}
\label{app:Mfds2LS}

The second term on the transport equation \eqref{eq:bettertransport} can be expanded using the definition in equation \eqref{eq:anisovector}
\begin{align*}
  A^{\alpha}\tilde{\nabla}_{\alpha}\phi 
  &= m^{\alpha\beta}\nabla_{i}\left(\dfrac{\partial \gamma}{\partial \nabla_{i}\varphi^{\beta}} + \gamma g^{iq}m_{\sigma\beta}\nabla_{q}\varphi^{\sigma}\right)\tilde{\nabla}_{\alpha}\phi\\
  &= m^{\alpha\beta}\left(\dfrac{\partial^{2} \gamma}{\partial \nabla_{j}\varphi^{\zeta}\partial \nabla_{i}\varphi^{\beta}}\nabla_{i}\nabla_{j}\varphi^{\zeta} + \nabla_{i}(\gamma g^{iq}m_{\sigma\beta}\nabla_{q}\varphi^{\sigma})\right)\tilde{\nabla}_{\alpha}\phi
\end{align*}
simplifying this equation one obtains 

\begin{align}
  A^{\alpha}\tilde{\nabla}_{\alpha}\phi = m^{\alpha\beta}\dfrac{\partial^{2} \gamma}{\partial \nabla_{j}\varphi^{\zeta}\partial \nabla_{i}\varphi^{\beta}}\nabla_{i}\nabla_{j}\varphi^{\zeta}\tilde{\nabla}_{\alpha}\phi + g^{iq}\nabla_{i}\gamma\nabla_{q}\varphi^{\alpha}\tilde{\nabla}_{\alpha}\phi - \gamma P^{\alpha\beta}\tilde{\nabla}_{\beta}\tilde{\nabla}_{\alpha}\phi
\label{eq:anisovectorxgradphi}
\end{align}

being $P \in \Gamma(T_{0}^{2}\mathcal{M}|_{\varphi(S)})$ a tangential projection tensor field 

\begin{gather}
  P^{\alpha\beta} = g^{ij}\nabla_{j}\varphi^{\alpha}\nabla_{i}\varphi^{\beta} = m^{\alpha\beta} - n^{\alpha}n^{\beta} \label{eq:projector}
\end{gather}

in addition, the second derivative term in equation~\eqref{eq:anisovector} can be reduced to 

\begin{align*}
  \dfrac{\partial \phi}{\partial t} + \mu\left(-\left(\dfrac{\partial^{2} \gamma}{\partial \tilde{\nabla}_{\beta}\phi \partial \tilde{\nabla}_{\alpha} \phi}  + \dfrac{\partial \gamma}{\partial \tilde{\nabla}_{\alpha}\phi}m^{\kappa\beta}\tilde{\nabla}_{\kappa}\phi\right)P^{\xi}_{\beta}P^{\sigma}_{\alpha}\tilde{\nabla}_{\sigma}\tilde{\nabla}_{\xi}\phi + P^{\alpha\beta}\tilde{\nabla}_{\beta}\gamma\tilde{\nabla}_{\alpha}\phi - \gamma P^{\alpha\beta}\tilde{\nabla}_{\alpha}\tilde{\nabla}_{\beta}\phi\right) = 0
\end{align*}
Considering $ n^{\alpha}\tilde{\nabla}_{\alpha}\tilde{\nabla}_{\beta}\phi = 0 $, the terms that involve a contraction between the tangential projection tensor field and the second derivative of the level set can be redefined as

\begin{align*}
  P_{\alpha}^{\sigma}\tilde{\nabla}_{\sigma}\tilde{\nabla}_{\xi}\phi &= (\delta_{\alpha}^{\sigma} - n^{\sigma}\nabla_{\alpha}\phi)\tilde{\nabla}_{\sigma}\tilde{\nabla}_{\xi}\phi\\
                                                                     &= \tilde{\nabla}_{\alpha}\tilde{\nabla}_{\xi}\phi
\end{align*}

and

\begin{align*}
  P^{\alpha\beta}\tilde{\nabla}_{\alpha}\tilde{\nabla}_{\beta}\phi &= m^{\alpha\beta}\tilde{\nabla}_{\alpha}\tilde{\nabla}_{\beta}\phi\\
                                                                   &= \Delta \phi
\end{align*}
being $\Delta$ the classical Laplacian operator in $\mathcal{M}$.  As such, one obtains the simplified level set transport equation

\begin{align}
  \dfrac{\partial \phi}{\partial t} + -\mu\left(\gamma m^{\alpha\beta} + \dfrac{\partial^{2} \gamma}{\partial \tilde{\nabla}_{\beta}\phi \partial \tilde{\nabla}_{\alpha} \phi}\right)\tilde{\nabla}_{\alpha}\tilde{\nabla}_{\beta}\phi = 0
\end{align}

\bibliographystyle{ieeetr}
\bibliography{Manuscript}

\end{document}